\documentclass[twocolumn,secnumarabic,amssymb, nobibnotes, aps, prl, superscriptaddress, nobalancelastpage,longbibliography,nofootinbib]{revtex4-2}
\usepackage{graphicx}
\usepackage{dcolumn}
\usepackage{bm}
\usepackage{siunitx}
\usepackage{xcolor}
\usepackage{placeins}
\usepackage{nicefrac}
\usepackage[final]{changes}
\usepackage{xr-hyper}
\usepackage{hyperref}
\hypersetup{breaklinks=true,colorlinks=true,linkcolor=blue,citecolor=blue,filecolor=magenta,urlcolor=cyan}

\graphicspath{ {Graphics/} }

\newcommand{\dmscr}{\ensuremath{\delta\!\left\langle r^2_{\rm c} \right \rangle}}

\def\orcid#1{\kern .08em\href{https://orcid.org/#1}{\includegraphics[keepaspectratio,width=0.7em]{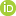}}}

\makeatletter
\def\@bibdataout@aps{%
\immediate\write\@bibdataout{%
@CONTROL{%
apsrev41Control%
\longbibliography@sw{%
    ,author="08",editor="1",pages="1",title="0",year="1"%
    }{%
    ,author="08",editor="1",pages="1",title="",year="1"%
    }%
  }%
}%
\if@filesw \immediate \write \@auxout {\string \citation {apsrev41Control}}\fi
}
\makeatother

\begin{document}

\preprint{APS/123-QED}

\title{Charge radii of $^{55,56}$Ni reveal a surprisingly similar behavior at $N=28$ in Ca and Ni isotopes}

\author{Felix Sommer\orcid{0000-0001-5084-0284}}
\email{fsommer@ikp.tu-darmstadt.de}
\affiliation{Institut f\"ur Kernphysik, Technische Universit\"at Darmstadt, 64289 Darmstadt, Germany}%

\author{Kristian K\"onig\orcid{0000-0001-9415-3208}}%
\affiliation{National Superconducting Cyclotron Laboratory, Michigan State University, East Lansing, Michigan 48824, USA}%

\author{Dominic M. Rossi\orcid{0000-0002-2461-0618}}
\affiliation{Institut f\"ur Kernphysik, Technische Universit\"at Darmstadt, 64289 Darmstadt, Germany}%
\affiliation{GSI Helmholtzzentrum f\"ur Schwerionenforschung GmbH, Planckstr. 1, 64291 Darmstadt, Germany}%

\author{Nathan Everett}%
\affiliation{National Superconducting Cyclotron Laboratory, Michigan State University, East Lansing, Michigan 48824, USA}%
\affiliation{Department of Physics and Astronomy, Michigan State University, East Lansing, Michigan 48824, USA}%

\author{David Garand}%
\affiliation{National Superconducting Cyclotron Laboratory, Michigan State University, East Lansing, Michigan 48824, USA}%

\author{Ruben P. de Groote}%
\affiliation{Department of Physics, University of Jyv\"askyl\"a, Survontie 9, Jyv\"askyl\"a, FI-40014, Finland}%

\author{Jason~D.~Holt\orcid{0000-0003-4833-7959}}%
\affiliation{TRIUMF 4004 Wesbrook Mall, Vancouver BC V6T 2A3, Canada}%
\affiliation{Department of Physics, McGill University, Montr\'eal, QC H3A 2T8, Canada}%

\author{Phillip Imgram}
\affiliation{Institut f\"ur Kernphysik, Technische Universit\"at Darmstadt, 64289 Darmstadt, Germany}%

\author{Anthony Incorvati}
\affiliation{National Superconducting Cyclotron Laboratory, Michigan State University, East Lansing, Michigan 48824, USA}%
\affiliation{Department of Physics and Astronomy, Michigan State University, East Lansing, Michigan 48824, USA}%

\author{Colton Kalman}%
\affiliation{National Superconducting Cyclotron Laboratory, Michigan State University, East Lansing, Michigan 48824, USA}%
\affiliation{Department of Chemistry, Michigan State University, East Lansing, Michigan 48824, USA}%

\author{Andrew Klose}%
\affiliation{Department of Chemistry, Augustana University, Sioux Falls, South Dakota 57197, USA}%

\author{Jeremy Lantis \orcid{0000-0002-8257-7852}}
\affiliation{National Superconducting Cyclotron Laboratory, Michigan State University, East Lansing, Michigan 48824, USA}%
\affiliation{Department of Chemistry, Michigan State University, East Lansing, Michigan 48824, USA}%

\author{Yuan Liu}%
\affiliation{Physics Division, Oak Ridge National Laboratory, Oak Ridge, Tennessee 37831, USA}%

\author{Andrew J. Miller}%
\affiliation{National Superconducting Cyclotron Laboratory, Michigan State University, East Lansing, Michigan 48824, USA}%
\affiliation{Department of Physics and Astronomy, Michigan State University, East Lansing, Michigan 48824, USA}%

\author{Kei Minamisono\orcid{0000-0003-2315-5032}}
\email{minamisono@nscl.msu.edu}
\affiliation{National Superconducting Cyclotron Laboratory, Michigan State University, East Lansing, Michigan 48824, USA}%
\affiliation{Department of Physics and Astronomy, Michigan State University, East Lansing, Michigan 48824, USA}%

\author{Takayuki Miyagi\orcid{0000-0002-6529-4164}}%
\affiliation{Institut f\"ur Kernphysik, Technische Universit\"at Darmstadt, 64289 Darmstadt, Germany}
\affiliation{ExtreMe Matter Institute EMMI, GSI Helmholtzzentrum f\"ur Schwerionenforschung GmbH, D-64291 Darmstadt, Germany}
\affiliation{TRIUMF 4004 Wesbrook Mall, Vancouver BC V6T 2A3, Canada}%

 \author{Witold Nazarewicz\orcid{0000-0002-8084-7425}}
 \affiliation{Facility for Rare Isotope Beams, Michigan State University, East Lansing, Michigan 48824, USA}
 \affiliation{Department of Physics and Astronomy, Michigan State University, East Lansing, Michigan 48824, USA}
 
\author{Wilfried N\"ortersh\"auser\orcid{0000-0001-7432-3687}}
\email{wnoertershaeuser@ikp.tu-darmstadt.de}
\affiliation{Institut f\"ur Kernphysik, Technische Universit\"at Darmstadt, 64289 Darmstadt, Germany}%
\affiliation{Helmholtz Research Academy Hesse for FAIR, Campus Darmstadt, 64289 Darmstadt, Germany}

\author{\firstname{Skyy V.} Pineda\orcid{0000-0003-1714-4628}}
\affiliation{National Superconducting Cyclotron Laboratory, Michigan State University, East Lansing, Michigan 48824, USA}%
\affiliation{Department of Chemistry, Michigan State University, East Lansing, Michigan 48824, USA}%

\author{Robert Powel}
\affiliation{National Superconducting Cyclotron Laboratory, Michigan State University, East Lansing, Michigan 48824, USA}%
\affiliation{Department of Physics and Astronomy, Michigan State University, East Lansing, Michigan 48824, USA}%

\author{Paul-Gerhard~Reinhard}
\affiliation{Institut f\"ur Theoretische Physik II, Universit\"at Erlangen-N\"urnberg, 91058 Erlangen, Germany}

\author{Laura Renth}
\affiliation{Institut f\"ur Kernphysik, Technische Universit\"at Darmstadt, 64289 Darmstadt, Germany}%

\author{Elisa Romero-Romero}%
\affiliation{Physics Division, Oak Ridge National Laboratory, Oak Ridge, Tennessee 37831, USA}%
\affiliation{Department of Physics and Astronomy, University of Tennessee, Knoxville, Knoxville, Tennessee 37996, USA}

\author{Robert Roth}%
\affiliation{Institut f\"ur Kernphysik, Technische Universit\"at Darmstadt, 64289 Darmstadt, Germany}
\affiliation{Helmholtz Research Academy Hesse for FAIR, Campus Darmstadt, 64289 Darmstadt, Germany}

\author{Achim~Schwenk\orcid{0000-0001-8027-4076}}
\affiliation{Institut f\"ur Kernphysik, Technische Universit\"at Darmstadt, 64289 Darmstadt, Germany}
\affiliation{ExtreMe Matter Institute EMMI, GSI Helmholtzzentrum f\"ur Schwerionenforschung GmbH, D-64291 Darmstadt, Germany}
\affiliation{Max-Planck-Institut f\"ur Kernphysik, D-69117 Heidelberg, Germany}

\author{Chandana Sumithrarachchi}
\affiliation{National Superconducting Cyclotron Laboratory, Michigan State University, East Lansing, Michigan 48824, USA}%

\author{Andrea Teigelh\"ofer}
\affiliation{TRIUMF 4004 Wesbrook Mall, Vancouver BC V6T 2A3, Canada}%

\date{\today}

\begin{abstract}
Nuclear charge radii of $^{55,56}$Ni were measured by collinear laser spectroscopy. The obtained information completes the behavior of the charge radii at the shell closure of the doubly magic nucleus $^{56}$Ni. The trend of charge radii across the shell closures in calcium and nickel is surprisingly similar 
despite the fact that the $^{56}$Ni core is supposed to be much softer than the $^{48}$Ca core. The very low magnetic moment $\mu(^{55}\text{Ni})=\SI{-1.108(20)}{\mu_N}$ indicates the impact of M1 excitations between spin-orbit partners across the $N,Z=28$ shell gaps.   Our charge-radii results are compared to \textit{ab initio} and nuclear density functional theory calculations, showing good agreement within theoretical uncertainties.
\end{abstract}

\maketitle

{\it Introduction. ---}
After seventy years, the concept of closed nuclear shells of protons and neutrons at so-called magic numbers is still a backbone of nuclear structure theory. The traditional magic numbers are based on properties of nuclei at or close to the valley of $\beta$-stability. With excursions into the exotic regions of the nuclear landscape, a modern understanding of magic numbers has been established. The evolution of shell gap sizes can lead to dramatic modifications of  magic numbers in isotopes with extreme neutron-to-proton ratios \cite{Dobaczewski2007,Sorlin.2008,Otsuka.2020}.

One of the fingerprints of a shell closure is a characteristic kink in the trend of charge radii along an isotopic chain. The origin of this kink and its relation to the strength of a shell closure is, however, still under debate \cite{Gorges.2019,DayGoodacre.2021,Goodacre.2021b,Reinhard.2021,Perera.2021}. Kinks in charge radii have been observed at all neutron shell closures for which data are available with the exception of the $N=20$ neutron shell closure, where it has been studied so far only for Ar, K and Ca \cite{Klein.1996,Rossi.2015,Miller.2019}. 
While $N=32$ in the Ca region has been proposed to become a magic number based on the observations of a sudden decrease in their binding energy beyond $N=32$ \cite{Wienholtz.2013,Rosenbusch.2015} and the high excitation energy of the first excited state in $^{52}$Ca \cite{Gade.2006}, this is not supported by the behavior of the charge radii in K across $N=32$ and binding energies \cite{Koszorus.2021}. Indeed, $N=32$ seems to be consistent with a local neutron sub-shell closure.

A comparison of the change in mean-square charge radius, \dmscr, across a neutron shell closure for several isotones reveals a remarkable similarity for the neutron shell closures 
at $N=28$,\,50,\,82, and 126.   \cite{Garcia.2020,Perera.2021}. The evolution of \dmscr\ above $N=28$ is already established for K, Ca, Mn and Fe isotopes \cite{Koszorus.2021,GarciaRuiz.2016,Heylen.2016b,Minamisono.2016} and are indeed very similar \cite{Kortelainen2022}. A measurement of the charge radius of $^{56}$Ni  provides essential data to study trends in $\dmscr$  for two doubly magic nuclei with the same neutron magic gap, of which the neutron-rich $^{48}$Ca is known to have  a fairly strong $N=28$ shell closure \cite{Otsuka.1998}. In contrast, the neutron-deficient $^{56}$Ni is believed to be a rather soft core because of its high $B(E2)$ value \cite{Kraus.1994,Otsuka.1998,Arnswald2021} and the nuclear magnetic moments of neighboring isotopes \cite{Berryman.2009,Callaghan.1973,Cocolios.2009,Ohtsubo.1996,Honma.2004,Vingerhoets.2010,Vingerhoets.2011} which are inconsistent with single-particle estimates. 
In fact, the measured  $B(E2; 2^+_1\rightarrow 0^+_1)$ value in $^{48}$Ca (1.7 W.u.) is significantly below that in $^{56}$Ni (7.1 W.u.) \cite{Pritychenko.2016}. 
The different nature of the proton shell-closure in Ca (the lower $\pi f$ spin-orbit partner is occupied; spin-unsaturated regime) and Ni (both $\pi f$ spin-orbit partners are occupied; spin-saturated regime), as well as different dynamics of the neutron single-particle energies caused by the tensor interaction \cite{Otsuka.2005} when filling the $\pi f_{\nicefrac{7}{2}}$ orbits between Ca and Ni \cite{Otsuka.2020} make the comparison between the charge radii and magnetic moments in these isotopic chains particularly interesting.  

Here, we report  the determination of the nuclear charge radii of $^{54,55,56}$Ni and the magnetic moment of $^{55}$Ni. In combination with previously published data  \cite{Kaufmann.2020,Pineda.2021,Malbrunot.2022}, this establishes the behavior of \dmscr\ at and across the $N=28$ shell closure. 
The measured magnetic moment of $^{55}$Ni corrects the previous $\beta$-NMR measurement~\cite{Berryman.2009}.

Similar to the doubly magic $^{40,48}$Ca (see, e.g., Ref.\, \cite{Hagen.2016b}), the nuclear charge radius of $^{56}$Ni is also an excellent benchmark for \textit{ab initio} nuclear structure theory. 
Different approaches have predicted the size of this nucleus \cite{Hagen.2016, Hoppe.2019, Huther.2020} and this Letter  contributes new results for this important observable.

{\it Experiment. ---}
Ions of $^{54,55,56}$Ni were produced at the National Superconducting Cyclotron Laboratory (NSCL) at Michigan State University (MSU) and collinear laser spectroscopy (CLS) was performed at the BECOLA facility \cite{Minamisono.2013}.
The radioactive nickel isotopes were produced through fragmentation of a \SI[per-mode=symbol]{160}{\MeV\per\atomicmassunit} primary $^{58}$Ni beam impinging on a Be target and separated from other reaction products in the A1900 fragment separator \cite{Morrisey.2003}. The particles were stopped and thermalized in a gas-stopper cell \cite{sum20}. The extracted Ni$^+$ ions were then accelerated to a kinetic beam energy of \SI{30}{\kilo\electronvolt} and transported to the BECOLA facility with rates of approximately \SI{4.5E3}{} and \SI{6E3}{ions\per\second} for $^{56}$Ni and $^{55}$Ni, respectively \cite{SM}. 
Here, a radio-frequency quadrupole cooler and buncher (RFQ) \cite{bar17} was used to trap and cool either the radioactive beam or, for reference measurements,  the stable nickel isotopes from a local penning ionization gauge (PIG) ion source \cite{Nouri.2010}. Bunches of ions were released from the RFQ into the CLS beamline with an efficiency of 70\% at ion energies of $E_\mathrm{ion}\approx\SI{29850}{\electronvolt}$ and were collinearly superimposed with the spectroscopy laser light and guided into the charge-exchange cell \cite{Klose.2012, Ryder.2015} loaded with sodium and heated to \SI{420}{\celsius}. Under these conditions, a neutralization efficiency of typically \SI{50}{\percent} was achieved of which an estimated fraction of 15\% populates the lower level of the atomic $3d^9\,4s\,^3\!D_3 \rightarrow 3d^9\,4p\,^3\!P_2$ transition at \SI{352}{\nano\meter}. Resonance spectra were recorded by changing a small voltage applied to the charge-exchange cell to Doppler-tune the laser frequency in the rest-frame of the atoms. The laser frequency was adjusted for each isotope 
to keep the central acceleration voltage almost identical. Fluorescence photons were detected with three consecutive photo-multiplier tubes mounted on chambers with different mirror geometries \cite{Minamisono.2015,Maass.2020}. The laser light of \SI{352}{\nano\meter} was generated in a frequency-doubling cavity (Wavetrain, Spectra Physics) from the output of a continuous-wave titanium-sapphire (Ti:Sa) laser (Matisse TS, Sirah Lasertechnik) operated at \SI{704}{\nano\meter}. The Ti:Sa output was measured and stabilized by a wavemeter (WSU30, HighFinesse), which in turn was calibrated to a frequency stabilized helium-neon laser (SL 03, SIOS Messtechnik) once every minute.

A pair of reference measurements of $^{58,60}$Ni isotopes from the off-line PIG source was conducted typically once every 6--12 hours. These reference measurements were used to determine the isotope shifts of the short-lived $^{54-56}$Ni isotopes with respect to $^{60}$Ni and also allowed to calibrate the ion energy to the known absolute transition frequency of $^{60}$Ni \cite{Konig.2021}.

{\it Results. ---}
\begin{figure}
    \centering
    \includegraphics[width=\columnwidth]{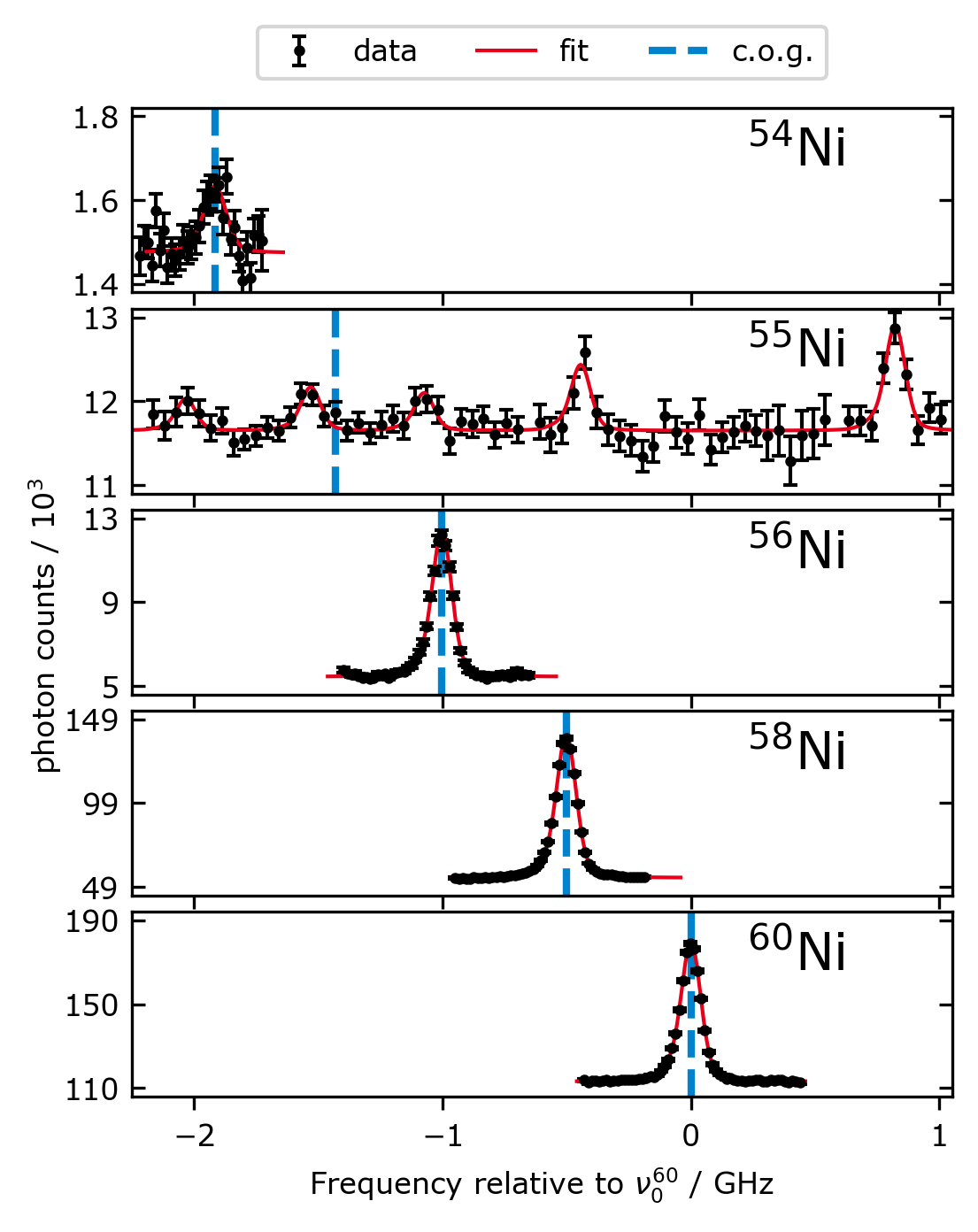}
    \caption{\label{fig:Spectra}Sum  spectra of proton-rich, radioactive nickel isotopes $^{54,55,56}$Ni and of the reference isotopes $^{58,60}$Ni measured at BECOLA. The solid red lines show the fits to the data and the centers of gravity (c.o.g.) for each spectrum are depicted as dashed blue lines. While the displayed counts are close to the actually observed numbers, deviations occur due to the normalization procedure used to combine all data of the beamtime in these spectra. Only the measured part of the $^{55}$Ni hyperfine spectrum is shown. For more details, see Suppl. Mat. \cite{SM}.}
\end{figure}
The resonance spectra of the measured nickel isotopes are shown in Fig.\,\ref{fig:Spectra}, together with a Voigt lineshape fitted to each dataset. Energy losses from inelastic collisions in the charge-exchange cell lead to an asymmetric lineshape that was modelled by including one additional, smaller Voigt profile into the fit function at a phenomenologically determined lower ion energy \cite{Klose.2012}. The spectra of stable $^{56,58,60}$Ni isotopes were fitted separately for each measurement, whereas the events of all $^{54,55}$Ni data sets were summed up before the fitting procedure due to lower production yields. Details regarding the fitting of the $^{55}$Ni spectrum are given in the Supplemental Material (Suppl. Mat.) \cite{SM}.

The isotope shift $\delta \nu^{A,60} = \nu^A - \nu^{60}$ for each isotope $^{A}$Ni relative to $^{60}$Ni was calculated from the extracted centroid frequencies. For the low-production isotopes $^{54}$Ni and $^{55}$Ni, the uncertainties of the isotope shifts are dominated by the fit uncertainty of their centroid positions. For $^{56}$Ni and the stable $^{58}$Ni, uncertainties of the frequency measurements \cite{Konig.2021} and an observed deviation between bunched-beam and continuous-beam measurements \cite{Konig.2021b} are the prevailing contributions to the isotope-shift uncertainties.

The differential mean-square (ms) charge radii \dmscr\ were determined as
\begin{equation}
    \label{eq: isotope shift}
    \dmscr^{A,A'} = \frac{\delta\nu^{A,A'} - K_{\alpha}\cdot\mu^{A,A'}}{F} + \alpha\cdot\mu^{A,A'},
\end{equation}
where $K_{\alpha}$ and $F$ are the so called mass- and field-shift factors, respectively, and $\mu^{A,A'}=(m_{A}-m_{A'})/(m_{A}+m_e)(m_{A'}+m_e)$ is the mass-scaling factor.
A constant factor $\alpha=\SI{388}{\giga\hertz\amu}$ shifts the abscissa to remove the correlation between $K$ and $F$ \cite{Hammen.2018}.
The factors $K_{\alpha}=\SI{954(4)}{\giga\hertz\amu}$ and $F=\SI[per-mode=symbol]{-805(66)}{\mega\hertz\per\femto\meter\squared}$ were determined in a King plot procedure by comparing the isotope shifts of stable nickel isotopes, measured off-line at BECOLA, with their known differential charge radii from literature \cite{FrickeHeilig.2004}. This King fit analysis is detailed in \cite{Konig.2021b}. The total root-mean-square (rms) charge radii $R_{\rm c}$ were then determined with respect to the reference value $R_{\rm c}(\mathrm{^{60}Ni})$ \cite{FrickeHeilig.2004}. The isotope shifts, differential ms charge radii, and rms charge radii are summarized in Tab.\,\ref{tab: final results radii}.

The values of charge radii along the $^{54}$Ni -- $^{58}$Ni isotopes establish the behavior of nickel charge radii across the $N=Z=28$ doubly-magic shell closure, and the value of $^{55}$Ni provides information on odd-even staggering in the neutron $f_{\nicefrac{7}{2}}$ shell. The charge radius of $^{57}$Ni, which could yield further insight to the odd-even staggering, has so far neither been obtained in literature nor has it been measured at BECOLA. Our result  for $^{58}$Ni agrees well with the previous measurements from Refs.\ \cite{Kaufmann.2020} and  \cite{Steudel.1980}.
Furthermore, the nuclear magnetic dipole moment of the $I=\frac{7}{2}$ \cite{Aysto.1984} isotope $^{55}$Ni was determined from the upper and lower  hyperfine $A$-factors as listed in Tab.\,\ref{tab:hyperfinestructure}. The Suppl. Mat.\ \cite{SM} contains a more detailed description of the fitting procedure for $^{55}$Ni and a discussion of the magnetic moment, which includes Refs.\,\cite{Berryman.2009,Honma.2004,Kaufmann.PhD,Stone.2019,Stone.2020,Georgiev.2002}. Our magnetic moment deviates significantly from the previously reported $\beta$-NMR value \cite{Berryman.2009}, which has been based on a single resonance point deviating  $3\sigma$ from the baseline. The very low magnetic moment being only $55\%$ of the single-particle $\nu f_{\nicefrac{7}{2}}$ value indicates the impact of M1 excitations between the $\nu f$ spin-orbit partners across the $N,Z=28$ shell gap. Our value is in good agreement with shell-model calculations with the GXPF1 interaction \cite{Berryman.2009}, which suggest a soft $^{56}$Ni core. This is in contrast with $^{47}$Ca, which has a magnetic moment very close to the effective $g$-value established in this region \cite{GarciaRuiz.2015}.
\begin{table}
\centering
\begingroup
\renewcommand{\arraystretch}{1.4}
    \caption[Results for isotope shifts and charge radii of Ni isotopes]{Isotope shifts, differential ms charge radii, and absolute rms charge radii for all nickel isotopes investigated at BECOLA. Uncertainties in parentheses denote combined uncorrelated uncertainties of statistical and systematic nature, whereas those in square brackets are correlated through the uncertainty of the King-plot parameters, which are taken from \cite{Konig.2021b} and given in the text.}
    \label{tab: final results radii}
        \begin{ruledtabular}
    \begin{tabular}{c c c c}
     & $\delta\nu^{A,60} / \si{\mega\hertz}$ & $\dmscr^{A,60} / \si{\femto\meter\squared}$ & $R_{\rm c}(\mathrm{^{A}Ni}) / \si{\femto\meter}$ \\
    \hline
    $^{54}$Ni & $-1919.7(7.9)$ & $-0.522(9)[19]$ & 3.7366(13)[31] \\
    $^{55}$Ni & $-1426.9(19.1)$   & $-0.607(23)[09]$ & 3.7252(32)[21] \\
    $^{56}$Ni & $-1002.7(3.8)$    & $-0.626(02)[16]$ & 3.7226(03)[27]\\
    $^{58}$Ni & $-506.3(2.5)$ & $-0.276(01)[06]$ & 3.7695(02)[19]\\
    $^{60}$Ni & 0 & 0 & 3.8059[17]
    \end{tabular}
        \end{ruledtabular}
\endgroup\\
\end{table}

\begin{table}[b]
\centering
\begingroup
\renewcommand{\arraystretch}{1.4}
    \caption[Results for the HFS parameters and moments of $^{55}$Ni]{$A$-parameters of the hyperfine structure that was fitted to the $^{55}$Ni spectrum. The nuclear magnetic dipole moment $\mu$ is the weighted average of the extraction using the upper and the lower $A$-factor based on the nuclear magnetic moment of $\mu(^{61}\mathrm{Ni})=\SI{-0.74965(5)}{\mu_\mathrm{N}}$ \cite{Stone.2014}.}
    \label{tab:hyperfinestructure}
    \begin{ruledtabular}
    \begin{tabular}{c c c c | c}

    \multicolumn{4}{c|}{This work} & Lit. \cite{Berryman.2009} \\  

    \hline
    $A_\mathrm{lo} / \si{\mega\hertz}$ & $A_\mathrm{up} / \si{\mega\hertz}$ & $A$-Ratio &  $\mu / \mu_N$ &  $\mu / \mu_N$  \\
    -288.4(5.6) & -112.1(4.9) & 0.389(19) & -1.108(20) & -0.976(26)\\

    \end{tabular}
    \end{ruledtabular}
\endgroup
\end{table}

{\it Theory. ---} The Ni chain and all medium-mass nuclei can be accessed by the \textit{ab initio} valence-space in-medium similarity renormalization group (VS-IMSRG)~\cite{Stroberg.2017,Stroberg.limits,Stroberg.2019}, which generates an approximate unitary transformation to decouple both a valence space and associated core from particle or hole excitations to outside configurations. The VS-IMSRG many-body calculations use the IMSRG code from~\cite{Stro17imsrg++} and follow those of  Ref.~\cite{Malbrunot.2022}, except that for three-nucleon (3N) matrix elements we use a sufficiently large truncation~\cite{Miyagi.2022}, so that energies and radii are converged with respect to the 3N basis size. Our calculations are based on two-nucleon (NN) and 3N interactions from chiral effective field theory (EFT). To assess the Hamiltonian dependence, the newly developed $\Delta$N2LO$_{\rm GO}$(394)~\cite{Jiang.2020} interaction from delta-full chiral EFT is employed, in addition to the N2LO$_{\rm sat}$ interaction that reproduces well the charge radii of neutron-rich Ni and Cu isotopes~\cite{Malbrunot.2022,Groote.2020}. 
In this work, we decouple the $pf$-shell valence space which enables full diagonalizations of the nuclei in consideration. The assessment of statistical and systematic uncertainties (along the lines of, e.g.,~\cite{Melendez.2019,Hu.2021}) as well as the inclusion of three-body contributions in the VS-IMSRG~\cite{Heinz.2021} are ongoing areas of theoretical development, not included in the results. Therefore, the VS-IMSRG uncertainties reported in the following stem from the model-space truncation and are extracted from the basis frequency dependence (as in~\cite{Malbrunot.2022}, but in a converged $E_\text{3max}$ space).

In addition, we employ a variant of the in-medium no-core shell model \cite{Gebrerufael.2016}  with a recent family of chiral NN+3N interactions up to N4LO in the NN interaction and N3LO in the 3N force \cite{Huther.2020}. For the description of nickel isotopes, the conventional $N_{\max}$-truncation, which derives from the harmonic oscillator basis, is not adequate anymore. Therefore, we employ a configuration-interaction-type (CI) active space with a particle-hole-type $T_{\max}$ truncation \cite{Stumpf.2015}. The underlying single-particle basis is constructed from perturbative natural orbitals \cite{Tichai.2018} and we consider the $pf$-shell valence space. The reference space for the multi-reference IMSRG decoupling is defined with a $T_{\max}=4$ truncation and the final CI calculation is performed for $T_{\max}=6$ ($4$ for $^{58}$Ni), lower $T_{\max}$ are used to assess convergence and many-body uncertainties. We also quantify the uncertainties resulting from the truncation of the chiral expansion of the interaction using a sequence of calculation from LO to N3LO and N4LO$^\prime$ using a pointwise Bayesian model \cite{Melendez.2019}. The error bars reported in the following are the sum of many-body and interaction uncertainties. All interactions use a cutoff $\Lambda=500\,$MeV and free-space SRG-evolution with flow-parameter $\alpha = 0.04\,\text{fm}^4$. For the calculation of radii, the translationally invariant radius operator is transformed consistently in the free-space and in-medium SRG. In the following, these calculations are referred to as in-medium configuration interaction (IM-CI).

Our nuclear density functional theory (DFT) calculations follow the methodology of Refs.~\cite{Malbrunot.2022,Kortelainen2022}. 
We use two non-relativistic energy density functionals (EDF), namely a Skyrme functional SV-min(HFB), a variant of SV-min \cite{Kluepfel2009}, and a Fayans functional  Fy($\Delta r$, HFB) \cite{Miller2019}.  Both have the same basic structure and both are calibrated to the same large body of nuclear ground-state data as described in Ref.\,\cite{Kluepfel2009}. 
In addition, for Fy($\Delta r$, HFB)  differential charge radii in the calcium chain were added to the optimization dataset.  
We emphasize that in both EDF pairing correlations are treated within the full Hartree-Fock-Bogoliubov (HFB) framework to properly handle proton continuum in the proton-rich Ni isotopes \cite{Miller2019,ReinhardNazarewicz2022}. Both parametrizations are fitted to empirical data which introduces statistical uncertainties, see \cite{Dobaczewski2014} for details. In addition, we consider a systematic error from collective quadrupole correlations which extends asymmetrically toward enhanced radii. In contrast to DFT, VS-IMSRG and IM-CI do not include these statistical uncertainties, but assess the systematic theoretical uncertainties from the truncation of the many-body expansion and, in the case of IM-CI, also the expansion of the interactions.
Moreover, the proton-rich  Ni isotopes have considerable zero-point quadrupole fluctuations which have been estimated as in \cite{Kluepfel2008,Kortelainen2022}; they provide an estimate of the systematic error for DFT.

{\it Discussion. ---}
%
\begin{figure}
    \includegraphics[width=\columnwidth]{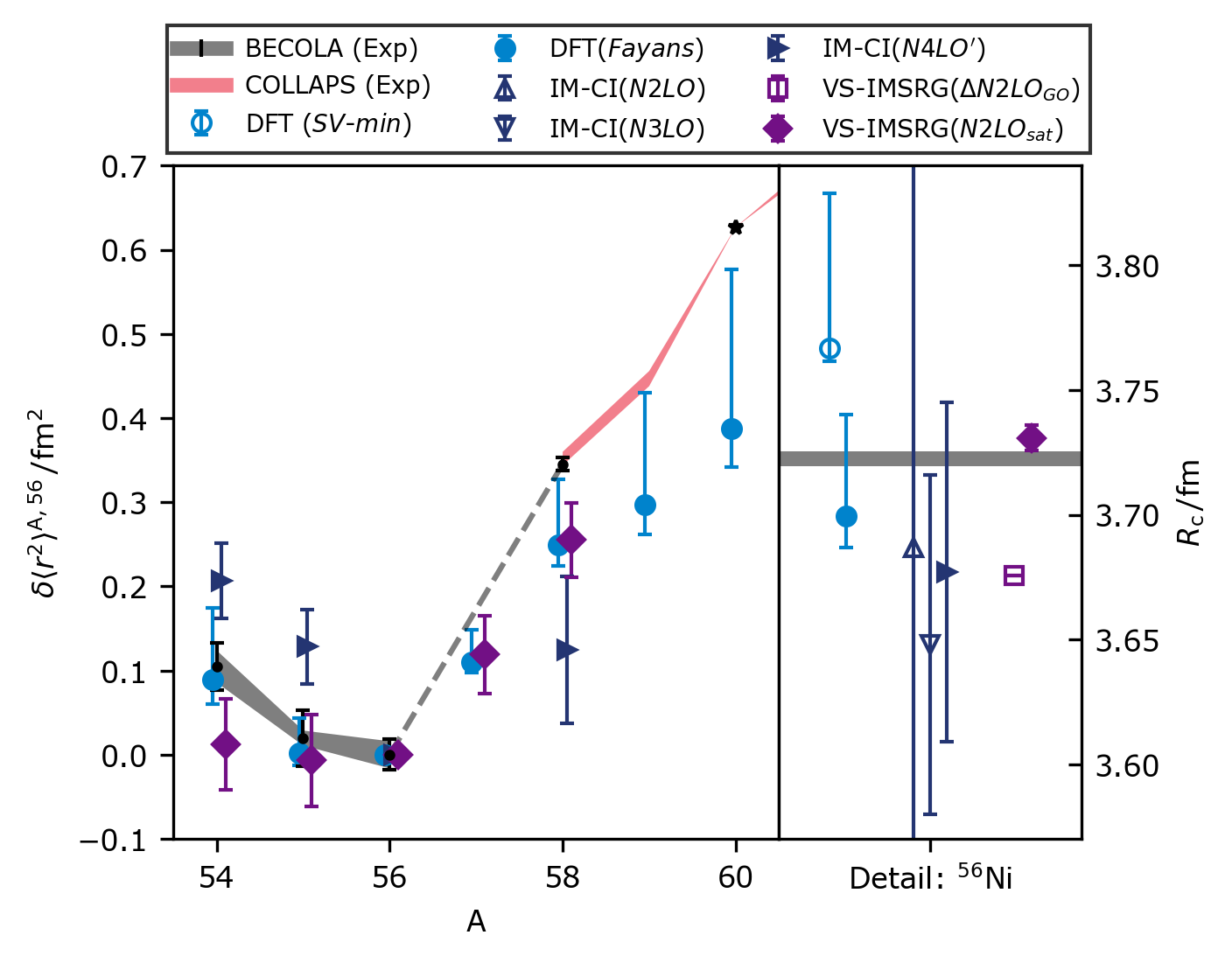}
    \caption{\label{fig:abs-radii}  Charge radii of nickel isotopes across the $N=28$ shell closure. Left panel: measured \dmscr\ from BECOLA (this work) and COLLAPS \cite{Malbrunot.2022} compared to  representative DFT, IM-CI, and VS-IMSRG results. The dashed line between $^{56,58}$Ni indicates that the charge radius of $^{57}$Ni has not been measured.
Right panel: summary of all theoretical results for the nuclear charge radius of the doubly magic nucleus $^{56}$Ni.
    }
\end{figure}
The spread of predictions for the absolute charge radius of $^{56}$Ni for all models explored in this Letter is shown in the right panel of Fig.\,\ref{fig:abs-radii} and compared to the experimental value (gray line). In DFT, the Fayans functional provides a better description than SV-min, similar to the quite accurate description of charge radii at this shell closure in K \cite{Koszorus.2021}, Ca \cite{GarciaRuiz.2016}, and Fe \cite{Minamisono.2016}. Charge radii from IM-CI with N2LO, N3LO, and N4LO$^\prime$ interactions scatter slightly but are compatible with the experiment within the error bars. For VS-IMSRG, the N2LO$_\mathrm{sat}$ interaction provides the best description and is in very good agreement with the experimental radius, while results using the $\Delta$N2LO$_\mathrm{GO}$ interaction are somewhat too small. 
The left panel of Fig.\,\ref{fig:abs-radii} shows \dmscr\ determined for $^{\text{54-60}}$Ni compared to DFT and VS-IMSRG results using the best-performing interaction or functional and to IM-CI for the highest-order chiral interaction.
The charge radii obtained with Fy($\Delta r$, HFB) describe the general trend quite well but are getting systematically too small particularly above $^{56}$Ni, which is in accordance with \cite{Malbrunot.2022}, where it was found that SV-min outperforms the Fayans functional along the chain of neutron-rich Ni isotopes. This is most likely caused by isovector components that are not yet included in the current Fayans functional but might become important with additional neutrons.  
The \dmscr\ results magnify the local trend and exhibit that the slope of the charge radii from IM-CI is overestimated below and underestimated above $^{56}$Ni. For VS-IMSRG, the differential ms radii are reasonably reproduced in the complete range $^{\mathrm{54-58}}$Ni, consistent with \cite{Malbrunot.2022}.
To facilitate further discussion of the kink in $R_{\rm c}$ seen in Fig.\,\ref{fig:abs-radii}, we introduce the two-neutron three-point indicator for a kink in the charge radius along an isotopic chain: 

\begin{equation}
\Delta^{(3)}_{2n}R_{\rm c}(N)\equiv\frac{1}{2}\left[R_{\rm c}(N+2)-2R_{\rm c}(N)+R_{\rm c}(N-2)\right].
\end{equation}
Figure \ref{fig:three-point-indicator} shows $\Delta^{(3)}_{2n}R_{\rm c}$ along the $N=28$ isotonic chain for all elements for which the charge radii have been measured for $N=26, 28, 30$, together with  the theoretical predictions for Ni and Ca. The uncertainty of the DFT calculation, both statistical and systematic, have been evaluated directly for the kink thus eliminating a common background error of the three involved nuclei. The experimental $\Delta^{(3)}_{2n}R_{\rm c}(28)$ are almost identical for K, Ca, and Ni and slightly larger for Mn and Fe. This is also clearly visible in the inset of Fig.\,\ref{fig:three-point-indicator} that shows \dmscr\ with respect to the neutron-magic nucleus in the chain. The larger values for Mn and Fe below $N=28$ can be explained by  contributions of ground-state quadrupole correlations in these open-proton-shell nuclei \cite{Minamisono.2016}. With the determination of the charge radius of $^{56}$Ni, we established the kink at $N=28$ in the nickel chain and can compare for the first time the 3-point charge radii differences for two doubly magic isotones, $^{48}$Ca and $^{56}$Ni. 
The equal size of $\Delta^{(3)}_{2n}R_{\rm c}(28)$ at $Z=20$ and $Z=28$, despite the quite different size of the neutron shell gap, different charges, and different types of shell closures, i.e.,  different proton-spin saturation in $^{48}$Ca and $^{56}$Ni, has been unanticipated. Still, the kink is reasonably reproduced by our DFT and VS-IMSRG calculations.

\begin{figure}
    \includegraphics[width=\columnwidth]{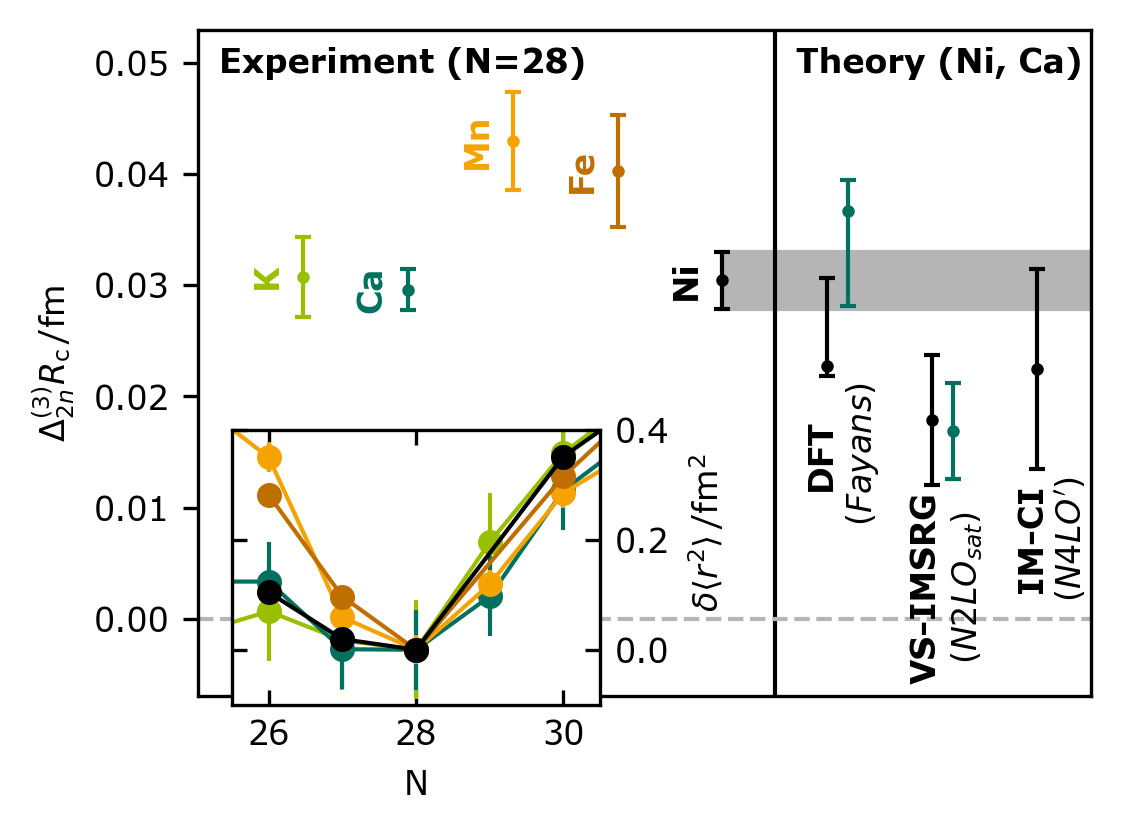}
    \caption{\label{fig:three-point-indicator}The three-point-indicator quantifies the strength the 'kink' at the $N=28$ shell closure. Experimental values are shown in the left part of the figure and their differential charge radii relative to the neutron-magic nucleus are detailed in the insert. Representative theory calculations for Ni (black) and Ca (green) are shown in the right part using DFT, VS-IMSRG, and IM-CI (only for Ni).
    Since the  charge radii calculated for different isotopes are strongly correlated, the estimated theoretical error bars of $\Delta^{(3)}_{2n}R_{\rm c}$ are smaller than those of $R_{\rm c}$ in Fig.\,\ref{fig:abs-radii}.
    }
\end{figure}

%

{\it Summary. ---} We have determined the nuclear charge radii of $^{54,55,56}$Ni and find the \dmscr\ from $^{54}$Ni to $^{58}$Ni to be within uncertainties identical to those of their Ca isotones. This is the first case where nuclear charge radii are available across two doubly-magic nuclei at the same neutron shell closure. The observed behavior is well reproduced by  \textit{ab initio} and DFT calculations based on realistic input.
Interestingly, the striking similarity 
of $\Delta^{(3)}_{2n}R_{\rm c}(N\text{=28})$ and fairly different $B(E2; 2^+_1\rightarrow 0^+_1)$ values \cite{Pritychenko.2016} for Ca and Ni, suggest that the kink in charge radii does not directly reflect the strength of a shell closure. \\ 

{\it Acknowledgement ---} This work was supported in part by the  Deutsche  Forschungsgemeinschaft  (DFG,  German Research Foundation) -- Project-ID 279384907 -- SFB 1245, by the U.S.\ Department of Energy, Office of Science, Office of Nuclear Physics under award numbers DE-SC0013365 and DE-SC0018083 (NUCLEI SciDAC-4 collaboration), by the National Science Foundation under grants PHY 19-13509, PHY-15-65546, and PHY-21-11185, and by NSERC under grants SAPIN-2018-00027 and RGPAS-2018-522453.
TRIUMF receives funding via a contribution through the National Research Council of Canada. Computations of VS-IMSRG were performed with an allocation of computing resources on Cedar at WestGrid and Compute Canada.
We also thank the RRZE computing center of the Friedrich-Alexander University Erlangen/N\"urnberg for supplying resources for this work.
\FloatBarrier

\bibliography{apssamp}

\begin{thebibliography}{79}%
\makeatletter
\providecommand \@ifxundefined [1]{%
 \@ifx{#1\undefined}
}%
\providecommand \@ifnum [1]{%
 \ifnum #1\expandafter \@firstoftwo
 \else \expandafter \@secondoftwo
 \fi
}%
\providecommand \@ifx [1]{%
 \ifx #1\expandafter \@firstoftwo
 \else \expandafter \@secondoftwo
 \fi
}%
\providecommand \natexlab [1]{#1}%
\providecommand \enquote  [1]{``#1''}%
\providecommand \bibnamefont  [1]{#1}%
\providecommand \bibfnamefont [1]{#1}%
\providecommand \citenamefont [1]{#1}%
\providecommand \href@noop [0]{\@secondoftwo}%
\providecommand \href [0]{\begingroup \@sanitize@url \@href}%
\providecommand \@href[1]{\@@startlink{#1}\@@href}%
\providecommand \@@href[1]{\endgroup#1\@@endlink}%
\providecommand \@sanitize@url [0]{\catcode `\\12\catcode `\$12\catcode
  `\&12\catcode `\#12\catcode `\^12\catcode `\_12\catcode `\%12\relax}%
\providecommand \@@startlink[1]{}%
\providecommand \@@endlink[0]{}%
\providecommand \url  [0]{\begingroup\@sanitize@url \@url }%
\providecommand \@url [1]{\endgroup\@href {#1}{\urlprefix }}%
\providecommand \urlprefix  [0]{URL }%
\providecommand \Eprint [0]{\href }%
\providecommand \doibase [0]{https://doi.org/}%
\providecommand \selectlanguage [0]{\@gobble}%
\providecommand \bibinfo  [0]{\@secondoftwo}%
\providecommand \bibfield  [0]{\@secondoftwo}%
\providecommand \translation [1]{[#1]}%
\providecommand \BibitemOpen [0]{}%
\providecommand \bibitemStop [0]{}%
\providecommand \bibitemNoStop [0]{.\EOS\space}%
\providecommand \EOS [0]{\spacefactor3000\relax}%
\providecommand \BibitemShut  [1]{\csname bibitem#1\endcsname}%
\let\auto@bib@innerbib\@empty
\bibitem [{\citenamefont {Dobaczewski}\ \emph {et~al.}(2007)\citenamefont
  {Dobaczewski}, \citenamefont {Michel}, \citenamefont {Nazarewicz},
  \citenamefont {Płoszajczak},\ and\ \citenamefont
  {Rotureau}}]{Dobaczewski2007}%
  \BibitemOpen
  \bibfield  {author} {\bibinfo {author} {\bibfnamefont {J.}~\bibnamefont
  {Dobaczewski}}, \bibinfo {author} {\bibfnamefont {N.}~\bibnamefont {Michel}},
  \bibinfo {author} {\bibfnamefont {W.}~\bibnamefont {Nazarewicz}}, \bibinfo
  {author} {\bibfnamefont {M.}~\bibnamefont {Płoszajczak}},\ and\ \bibinfo
  {author} {\bibfnamefont {J.}~\bibnamefont {Rotureau}},\ }\bibfield  {title}
  {\bibinfo {title} {Shell structure of exotic nuclei},\ }\href
  {https://doi.org/10.1016/j.ppnp.2007.01.022} {\bibfield  {journal} {\bibinfo
  {journal} {Prog. Part. Nucl. Phys.}\ }\textbf {\bibinfo {volume} {59}},\
  \bibinfo {pages} {432–445} (\bibinfo {year} {2007})}\BibitemShut {NoStop}%
\bibitem [{\citenamefont {Sorlin}\ and\ \citenamefont
  {Porquet}(2008)}]{Sorlin.2008}%
  \BibitemOpen
  \bibfield  {author} {\bibinfo {author} {\bibfnamefont {O.}~\bibnamefont
  {Sorlin}}\ and\ \bibinfo {author} {\bibfnamefont {M.-G.}\ \bibnamefont
  {Porquet}},\ }\bibfield  {title} {\bibinfo {title} {Nuclear magic numbers:
  New features far from stability},\ }\href
  {https://doi.org/https://doi.org/10.1016/j.ppnp.2008.05.001} {\bibfield
  {journal} {\bibinfo  {journal} {Prog. Part. Nucl. Phys.}\ }\textbf {\bibinfo
  {volume} {61}},\ \bibinfo {pages} {602--673} (\bibinfo {year}
  {2008})}\BibitemShut {NoStop}%
\bibitem [{\citenamefont {Otsuka}\ \emph {et~al.}(2020)\citenamefont {Otsuka},
  \citenamefont {Gade}, \citenamefont {Sorlin}, \citenamefont {Suzuki},\ and\
  \citenamefont {Utsuno}}]{Otsuka.2020}%
  \BibitemOpen
  \bibfield  {author} {\bibinfo {author} {\bibfnamefont {T.}~\bibnamefont
  {Otsuka}}, \bibinfo {author} {\bibfnamefont {A.}~\bibnamefont {Gade}},
  \bibinfo {author} {\bibfnamefont {O.}~\bibnamefont {Sorlin}}, \bibinfo
  {author} {\bibfnamefont {T.}~\bibnamefont {Suzuki}},\ and\ \bibinfo {author}
  {\bibfnamefont {Y.}~\bibnamefont {Utsuno}},\ }\bibfield  {title} {\bibinfo
  {title} {Evolution of shell structure in exotic nuclei},\ }\href
  {https://doi.org/10.1103/RevModPhys.92.015002} {\bibfield  {journal}
  {\bibinfo  {journal} {Rev. Mod. Phys.}\ }\textbf {\bibinfo {volume} {92}},\
  \bibinfo {pages} {015002} (\bibinfo {year} {2020})}\BibitemShut {NoStop}%
\bibitem [{\citenamefont {Gorges}\ \emph {et~al.}(2019)\citenamefont {Gorges},
  \citenamefont {Rodr{\'i}guez}, \citenamefont {Balabanski}, \citenamefont
  {Bissell}, \citenamefont {Blaum}, \citenamefont {Cheal}, \citenamefont
  {{Garcia Ruiz}}, \citenamefont {Georgiev}, \citenamefont {Gins},
  \citenamefont {Heylen}, \citenamefont {Kanellakopoulos}, \citenamefont
  {Kaufmann}, \citenamefont {Kowalska}, \citenamefont {Lagaki}, \citenamefont
  {Lechner}, \citenamefont {Maa{\ss}}, \citenamefont {Malbrunot-Ettenauer},
  \citenamefont {Nazarewicz}, \citenamefont {Neugart}, \citenamefont {Neyens},
  \citenamefont {N{\"o}rtersh{\"a}user}, \citenamefont {Reinhard},
  \citenamefont {Sailer}, \citenamefont {S{\'a}nchez}, \citenamefont {Schmidt},
  \citenamefont {Wehner}, \citenamefont {Wraith}, \citenamefont {Xie},
  \citenamefont {Xu}, \citenamefont {Yang},\ and\ \citenamefont
  {Yordanov}}]{Gorges.2019}%
  \BibitemOpen
  \bibfield  {author} {\bibinfo {author} {\bibfnamefont {C.}~\bibnamefont
  {Gorges}}, \bibinfo {author} {\bibfnamefont {L.~V.}\ \bibnamefont
  {Rodr{\'i}guez}}, \bibinfo {author} {\bibfnamefont {D.~L.}\ \bibnamefont
  {Balabanski}}, \bibinfo {author} {\bibfnamefont {M.~L.}\ \bibnamefont
  {Bissell}}, \bibinfo {author} {\bibfnamefont {K.}~\bibnamefont {Blaum}},
  \bibinfo {author} {\bibfnamefont {B.}~\bibnamefont {Cheal}}, \bibinfo
  {author} {\bibfnamefont {R.~F.}\ \bibnamefont {{Garcia Ruiz}}}, \bibinfo
  {author} {\bibfnamefont {G.}~\bibnamefont {Georgiev}}, \bibinfo {author}
  {\bibfnamefont {W.}~\bibnamefont {Gins}}, \bibinfo {author} {\bibfnamefont
  {H.}~\bibnamefont {Heylen}}, \bibinfo {author} {\bibfnamefont
  {A.}~\bibnamefont {Kanellakopoulos}}, \bibinfo {author} {\bibfnamefont
  {S.}~\bibnamefont {Kaufmann}}, \bibinfo {author} {\bibfnamefont
  {M.}~\bibnamefont {Kowalska}}, \bibinfo {author} {\bibfnamefont
  {V.}~\bibnamefont {Lagaki}}, \bibinfo {author} {\bibfnamefont
  {S.}~\bibnamefont {Lechner}}, \bibinfo {author} {\bibfnamefont
  {B.}~\bibnamefont {Maa{\ss}}}, \bibinfo {author} {\bibfnamefont
  {S.}~\bibnamefont {Malbrunot-Ettenauer}}, \bibinfo {author} {\bibfnamefont
  {W.}~\bibnamefont {Nazarewicz}}, \bibinfo {author} {\bibfnamefont
  {R.}~\bibnamefont {Neugart}}, \bibinfo {author} {\bibfnamefont
  {G.}~\bibnamefont {Neyens}}, \bibinfo {author} {\bibfnamefont
  {W.}~\bibnamefont {N{\"o}rtersh{\"a}user}}, \bibinfo {author} {\bibfnamefont
  {P.-G.}\ \bibnamefont {Reinhard}}, \bibinfo {author} {\bibfnamefont
  {S.}~\bibnamefont {Sailer}}, \bibinfo {author} {\bibfnamefont
  {R.}~\bibnamefont {S{\'a}nchez}}, \bibinfo {author} {\bibfnamefont
  {S.}~\bibnamefont {Schmidt}}, \bibinfo {author} {\bibfnamefont
  {L.}~\bibnamefont {Wehner}}, \bibinfo {author} {\bibfnamefont
  {C.}~\bibnamefont {Wraith}}, \bibinfo {author} {\bibfnamefont
  {L.}~\bibnamefont {Xie}}, \bibinfo {author} {\bibfnamefont {Z.~Y.}\
  \bibnamefont {Xu}}, \bibinfo {author} {\bibfnamefont {X.~F.}\ \bibnamefont
  {Yang}},\ and\ \bibinfo {author} {\bibfnamefont {D.~T.}\ \bibnamefont
  {Yordanov}},\ }\bibfield  {title} {\bibinfo {title} {{Laser Spectroscopy of
  Neutron-Rich Tin Isotopes: A Discontinuity in Charge Radii across the $N$=82
  Shell Closure}},\ }\href {https://doi.org/10.1103/PhysRevLett.122.192502}
  {\bibfield  {journal} {\bibinfo  {journal} {Phys.\ Rev.\ Lett.}\ }\textbf
  {\bibinfo {volume} {122}},\ \bibinfo {pages} {192502} (\bibinfo {year}
  {2019})}\BibitemShut {NoStop}%
\bibitem [{\citenamefont {{Day Goodacre}}\ \emph {et~al.}(2021)\citenamefont
  {{Day Goodacre}}, \citenamefont {Afanasjev}, \citenamefont {Barzakh},
  \citenamefont {Marsh}, \citenamefont {Sels}, \citenamefont {Ring},
  \citenamefont {Nakada}, \citenamefont {Andreyev}, \citenamefont {{van
  Duppen}}, \citenamefont {Althubiti}, \citenamefont {Andel}, \citenamefont
  {Atanasov}, \citenamefont {Billowes}, \citenamefont {Blaum}, \citenamefont
  {Cocolios}, \citenamefont {Cubiss}, \citenamefont {Farooq-Smith},
  \citenamefont {Fedorov}, \citenamefont {Fedosseev}, \citenamefont {Flanagan},
  \citenamefont {Gaffney}, \citenamefont {Ghys}, \citenamefont {Huyse},
  \citenamefont {Kreim}, \citenamefont {Lunney}, \citenamefont {Lynch},
  \citenamefont {Manea}, \citenamefont {{Martinez Palenzuela}}, \citenamefont
  {Molkanov}, \citenamefont {Rosenbusch}, \citenamefont {Rossel}, \citenamefont
  {Rothe}, \citenamefont {Schweikhard}, \citenamefont {Seliverstov},
  \citenamefont {Spagnoletti}, \citenamefont {{van Beveren}}, \citenamefont
  {Veinhard}, \citenamefont {Verstraelen}, \citenamefont {Welker},
  \citenamefont {Wendt}, \citenamefont {Wienholtz}, \citenamefont {Wolf},
  \citenamefont {Zadvornaya},\ and\ \citenamefont {Zuber}}]{DayGoodacre.2021}%
  \BibitemOpen
  \bibfield  {author} {\bibinfo {author} {\bibfnamefont {T.}~\bibnamefont {{Day
  Goodacre}}}, \bibinfo {author} {\bibfnamefont {A.~V.}\ \bibnamefont
  {Afanasjev}}, \bibinfo {author} {\bibfnamefont {A.~E.}\ \bibnamefont
  {Barzakh}}, \bibinfo {author} {\bibfnamefont {B.~A.}\ \bibnamefont {Marsh}},
  \bibinfo {author} {\bibfnamefont {S.}~\bibnamefont {Sels}}, \bibinfo {author}
  {\bibfnamefont {P.}~\bibnamefont {Ring}}, \bibinfo {author} {\bibfnamefont
  {H.}~\bibnamefont {Nakada}}, \bibinfo {author} {\bibfnamefont {A.~N.}\
  \bibnamefont {Andreyev}}, \bibinfo {author} {\bibfnamefont {P.}~\bibnamefont
  {{van Duppen}}}, \bibinfo {author} {\bibfnamefont {N.~A.}\ \bibnamefont
  {Althubiti}}, \bibinfo {author} {\bibfnamefont {B.}~\bibnamefont {Andel}},
  \bibinfo {author} {\bibfnamefont {D.}~\bibnamefont {Atanasov}}, \bibinfo
  {author} {\bibfnamefont {J.}~\bibnamefont {Billowes}}, \bibinfo {author}
  {\bibfnamefont {K.}~\bibnamefont {Blaum}}, \bibinfo {author} {\bibfnamefont
  {T.~E.}\ \bibnamefont {Cocolios}}, \bibinfo {author} {\bibfnamefont {J.~G.}\
  \bibnamefont {Cubiss}}, \bibinfo {author} {\bibfnamefont {G.~J.}\
  \bibnamefont {Farooq-Smith}}, \bibinfo {author} {\bibfnamefont {D.~V.}\
  \bibnamefont {Fedorov}}, \bibinfo {author} {\bibfnamefont {V.~N.}\
  \bibnamefont {Fedosseev}}, \bibinfo {author} {\bibfnamefont {K.~T.}\
  \bibnamefont {Flanagan}}, \bibinfo {author} {\bibfnamefont {L.~P.}\
  \bibnamefont {Gaffney}}, \bibinfo {author} {\bibfnamefont {L.}~\bibnamefont
  {Ghys}}, \bibinfo {author} {\bibfnamefont {M.}~\bibnamefont {Huyse}},
  \bibinfo {author} {\bibfnamefont {S.}~\bibnamefont {Kreim}}, \bibinfo
  {author} {\bibfnamefont {D.}~\bibnamefont {Lunney}}, \bibinfo {author}
  {\bibfnamefont {K.~M.}\ \bibnamefont {Lynch}}, \bibinfo {author}
  {\bibfnamefont {V.}~\bibnamefont {Manea}}, \bibinfo {author} {\bibfnamefont
  {Y.}~\bibnamefont {{Martinez Palenzuela}}}, \bibinfo {author} {\bibfnamefont
  {P.~L.}\ \bibnamefont {Molkanov}}, \bibinfo {author} {\bibfnamefont
  {M.}~\bibnamefont {Rosenbusch}}, \bibinfo {author} {\bibfnamefont {R.~E.}\
  \bibnamefont {Rossel}}, \bibinfo {author} {\bibfnamefont {S.}~\bibnamefont
  {Rothe}}, \bibinfo {author} {\bibfnamefont {L.}~\bibnamefont {Schweikhard}},
  \bibinfo {author} {\bibfnamefont {M.~D.}\ \bibnamefont {Seliverstov}},
  \bibinfo {author} {\bibfnamefont {P.}~\bibnamefont {Spagnoletti}}, \bibinfo
  {author} {\bibfnamefont {C.}~\bibnamefont {{van Beveren}}}, \bibinfo {author}
  {\bibfnamefont {M.}~\bibnamefont {Veinhard}}, \bibinfo {author}
  {\bibfnamefont {E.}~\bibnamefont {Verstraelen}}, \bibinfo {author}
  {\bibfnamefont {A.}~\bibnamefont {Welker}}, \bibinfo {author} {\bibfnamefont
  {K.}~\bibnamefont {Wendt}}, \bibinfo {author} {\bibfnamefont
  {F.}~\bibnamefont {Wienholtz}}, \bibinfo {author} {\bibfnamefont {R.~N.}\
  \bibnamefont {Wolf}}, \bibinfo {author} {\bibfnamefont {A.}~\bibnamefont
  {Zadvornaya}},\ and\ \bibinfo {author} {\bibfnamefont {K.}~\bibnamefont
  {Zuber}},\ }\bibfield  {title} {\bibinfo {title} {{Laser Spectroscopy of
  Neutron-Rich $^{207,208}$Hg Isotopes: Illuminating the Kink and Odd-Even
  Staggering in Charge Radii across the $N=126$ Shell Closure}},\ }\href
  {https://doi.org/10.1103/PhysRevLett.126.032502} {\bibfield  {journal}
  {\bibinfo  {journal} {Phys.\ Rev.\ Lett.}\ }\textbf {\bibinfo {volume}
  {126}},\ \bibinfo {pages} {032502} (\bibinfo {year} {2021})}\BibitemShut
  {NoStop}%
\bibitem [{\citenamefont {Day~Goodacre}\ \emph {et~al.}(2021)\citenamefont
  {Day~Goodacre}, \citenamefont {Afanasjev}, \citenamefont {Barzakh},
  \citenamefont {Nies}, \citenamefont {Marsh}, \citenamefont {Sels},
  \citenamefont {Perera}, \citenamefont {Ring}, \citenamefont {Wienholtz},
  \citenamefont {Andreyev}, \citenamefont {Van~Duppen}, \citenamefont
  {Althubiti}, \citenamefont {Andel}, \citenamefont {Atanasov}, \citenamefont
  {Augusto}, \citenamefont {Billowes}, \citenamefont {Blaum}, \citenamefont
  {Cocolios}, \citenamefont {Cubiss}, \citenamefont {Farooq-Smith},
  \citenamefont {Fedorov}, \citenamefont {Fedosseev}, \citenamefont {Flanagan},
  \citenamefont {Gaffney}, \citenamefont {Ghys}, \citenamefont {Gottberg},
  \citenamefont {Huyse}, \citenamefont {Kreim}, \citenamefont {Kunz},
  \citenamefont {Lunney}, \citenamefont {Lynch}, \citenamefont {Manea},
  \citenamefont {Palenzuela}, \citenamefont {Medonca}, \citenamefont
  {Molkanov}, \citenamefont {Mougeot}, \citenamefont {Ramos}, \citenamefont
  {Rosenbusch}, \citenamefont {Rossel}, \citenamefont {Rothe}, \citenamefont
  {Schweikhard}, \citenamefont {Seliverstov}, \citenamefont {Spagnoletti},
  \citenamefont {Van~Beveren}, \citenamefont {Veinhard}, \citenamefont
  {Verstraelen}, \citenamefont {Welker}, \citenamefont {Wendt}, \citenamefont
  {Wolf}, \citenamefont {Zadvornaya},\ and\ \citenamefont
  {Zuber}}]{Goodacre.2021b}%
  \BibitemOpen
  \bibfield  {author} {\bibinfo {author} {\bibfnamefont {T.}~\bibnamefont
  {Day~Goodacre}}, \bibinfo {author} {\bibfnamefont {A.~V.}\ \bibnamefont
  {Afanasjev}}, \bibinfo {author} {\bibfnamefont {A.~E.}\ \bibnamefont
  {Barzakh}}, \bibinfo {author} {\bibfnamefont {L.}~\bibnamefont {Nies}},
  \bibinfo {author} {\bibfnamefont {B.~A.}\ \bibnamefont {Marsh}}, \bibinfo
  {author} {\bibfnamefont {S.}~\bibnamefont {Sels}}, \bibinfo {author}
  {\bibfnamefont {U.~C.}\ \bibnamefont {Perera}}, \bibinfo {author}
  {\bibfnamefont {P.}~\bibnamefont {Ring}}, \bibinfo {author} {\bibfnamefont
  {F.}~\bibnamefont {Wienholtz}}, \bibinfo {author} {\bibfnamefont {A.~N.}\
  \bibnamefont {Andreyev}}, \bibinfo {author} {\bibfnamefont {P.}~\bibnamefont
  {Van~Duppen}}, \bibinfo {author} {\bibfnamefont {N.~A.}\ \bibnamefont
  {Althubiti}}, \bibinfo {author} {\bibfnamefont {B.}~\bibnamefont {Andel}},
  \bibinfo {author} {\bibfnamefont {D.}~\bibnamefont {Atanasov}}, \bibinfo
  {author} {\bibfnamefont {R.~S.}\ \bibnamefont {Augusto}}, \bibinfo {author}
  {\bibfnamefont {J.}~\bibnamefont {Billowes}}, \bibinfo {author}
  {\bibfnamefont {K.}~\bibnamefont {Blaum}}, \bibinfo {author} {\bibfnamefont
  {T.~E.}\ \bibnamefont {Cocolios}}, \bibinfo {author} {\bibfnamefont {J.~G.}\
  \bibnamefont {Cubiss}}, \bibinfo {author} {\bibfnamefont {G.~J.}\
  \bibnamefont {Farooq-Smith}}, \bibinfo {author} {\bibfnamefont {D.~V.}\
  \bibnamefont {Fedorov}}, \bibinfo {author} {\bibfnamefont {V.~N.}\
  \bibnamefont {Fedosseev}}, \bibinfo {author} {\bibfnamefont {K.~T.}\
  \bibnamefont {Flanagan}}, \bibinfo {author} {\bibfnamefont {L.~P.}\
  \bibnamefont {Gaffney}}, \bibinfo {author} {\bibfnamefont {L.}~\bibnamefont
  {Ghys}}, \bibinfo {author} {\bibfnamefont {A.}~\bibnamefont {Gottberg}},
  \bibinfo {author} {\bibfnamefont {M.}~\bibnamefont {Huyse}}, \bibinfo
  {author} {\bibfnamefont {S.}~\bibnamefont {Kreim}}, \bibinfo {author}
  {\bibfnamefont {P.}~\bibnamefont {Kunz}}, \bibinfo {author} {\bibfnamefont
  {D.}~\bibnamefont {Lunney}}, \bibinfo {author} {\bibfnamefont {K.~M.}\
  \bibnamefont {Lynch}}, \bibinfo {author} {\bibfnamefont {V.}~\bibnamefont
  {Manea}}, \bibinfo {author} {\bibfnamefont {Y.~M.}\ \bibnamefont
  {Palenzuela}}, \bibinfo {author} {\bibfnamefont {T.~M.}\ \bibnamefont
  {Medonca}}, \bibinfo {author} {\bibfnamefont {P.~L.}\ \bibnamefont
  {Molkanov}}, \bibinfo {author} {\bibfnamefont {M.}~\bibnamefont {Mougeot}},
  \bibinfo {author} {\bibfnamefont {J.~P.}\ \bibnamefont {Ramos}}, \bibinfo
  {author} {\bibfnamefont {M.}~\bibnamefont {Rosenbusch}}, \bibinfo {author}
  {\bibfnamefont {R.~E.}\ \bibnamefont {Rossel}}, \bibinfo {author}
  {\bibfnamefont {S.}~\bibnamefont {Rothe}}, \bibinfo {author} {\bibfnamefont
  {L.}~\bibnamefont {Schweikhard}}, \bibinfo {author} {\bibfnamefont {M.~D.}\
  \bibnamefont {Seliverstov}}, \bibinfo {author} {\bibfnamefont
  {P.}~\bibnamefont {Spagnoletti}}, \bibinfo {author} {\bibfnamefont
  {C.}~\bibnamefont {Van~Beveren}}, \bibinfo {author} {\bibfnamefont
  {M.}~\bibnamefont {Veinhard}}, \bibinfo {author} {\bibfnamefont
  {E.}~\bibnamefont {Verstraelen}}, \bibinfo {author} {\bibfnamefont
  {A.}~\bibnamefont {Welker}}, \bibinfo {author} {\bibfnamefont
  {K.}~\bibnamefont {Wendt}}, \bibinfo {author} {\bibfnamefont {R.~N.}\
  \bibnamefont {Wolf}}, \bibinfo {author} {\bibfnamefont {A.}~\bibnamefont
  {Zadvornaya}},\ and\ \bibinfo {author} {\bibfnamefont {K.}~\bibnamefont
  {Zuber}},\ }\bibfield  {title} {\bibinfo {title} {{Charge radii, moments, and
  masses of mercury isotopes across the $N=126$ shell closure}},\ }\href
  {https://doi.org/10.1103/PhysRevC.104.054322} {\bibfield  {journal} {\bibinfo
   {journal} {Phys.\ Rev.\ C}\ }\textbf {\bibinfo {volume} {104}},\ \bibinfo
  {pages} {054322} (\bibinfo {year} {2021})}\BibitemShut {NoStop}%
\bibitem [{\citenamefont {Reinhard}\ and\ \citenamefont
  {Nazarewicz}(2021)}]{Reinhard.2021}%
  \BibitemOpen
  \bibfield  {author} {\bibinfo {author} {\bibfnamefont {P.-G.}\ \bibnamefont
  {Reinhard}}\ and\ \bibinfo {author} {\bibfnamefont {W.}~\bibnamefont
  {Nazarewicz}},\ }\bibfield  {title} {\bibinfo {title} {Nuclear charge
  densities in spherical and deformed nuclei: Toward precise calculations of
  charge radii},\ }\href {https://doi.org/10.1103/PhysRevC.103.054310}
  {\bibfield  {journal} {\bibinfo  {journal} {Phys.\ Rev.\ C}\ }\textbf
  {\bibinfo {volume} {103}},\ \bibinfo {pages} {054310} (\bibinfo {year}
  {2021})}\BibitemShut {NoStop}%
\bibitem [{\citenamefont {Perera}\ \emph {et~al.}(2021)\citenamefont {Perera},
  \citenamefont {Afanasjev},\ and\ \citenamefont {Ring}}]{Perera.2021}%
  \BibitemOpen
  \bibfield  {author} {\bibinfo {author} {\bibfnamefont {U.~C.}\ \bibnamefont
  {Perera}}, \bibinfo {author} {\bibfnamefont {A.~V.}\ \bibnamefont
  {Afanasjev}},\ and\ \bibinfo {author} {\bibfnamefont {P.}~\bibnamefont
  {Ring}},\ }\bibfield  {title} {\bibinfo {title} {Charge radii in covariant
  density functional theory:{A} global view},\ }\href
  {https://doi.org/10.1103/PhysRevC.104.064313} {\bibfield  {journal} {\bibinfo
   {journal} {Phys.\ Rev.\ C}\ }\textbf {\bibinfo {volume} {104}},\ \bibinfo
  {pages} {064313} (\bibinfo {year} {2021})}\BibitemShut {NoStop}%
\bibitem [{\citenamefont {Klein}\ \emph {et~al.}(1996)\citenamefont {Klein},
  \citenamefont {Brown}, \citenamefont {Georg}, \citenamefont {Keim},
  \citenamefont {Lievens}, \citenamefont {Neugart}, \citenamefont {Neuroth},
  \citenamefont {Silverans}, \citenamefont {Vermeeren},\ and\ \citenamefont
  {Collaboration}}]{Klein.1996}%
  \BibitemOpen
  \bibfield  {author} {\bibinfo {author} {\bibfnamefont {A.}~\bibnamefont
  {Klein}}, \bibinfo {author} {\bibfnamefont {B.~A.}\ \bibnamefont {Brown}},
  \bibinfo {author} {\bibfnamefont {U.}~\bibnamefont {Georg}}, \bibinfo
  {author} {\bibfnamefont {M.}~\bibnamefont {Keim}}, \bibinfo {author}
  {\bibfnamefont {P.}~\bibnamefont {Lievens}}, \bibinfo {author} {\bibfnamefont
  {R.}~\bibnamefont {Neugart}}, \bibinfo {author} {\bibfnamefont
  {M.}~\bibnamefont {Neuroth}}, \bibinfo {author} {\bibfnamefont {R.~E.}\
  \bibnamefont {Silverans}}, \bibinfo {author} {\bibfnamefont {L.}~\bibnamefont
  {Vermeeren}},\ and\ \bibinfo {author} {\bibfnamefont {I.}~\bibnamefont
  {Collaboration}},\ }\bibfield  {title} {\bibinfo {title} {Moments and mean
  square charge radii of short-lived argon isotopes},\ }\href
  {https://doi.org/10.1016/0375-9474(96)00192-3} {\bibfield  {journal}
  {\bibinfo  {journal} {Nucl. Phys. A}\ }\textbf {\bibinfo {volume} {607}},\
  \bibinfo {pages} {1--22} (\bibinfo {year} {1996})}\BibitemShut {NoStop}%
\bibitem [{\citenamefont {Rossi}\ \emph {et~al.}(2015)\citenamefont {Rossi},
  \citenamefont {Minamisono}, \citenamefont {Asberry}, \citenamefont {Bollen},
  \citenamefont {Brown}, \citenamefont {Cooper}, \citenamefont {Isherwood},
  \citenamefont {Mantica}, \citenamefont {Miller}, \citenamefont {Morrissey},
  \citenamefont {Ringle}, \citenamefont {Rodriguez}, \citenamefont {Ryder},
  \citenamefont {Smith}, \citenamefont {Strum},\ and\ \citenamefont
  {Sumithrarachchi}}]{Rossi.2015}%
  \BibitemOpen
  \bibfield  {author} {\bibinfo {author} {\bibfnamefont {D.~M.}\ \bibnamefont
  {Rossi}}, \bibinfo {author} {\bibfnamefont {K.}~\bibnamefont {Minamisono}},
  \bibinfo {author} {\bibfnamefont {H.~B.}\ \bibnamefont {Asberry}}, \bibinfo
  {author} {\bibfnamefont {G.}~\bibnamefont {Bollen}}, \bibinfo {author}
  {\bibfnamefont {B.~A.}\ \bibnamefont {Brown}}, \bibinfo {author}
  {\bibfnamefont {K.}~\bibnamefont {Cooper}}, \bibinfo {author} {\bibfnamefont
  {B.}~\bibnamefont {Isherwood}}, \bibinfo {author} {\bibfnamefont {P.~F.}\
  \bibnamefont {Mantica}}, \bibinfo {author} {\bibfnamefont {A.}~\bibnamefont
  {Miller}}, \bibinfo {author} {\bibfnamefont {D.~J.}\ \bibnamefont
  {Morrissey}}, \bibinfo {author} {\bibfnamefont {R.}~\bibnamefont {Ringle}},
  \bibinfo {author} {\bibfnamefont {J.~A.}\ \bibnamefont {Rodriguez}}, \bibinfo
  {author} {\bibfnamefont {C.~A.}\ \bibnamefont {Ryder}}, \bibinfo {author}
  {\bibfnamefont {A.}~\bibnamefont {Smith}}, \bibinfo {author} {\bibfnamefont
  {R.}~\bibnamefont {Strum}},\ and\ \bibinfo {author} {\bibfnamefont
  {C.}~\bibnamefont {Sumithrarachchi}},\ }\bibfield  {title} {\bibinfo {title}
  {{Charge radii of neutron-deficient $^{36}$K and $^{37}$K}},\ }\href
  {https://doi.org/10.1103/PhysRevC.92.014305} {\bibfield  {journal} {\bibinfo
  {journal} {Phys.\ Rev.\ C}\ }\textbf {\bibinfo {volume} {92}},\ \bibinfo
  {pages} {014305} (\bibinfo {year} {2015})}\BibitemShut {NoStop}%
\bibitem [{\citenamefont {Miller}\ \emph
  {et~al.}(2019{\natexlab{a}})\citenamefont {Miller}, \citenamefont
  {Minamisono}, \citenamefont {Klose}, \citenamefont {Garand}, \citenamefont
  {Kujawa}, \citenamefont {Lantis}, \citenamefont {Liu}, \citenamefont
  {Maa{\ss}}, \citenamefont {Mantica}, \citenamefont {Nazarewicz},
  \citenamefont {N{\"o}rtersh{\"a}user}, \citenamefont {Pineda}, \citenamefont
  {Reinhard}, \citenamefont {Rossi}, \citenamefont {Sommer}, \citenamefont
  {Sumithrarachchi}, \citenamefont {Teigelh{\"o}fer},\ and\ \citenamefont
  {Watkins}}]{Miller.2019}%
  \BibitemOpen
  \bibfield  {author} {\bibinfo {author} {\bibfnamefont {A.~J.}\ \bibnamefont
  {Miller}}, \bibinfo {author} {\bibfnamefont {K.}~\bibnamefont {Minamisono}},
  \bibinfo {author} {\bibfnamefont {A.}~\bibnamefont {Klose}}, \bibinfo
  {author} {\bibfnamefont {D.}~\bibnamefont {Garand}}, \bibinfo {author}
  {\bibfnamefont {C.}~\bibnamefont {Kujawa}}, \bibinfo {author} {\bibfnamefont
  {J.~D.}\ \bibnamefont {Lantis}}, \bibinfo {author} {\bibfnamefont
  {Y.}~\bibnamefont {Liu}}, \bibinfo {author} {\bibfnamefont {B.}~\bibnamefont
  {Maa{\ss}}}, \bibinfo {author} {\bibfnamefont {P.~F.}\ \bibnamefont
  {Mantica}}, \bibinfo {author} {\bibfnamefont {W.}~\bibnamefont {Nazarewicz}},
  \bibinfo {author} {\bibfnamefont {W.}~\bibnamefont {N{\"o}rtersh{\"a}user}},
  \bibinfo {author} {\bibfnamefont {S.~V.}\ \bibnamefont {Pineda}}, \bibinfo
  {author} {\bibfnamefont {P.-G.}\ \bibnamefont {Reinhard}}, \bibinfo {author}
  {\bibfnamefont {D.~M.}\ \bibnamefont {Rossi}}, \bibinfo {author}
  {\bibfnamefont {F.}~\bibnamefont {Sommer}}, \bibinfo {author} {\bibfnamefont
  {C.}~\bibnamefont {Sumithrarachchi}}, \bibinfo {author} {\bibfnamefont
  {A.}~\bibnamefont {Teigelh{\"o}fer}},\ and\ \bibinfo {author} {\bibfnamefont
  {J.}~\bibnamefont {Watkins}},\ }\bibfield  {title} {\bibinfo {title} {Proton
  superfluidity and charge radii in proton-rich calcium isotopes},\ }\href
  {https://doi.org/10.1038/s41567-019-0416-9} {\bibfield  {journal} {\bibinfo
  {journal} {Nat. Phys.}\ }\textbf {\bibinfo {volume} {15}},\ \bibinfo {pages}
  {432--436} (\bibinfo {year} {2019}{\natexlab{a}})}\BibitemShut {NoStop}%
\bibitem [{\citenamefont {Wienholtz}\ \emph {et~al.}(2013)\citenamefont
  {Wienholtz}, \citenamefont {Beck}, \citenamefont {Blaum}, \citenamefont
  {Borgmann}, \citenamefont {Breitenfeldt}, \citenamefont {Cakirli},
  \citenamefont {George}, \citenamefont {Herfurth}, \citenamefont {Holt},
  \citenamefont {Kowalska}, \citenamefont {Kreim}, \citenamefont {Lunney},
  \citenamefont {Manea}, \citenamefont {Menendez}, \citenamefont {Neidherr},
  \citenamefont {Rosenbusch}, \citenamefont {Schweikhard}, \citenamefont
  {Schwenk}, \citenamefont {Simonis}, \citenamefont {Stanja}, \citenamefont
  {Wolf},\ and\ \citenamefont {Zuber}}]{Wienholtz.2013}%
  \BibitemOpen
  \bibfield  {author} {\bibinfo {author} {\bibfnamefont {F.}~\bibnamefont
  {Wienholtz}}, \bibinfo {author} {\bibfnamefont {D.}~\bibnamefont {Beck}},
  \bibinfo {author} {\bibfnamefont {K.}~\bibnamefont {Blaum}}, \bibinfo
  {author} {\bibfnamefont {C.}~\bibnamefont {Borgmann}}, \bibinfo {author}
  {\bibfnamefont {M.}~\bibnamefont {Breitenfeldt}}, \bibinfo {author}
  {\bibfnamefont {R.~B.}\ \bibnamefont {Cakirli}}, \bibinfo {author}
  {\bibfnamefont {S.}~\bibnamefont {George}}, \bibinfo {author} {\bibfnamefont
  {F.}~\bibnamefont {Herfurth}}, \bibinfo {author} {\bibfnamefont {J.~D.}\
  \bibnamefont {Holt}}, \bibinfo {author} {\bibfnamefont {M.}~\bibnamefont
  {Kowalska}}, \bibinfo {author} {\bibfnamefont {S.}~\bibnamefont {Kreim}},
  \bibinfo {author} {\bibfnamefont {D.}~\bibnamefont {Lunney}}, \bibinfo
  {author} {\bibfnamefont {V.}~\bibnamefont {Manea}}, \bibinfo {author}
  {\bibfnamefont {J.}~\bibnamefont {Menendez}}, \bibinfo {author}
  {\bibfnamefont {D.}~\bibnamefont {Neidherr}}, \bibinfo {author}
  {\bibfnamefont {M.}~\bibnamefont {Rosenbusch}}, \bibinfo {author}
  {\bibfnamefont {L.}~\bibnamefont {Schweikhard}}, \bibinfo {author}
  {\bibfnamefont {A.}~\bibnamefont {Schwenk}}, \bibinfo {author} {\bibfnamefont
  {J.}~\bibnamefont {Simonis}}, \bibinfo {author} {\bibfnamefont
  {J.}~\bibnamefont {Stanja}}, \bibinfo {author} {\bibfnamefont {R.~N.}\
  \bibnamefont {Wolf}},\ and\ \bibinfo {author} {\bibfnamefont
  {K.}~\bibnamefont {Zuber}},\ }\bibfield  {title} {\bibinfo {title} {Masses of
  exotic calcium isotopes pin down nuclear forces},\ }\href
  {https://doi.org/10.1038/nature12226} {\bibfield  {journal} {\bibinfo
  {journal} {Nature}\ }\textbf {\bibinfo {volume} {498}},\ \bibinfo {pages}
  {346--349} (\bibinfo {year} {2013})}\BibitemShut {NoStop}%
\bibitem [{\citenamefont {Rosenbusch}\ \emph {et~al.}(2015)\citenamefont
  {Rosenbusch}, \citenamefont {Ascher}, \citenamefont {Atanasov}, \citenamefont
  {Barbieri}, \citenamefont {Beck}, \citenamefont {Blaum}, \citenamefont
  {Borgmann}, \citenamefont {Breitenfeldt}, \citenamefont {Cakirli},
  \citenamefont {Cipollone}, \citenamefont {George}, \citenamefont {Herfurth},
  \citenamefont {Kowalska}, \citenamefont {Kreim}, \citenamefont {Lunney},
  \citenamefont {Manea}, \citenamefont {Navr\'atil}, \citenamefont {Neidherr},
  \citenamefont {Schweikhard}, \citenamefont {Som\`a}, \citenamefont {Stanja},
  \citenamefont {Wienholtz}, \citenamefont {Wolf},\ and\ \citenamefont
  {Zuber}}]{Rosenbusch.2015}%
  \BibitemOpen
  \bibfield  {author} {\bibinfo {author} {\bibfnamefont {M.}~\bibnamefont
  {Rosenbusch}}, \bibinfo {author} {\bibfnamefont {P.}~\bibnamefont {Ascher}},
  \bibinfo {author} {\bibfnamefont {D.}~\bibnamefont {Atanasov}}, \bibinfo
  {author} {\bibfnamefont {C.}~\bibnamefont {Barbieri}}, \bibinfo {author}
  {\bibfnamefont {D.}~\bibnamefont {Beck}}, \bibinfo {author} {\bibfnamefont
  {K.}~\bibnamefont {Blaum}}, \bibinfo {author} {\bibfnamefont
  {C.}~\bibnamefont {Borgmann}}, \bibinfo {author} {\bibfnamefont
  {M.}~\bibnamefont {Breitenfeldt}}, \bibinfo {author} {\bibfnamefont {R.~B.}\
  \bibnamefont {Cakirli}}, \bibinfo {author} {\bibfnamefont {A.}~\bibnamefont
  {Cipollone}}, \bibinfo {author} {\bibfnamefont {S.}~\bibnamefont {George}},
  \bibinfo {author} {\bibfnamefont {F.}~\bibnamefont {Herfurth}}, \bibinfo
  {author} {\bibfnamefont {M.}~\bibnamefont {Kowalska}}, \bibinfo {author}
  {\bibfnamefont {S.}~\bibnamefont {Kreim}}, \bibinfo {author} {\bibfnamefont
  {D.}~\bibnamefont {Lunney}}, \bibinfo {author} {\bibfnamefont
  {V.}~\bibnamefont {Manea}}, \bibinfo {author} {\bibfnamefont
  {P.}~\bibnamefont {Navr\'atil}}, \bibinfo {author} {\bibfnamefont
  {D.}~\bibnamefont {Neidherr}}, \bibinfo {author} {\bibfnamefont
  {L.}~\bibnamefont {Schweikhard}}, \bibinfo {author} {\bibfnamefont
  {V.}~\bibnamefont {Som\`a}}, \bibinfo {author} {\bibfnamefont
  {J.}~\bibnamefont {Stanja}}, \bibinfo {author} {\bibfnamefont
  {F.}~\bibnamefont {Wienholtz}}, \bibinfo {author} {\bibfnamefont {R.~N.}\
  \bibnamefont {Wolf}},\ and\ \bibinfo {author} {\bibfnamefont
  {K.}~\bibnamefont {Zuber}},\ }\bibfield  {title} {\bibinfo {title} {{Probing
  the $N=32$ Shell Closure below the Magic Proton Number $Z=20$: Mass
  Measurements of the Exotic Isotopes $^{52,53}\mathrm{K}$}},\ }\href
  {https://doi.org/10.1103/PhysRevLett.114.202501} {\bibfield  {journal}
  {\bibinfo  {journal} {Phys. Rev. Lett.}\ }\textbf {\bibinfo {volume} {114}},\
  \bibinfo {pages} {202501} (\bibinfo {year} {2015})}\BibitemShut {NoStop}%
\bibitem [{\citenamefont {Gade}\ \emph {et~al.}(2006)\citenamefont {Gade},
  \citenamefont {Janssens}, \citenamefont {Bazin}, \citenamefont {Broda},
  \citenamefont {Brown}, \citenamefont {Campbell}, \citenamefont {Carpenter},
  \citenamefont {Cook}, \citenamefont {Deacon}, \citenamefont {Dinca},
  \citenamefont {Fornal}, \citenamefont {Freeman}, \citenamefont {Glasmacher},
  \citenamefont {Hansen}, \citenamefont {Kay}, \citenamefont {Mantica},
  \citenamefont {Mueller}, \citenamefont {Terry}, \citenamefont {Tostevin},\
  and\ \citenamefont {Zhu}}]{Gade.2006}%
  \BibitemOpen
  \bibfield  {author} {\bibinfo {author} {\bibfnamefont {A.}~\bibnamefont
  {Gade}}, \bibinfo {author} {\bibfnamefont {R.~V.~F.}\ \bibnamefont
  {Janssens}}, \bibinfo {author} {\bibfnamefont {D.}~\bibnamefont {Bazin}},
  \bibinfo {author} {\bibfnamefont {R.}~\bibnamefont {Broda}}, \bibinfo
  {author} {\bibfnamefont {B.~A.}\ \bibnamefont {Brown}}, \bibinfo {author}
  {\bibfnamefont {C.~M.}\ \bibnamefont {Campbell}}, \bibinfo {author}
  {\bibfnamefont {M.~P.}\ \bibnamefont {Carpenter}}, \bibinfo {author}
  {\bibfnamefont {J.~M.}\ \bibnamefont {Cook}}, \bibinfo {author}
  {\bibfnamefont {A.~N.}\ \bibnamefont {Deacon}}, \bibinfo {author}
  {\bibfnamefont {D.-C.}\ \bibnamefont {Dinca}}, \bibinfo {author}
  {\bibfnamefont {B.}~\bibnamefont {Fornal}}, \bibinfo {author} {\bibfnamefont
  {S.~J.}\ \bibnamefont {Freeman}}, \bibinfo {author} {\bibfnamefont
  {T.}~\bibnamefont {Glasmacher}}, \bibinfo {author} {\bibfnamefont {P.~G.}\
  \bibnamefont {Hansen}}, \bibinfo {author} {\bibfnamefont {B.~P.}\
  \bibnamefont {Kay}}, \bibinfo {author} {\bibfnamefont {P.~F.}\ \bibnamefont
  {Mantica}}, \bibinfo {author} {\bibfnamefont {W.~F.}\ \bibnamefont
  {Mueller}}, \bibinfo {author} {\bibfnamefont {J.~R.}\ \bibnamefont {Terry}},
  \bibinfo {author} {\bibfnamefont {J.~A.}\ \bibnamefont {Tostevin}},\ and\
  \bibinfo {author} {\bibfnamefont {S.}~\bibnamefont {Zhu}},\ }\bibfield
  {title} {\bibinfo {title} {{Cross-shell excitation in two-proton knockout:
  Structure of $^{52}$Ca}},\ }\href
  {https://doi.org/10.1103/PhysRevC.74.021302} {\bibfield  {journal} {\bibinfo
  {journal} {Phys.\ Rev.\ C}\ }\textbf {\bibinfo {volume} {74}},\ \bibinfo
  {pages} {021302} (\bibinfo {year} {2006})}\BibitemShut {NoStop}%
\bibitem [{\citenamefont {Koszor{\'u}s}\ \emph {et~al.}(2021)\citenamefont
  {Koszor{\'u}s}, \citenamefont {Yang}, \citenamefont {Jiang}, \citenamefont
  {Novario}, \citenamefont {Bai}, \citenamefont {Billowes}, \citenamefont
  {Binnersley}, \citenamefont {Bissell}, \citenamefont {Cocolios},
  \citenamefont {Cooper}, \citenamefont {de~Groote}, \citenamefont
  {Ekstr{\"o}m}, \citenamefont {Flanagan}, \citenamefont {Forss{\'e}n},
  \citenamefont {Franchoo}, \citenamefont {Ruiz}, \citenamefont {Gustafsson},
  \citenamefont {Hagen}, \citenamefont {Jansen}, \citenamefont
  {Kanellakopoulos}, \citenamefont {Kortelainen}, \citenamefont {Nazarewicz},
  \citenamefont {Neyens}, \citenamefont {Papenbrock}, \citenamefont {Reinhard},
  \citenamefont {Ricketts}, \citenamefont {Sahoo}, \citenamefont {Vernon},\
  and\ \citenamefont {Wilkins}}]{Koszorus.2021}%
  \BibitemOpen
  \bibfield  {author} {\bibinfo {author} {\bibfnamefont {{\'A}.}~\bibnamefont
  {Koszor{\'u}s}}, \bibinfo {author} {\bibfnamefont {X.~F.}\ \bibnamefont
  {Yang}}, \bibinfo {author} {\bibfnamefont {W.~G.}\ \bibnamefont {Jiang}},
  \bibinfo {author} {\bibfnamefont {S.~J.}\ \bibnamefont {Novario}}, \bibinfo
  {author} {\bibfnamefont {S.~W.}\ \bibnamefont {Bai}}, \bibinfo {author}
  {\bibfnamefont {J.}~\bibnamefont {Billowes}}, \bibinfo {author}
  {\bibfnamefont {C.~L.}\ \bibnamefont {Binnersley}}, \bibinfo {author}
  {\bibfnamefont {M.~L.}\ \bibnamefont {Bissell}}, \bibinfo {author}
  {\bibfnamefont {T.~E.}\ \bibnamefont {Cocolios}}, \bibinfo {author}
  {\bibfnamefont {B.~S.}\ \bibnamefont {Cooper}}, \bibinfo {author}
  {\bibfnamefont {R.~P.}\ \bibnamefont {de~Groote}}, \bibinfo {author}
  {\bibfnamefont {A.}~\bibnamefont {Ekstr{\"o}m}}, \bibinfo {author}
  {\bibfnamefont {K.~T.}\ \bibnamefont {Flanagan}}, \bibinfo {author}
  {\bibfnamefont {C.}~\bibnamefont {Forss{\'e}n}}, \bibinfo {author}
  {\bibfnamefont {S.}~\bibnamefont {Franchoo}}, \bibinfo {author}
  {\bibfnamefont {R.~F.~G.}\ \bibnamefont {Ruiz}}, \bibinfo {author}
  {\bibfnamefont {F.~P.}\ \bibnamefont {Gustafsson}}, \bibinfo {author}
  {\bibfnamefont {G.}~\bibnamefont {Hagen}}, \bibinfo {author} {\bibfnamefont
  {G.~R.}\ \bibnamefont {Jansen}}, \bibinfo {author} {\bibfnamefont
  {A.}~\bibnamefont {Kanellakopoulos}}, \bibinfo {author} {\bibfnamefont
  {M.}~\bibnamefont {Kortelainen}}, \bibinfo {author} {\bibfnamefont
  {W.}~\bibnamefont {Nazarewicz}}, \bibinfo {author} {\bibfnamefont
  {G.}~\bibnamefont {Neyens}}, \bibinfo {author} {\bibfnamefont
  {T.}~\bibnamefont {Papenbrock}}, \bibinfo {author} {\bibfnamefont {P.-G.}\
  \bibnamefont {Reinhard}}, \bibinfo {author} {\bibfnamefont {C.~M.}\
  \bibnamefont {Ricketts}}, \bibinfo {author} {\bibfnamefont {B.~K.}\
  \bibnamefont {Sahoo}}, \bibinfo {author} {\bibfnamefont {A.~R.}\ \bibnamefont
  {Vernon}},\ and\ \bibinfo {author} {\bibfnamefont {S.~G.}\ \bibnamefont
  {Wilkins}},\ }\bibfield  {title} {\bibinfo {title} {{Charge radii of exotic
  potassium isotopes challenge nuclear theory and the magic character of $N =
  32$}},\ }\href {https://doi.org/10.1038/s41567-020-01136-5} {\bibfield
  {journal} {\bibinfo  {journal} {Nat. Phys.}\ }\textbf {\bibinfo {volume}
  {17}},\ \bibinfo {pages} {439--443} (\bibinfo {year} {2021})}\BibitemShut
  {NoStop}%
\bibitem [{\citenamefont {{Garcia Ruiz}}\ and\ \citenamefont
  {Vernon}(2020)}]{Garcia.2020}%
  \BibitemOpen
  \bibfield  {author} {\bibinfo {author} {\bibfnamefont {R.~F.}\ \bibnamefont
  {{Garcia Ruiz}}}\ and\ \bibinfo {author} {\bibfnamefont {A.~R.}\ \bibnamefont
  {Vernon}},\ }\bibfield  {title} {\bibinfo {title} {Emergence of simple
  patterns in many-body systems: from macroscopic objects to the atomic
  nucleus},\ }\href {https://doi.org/10.1140/epja/s10050-020-00134-8}
  {\bibfield  {journal} {\bibinfo  {journal} {Eur. Phys. J. A}\ }\textbf
  {\bibinfo {volume} {56}},\ \bibinfo {pages} {136} (\bibinfo {year}
  {2020})}\BibitemShut {NoStop}%
\bibitem [{\citenamefont {{Garcia Ruiz}}\ \emph {et~al.}(2016)\citenamefont
  {{Garcia Ruiz}}, \citenamefont {Bissell}, \citenamefont {Blaum},
  \citenamefont {Ekstr{\"o}m}, \citenamefont {Fr{\"o}mmgen}, \citenamefont
  {Hagen}, \citenamefont {Hammen}, \citenamefont {Hebeler}, \citenamefont
  {Holt}, \citenamefont {Jansen}, \citenamefont {Kowalska}, \citenamefont
  {Kreim}, \citenamefont {Nazarewicz}, \citenamefont {Neugart}, \citenamefont
  {Neyens}, \citenamefont {N{\"o}rtersh{\"a}user}, \citenamefont {Papenbrock},
  \citenamefont {Papuga}, \citenamefont {Schwenk}, \citenamefont {Simonis},
  \citenamefont {Wendt},\ and\ \citenamefont {Yordanov}}]{GarciaRuiz.2016}%
  \BibitemOpen
  \bibfield  {author} {\bibinfo {author} {\bibfnamefont {R.~F.}\ \bibnamefont
  {{Garcia Ruiz}}}, \bibinfo {author} {\bibfnamefont {M.~L.}\ \bibnamefont
  {Bissell}}, \bibinfo {author} {\bibfnamefont {K.}~\bibnamefont {Blaum}},
  \bibinfo {author} {\bibfnamefont {A.}~\bibnamefont {Ekstr{\"o}m}}, \bibinfo
  {author} {\bibfnamefont {N.}~\bibnamefont {Fr{\"o}mmgen}}, \bibinfo {author}
  {\bibfnamefont {G.}~\bibnamefont {Hagen}}, \bibinfo {author} {\bibfnamefont
  {M.}~\bibnamefont {Hammen}}, \bibinfo {author} {\bibfnamefont
  {K.}~\bibnamefont {Hebeler}}, \bibinfo {author} {\bibfnamefont {J.~D.}\
  \bibnamefont {Holt}}, \bibinfo {author} {\bibfnamefont {G.~R.}\ \bibnamefont
  {Jansen}}, \bibinfo {author} {\bibfnamefont {M.}~\bibnamefont {Kowalska}},
  \bibinfo {author} {\bibfnamefont {K.}~\bibnamefont {Kreim}}, \bibinfo
  {author} {\bibfnamefont {W.}~\bibnamefont {Nazarewicz}}, \bibinfo {author}
  {\bibfnamefont {R.}~\bibnamefont {Neugart}}, \bibinfo {author} {\bibfnamefont
  {G.}~\bibnamefont {Neyens}}, \bibinfo {author} {\bibfnamefont
  {W.}~\bibnamefont {N{\"o}rtersh{\"a}user}}, \bibinfo {author} {\bibfnamefont
  {T.}~\bibnamefont {Papenbrock}}, \bibinfo {author} {\bibfnamefont
  {J.}~\bibnamefont {Papuga}}, \bibinfo {author} {\bibfnamefont
  {A.}~\bibnamefont {Schwenk}}, \bibinfo {author} {\bibfnamefont
  {J.}~\bibnamefont {Simonis}}, \bibinfo {author} {\bibfnamefont {K.~A.}\
  \bibnamefont {Wendt}},\ and\ \bibinfo {author} {\bibfnamefont {D.~T.}\
  \bibnamefont {Yordanov}},\ }\bibfield  {title} {\bibinfo {title}
  {Unexpectedly large charge radii of neutron-rich calcium isotopes},\ }\href
  {https://doi.org/10.1038/nphys3645} {\bibfield  {journal} {\bibinfo
  {journal} {Nat. Phys.}\ }\textbf {\bibinfo {volume} {12}},\ \bibinfo {pages}
  {594--598} (\bibinfo {year} {2016})}\BibitemShut {NoStop}%
\bibitem [{\citenamefont {Heylen}\ \emph {et~al.}(2016)\citenamefont {Heylen},
  \citenamefont {Babcock}, \citenamefont {Beerwerth}, \citenamefont {Billowes},
  \citenamefont {Bissell}, \citenamefont {Blaum}, \citenamefont {Bonnard},
  \citenamefont {Campbell}, \citenamefont {Cheal}, \citenamefont {{Day
  Goodacre}}, \citenamefont {Fedorov}, \citenamefont {Fritzsche}, \citenamefont
  {{Garcia Ruiz}}, \citenamefont {Geithner}, \citenamefont {Geppert},
  \citenamefont {Gins}, \citenamefont {Grob}, \citenamefont {Kowalska},
  \citenamefont {Kreim}, \citenamefont {Lenzi}, \citenamefont {Moore},
  \citenamefont {Maass}, \citenamefont {Malbrunot-Ettenauer}, \citenamefont
  {Marsh}, \citenamefont {Neugart}, \citenamefont {Neyens}, \citenamefont
  {N{\"o}rtersh{\"a}user}, \citenamefont {Otsuka}, \citenamefont {Papuga},
  \citenamefont {Rossel}, \citenamefont {Rothe}, \citenamefont {S{\'a}nchez},
  \citenamefont {Tsunoda}, \citenamefont {Wraith}, \citenamefont {Xie},
  \citenamefont {Yang},\ and\ \citenamefont {Yordanov}}]{Heylen.2016b}%
  \BibitemOpen
  \bibfield  {author} {\bibinfo {author} {\bibfnamefont {H.}~\bibnamefont
  {Heylen}}, \bibinfo {author} {\bibfnamefont {C.}~\bibnamefont {Babcock}},
  \bibinfo {author} {\bibfnamefont {R.}~\bibnamefont {Beerwerth}}, \bibinfo
  {author} {\bibfnamefont {J.}~\bibnamefont {Billowes}}, \bibinfo {author}
  {\bibfnamefont {M.~L.}\ \bibnamefont {Bissell}}, \bibinfo {author}
  {\bibfnamefont {K.}~\bibnamefont {Blaum}}, \bibinfo {author} {\bibfnamefont
  {J.}~\bibnamefont {Bonnard}}, \bibinfo {author} {\bibfnamefont
  {P.}~\bibnamefont {Campbell}}, \bibinfo {author} {\bibfnamefont
  {B.}~\bibnamefont {Cheal}}, \bibinfo {author} {\bibfnamefont
  {T.}~\bibnamefont {{Day Goodacre}}}, \bibinfo {author} {\bibfnamefont
  {D.}~\bibnamefont {Fedorov}}, \bibinfo {author} {\bibfnamefont
  {S.}~\bibnamefont {Fritzsche}}, \bibinfo {author} {\bibfnamefont {R.~F.}\
  \bibnamefont {{Garcia Ruiz}}}, \bibinfo {author} {\bibfnamefont
  {W.}~\bibnamefont {Geithner}}, \bibinfo {author} {\bibfnamefont
  {C.}~\bibnamefont {Geppert}}, \bibinfo {author} {\bibfnamefont
  {W.}~\bibnamefont {Gins}}, \bibinfo {author} {\bibfnamefont {L.~K.}\
  \bibnamefont {Grob}}, \bibinfo {author} {\bibfnamefont {M.}~\bibnamefont
  {Kowalska}}, \bibinfo {author} {\bibfnamefont {K.}~\bibnamefont {Kreim}},
  \bibinfo {author} {\bibfnamefont {S.~M.}\ \bibnamefont {Lenzi}}, \bibinfo
  {author} {\bibfnamefont {I.~D.}\ \bibnamefont {Moore}}, \bibinfo {author}
  {\bibfnamefont {B.}~\bibnamefont {Maass}}, \bibinfo {author} {\bibfnamefont
  {S.}~\bibnamefont {Malbrunot-Ettenauer}}, \bibinfo {author} {\bibfnamefont
  {B.}~\bibnamefont {Marsh}}, \bibinfo {author} {\bibfnamefont
  {R.}~\bibnamefont {Neugart}}, \bibinfo {author} {\bibfnamefont
  {G.}~\bibnamefont {Neyens}}, \bibinfo {author} {\bibfnamefont
  {W.}~\bibnamefont {N{\"o}rtersh{\"a}user}}, \bibinfo {author} {\bibfnamefont
  {T.}~\bibnamefont {Otsuka}}, \bibinfo {author} {\bibfnamefont
  {J.}~\bibnamefont {Papuga}}, \bibinfo {author} {\bibfnamefont
  {R.}~\bibnamefont {Rossel}}, \bibinfo {author} {\bibfnamefont
  {S.}~\bibnamefont {Rothe}}, \bibinfo {author} {\bibfnamefont
  {R.}~\bibnamefont {S{\'a}nchez}}, \bibinfo {author} {\bibfnamefont
  {Y.}~\bibnamefont {Tsunoda}}, \bibinfo {author} {\bibfnamefont
  {C.}~\bibnamefont {Wraith}}, \bibinfo {author} {\bibfnamefont
  {L.}~\bibnamefont {Xie}}, \bibinfo {author} {\bibfnamefont {X.~F.}\
  \bibnamefont {Yang}},\ and\ \bibinfo {author} {\bibfnamefont {D.~T.}\
  \bibnamefont {Yordanov}},\ }\bibfield  {title} {\bibinfo {title} {{Changes in
  nuclear structure along the Mn isotopic chain studied via charge radii}},\
  }\href {https://doi.org/10.1103/PhysRevC.94.054321} {\bibfield  {journal}
  {\bibinfo  {journal} {Phys.\ Rev.\ C}\ }\textbf {\bibinfo {volume} {94}},\
  \bibinfo {pages} {054321} (\bibinfo {year} {2016})}\BibitemShut {NoStop}%
\bibitem [{\citenamefont {Minamisono}\ \emph {et~al.}(2016)\citenamefont
  {Minamisono}, \citenamefont {Rossi}, \citenamefont {Beerwerth}, \citenamefont
  {Fritzsche}, \citenamefont {Garand}, \citenamefont {Klose}, \citenamefont
  {Liu}, \citenamefont {Maa{\ss}}, \citenamefont {Mantica}, \citenamefont
  {Miller}, \citenamefont {M{\"u}ller}, \citenamefont {Nazarewicz},
  \citenamefont {N{\"o}rtersh{\"a}user}, \citenamefont {Olsen}, \citenamefont
  {Pearson}, \citenamefont {Reinhard}, \citenamefont {Saperstein},
  \citenamefont {Sumithrarachchi},\ and\ \citenamefont
  {Tolokonnikov}}]{Minamisono.2016}%
  \BibitemOpen
  \bibfield  {author} {\bibinfo {author} {\bibfnamefont {K.}~\bibnamefont
  {Minamisono}}, \bibinfo {author} {\bibfnamefont {D.~M.}\ \bibnamefont
  {Rossi}}, \bibinfo {author} {\bibfnamefont {R.}~\bibnamefont {Beerwerth}},
  \bibinfo {author} {\bibfnamefont {S.}~\bibnamefont {Fritzsche}}, \bibinfo
  {author} {\bibfnamefont {D.}~\bibnamefont {Garand}}, \bibinfo {author}
  {\bibfnamefont {A.}~\bibnamefont {Klose}}, \bibinfo {author} {\bibfnamefont
  {Y.}~\bibnamefont {Liu}}, \bibinfo {author} {\bibfnamefont {B.}~\bibnamefont
  {Maa{\ss}}}, \bibinfo {author} {\bibfnamefont {P.~F.}\ \bibnamefont
  {Mantica}}, \bibinfo {author} {\bibfnamefont {A.~J.}\ \bibnamefont {Miller}},
  \bibinfo {author} {\bibfnamefont {P.}~\bibnamefont {M{\"u}ller}}, \bibinfo
  {author} {\bibfnamefont {W.}~\bibnamefont {Nazarewicz}}, \bibinfo {author}
  {\bibfnamefont {W.}~\bibnamefont {N{\"o}rtersh{\"a}user}}, \bibinfo {author}
  {\bibfnamefont {E.}~\bibnamefont {Olsen}}, \bibinfo {author} {\bibfnamefont
  {M.~R.}\ \bibnamefont {Pearson}}, \bibinfo {author} {\bibfnamefont {P.-G.}\
  \bibnamefont {Reinhard}}, \bibinfo {author} {\bibfnamefont {E.~E.}\
  \bibnamefont {Saperstein}}, \bibinfo {author} {\bibfnamefont
  {C.}~\bibnamefont {Sumithrarachchi}},\ and\ \bibinfo {author} {\bibfnamefont
  {S.~V.}\ \bibnamefont {Tolokonnikov}},\ }\bibfield  {title} {\bibinfo {title}
  {{Charge Radii of Neutron Deficient \textsuperscript{52,53}Fe Produced by
  Projectile Fragmentation}},\ }\href
  {https://doi.org/10.1103/PhysRevLett.117.252501} {\bibfield  {journal}
  {\bibinfo  {journal} {Phys. Rev. Lett.}\ }\textbf {\bibinfo {volume} {117}},\
  \bibinfo {pages} {252501} (\bibinfo {year} {2016})}\BibitemShut {NoStop}%
\bibitem [{\citenamefont {Kortelainen}\ \emph {et~al.}(2022)\citenamefont
  {Kortelainen}, \citenamefont {Sun}, \citenamefont {Hagen}, \citenamefont
  {Nazarewicz}, \citenamefont {Papenbrock},\ and\ \citenamefont
  {Reinhard}}]{Kortelainen2022}%
  \BibitemOpen
  \bibfield  {author} {\bibinfo {author} {\bibfnamefont {M.}~\bibnamefont
  {Kortelainen}}, \bibinfo {author} {\bibfnamefont {Z.}~\bibnamefont {Sun}},
  \bibinfo {author} {\bibfnamefont {G.}~\bibnamefont {Hagen}}, \bibinfo
  {author} {\bibfnamefont {W.}~\bibnamefont {Nazarewicz}}, \bibinfo {author}
  {\bibfnamefont {T.}~\bibnamefont {Papenbrock}},\ and\ \bibinfo {author}
  {\bibfnamefont {P.-G.}\ \bibnamefont {Reinhard}},\ }\bibfield  {title}
  {\bibinfo {title} {{Universal trend of charge radii of even-even Ca--Zn
  nuclei}},\ }\href {https://doi.org/10.1103/PhysRevC.105.L021303} {\bibfield
  {journal} {\bibinfo  {journal} {Phys.\ Rev.\ C}\ }\textbf {\bibinfo {volume}
  {105}},\ \bibinfo {pages} {L021303} (\bibinfo {year} {2022})}\BibitemShut
  {NoStop}%
\bibitem [{\citenamefont {Otsuka}\ \emph {et~al.}(1998)\citenamefont {Otsuka},
  \citenamefont {Honma},\ and\ \citenamefont {Mizusaki}}]{Otsuka.1998}%
  \BibitemOpen
  \bibfield  {author} {\bibinfo {author} {\bibfnamefont {T.}~\bibnamefont
  {Otsuka}}, \bibinfo {author} {\bibfnamefont {M.}~\bibnamefont {Honma}},\ and\
  \bibinfo {author} {\bibfnamefont {T.}~\bibnamefont {Mizusaki}},\ }\bibfield
  {title} {\bibinfo {title} {{Structure of the $N=Z=28$ Closed Shell Studied by
  Monte Carlo Shell Model Calculation}},\ }\href
  {https://doi.org/10.1103/PhysRevLett.81.1588} {\bibfield  {journal} {\bibinfo
   {journal} {Phys.\ Rev.\ Lett.}\ }\textbf {\bibinfo {volume} {81}},\ \bibinfo
  {pages} {1588--1591} (\bibinfo {year} {1998})}\BibitemShut {NoStop}%
\bibitem [{\citenamefont {Kraus}\ \emph {et~al.}(1994)\citenamefont {Kraus},
  \citenamefont {Egelhof}, \citenamefont {Fischer}, \citenamefont {Geissel},
  \citenamefont {Himmler}, \citenamefont {Nickel}, \citenamefont
  {M\"unzenberg}, \citenamefont {Schwab}, \citenamefont {Weiss}, \citenamefont
  {Friese}, \citenamefont {Gillitzer}, \citenamefont {K\"orner}, \citenamefont
  {Peter}, \citenamefont {Henning}, \citenamefont {Schiffer}, \citenamefont
  {Kratz}, \citenamefont {Chulkov}, \citenamefont {Golovkov}, \citenamefont
  {Ogloblin},\ and\ \citenamefont {Brown}}]{Kraus.1994}%
  \BibitemOpen
  \bibfield  {author} {\bibinfo {author} {\bibfnamefont {G.}~\bibnamefont
  {Kraus}}, \bibinfo {author} {\bibfnamefont {P.}~\bibnamefont {Egelhof}},
  \bibinfo {author} {\bibfnamefont {C.}~\bibnamefont {Fischer}}, \bibinfo
  {author} {\bibfnamefont {H.}~\bibnamefont {Geissel}}, \bibinfo {author}
  {\bibfnamefont {A.}~\bibnamefont {Himmler}}, \bibinfo {author} {\bibfnamefont
  {F.}~\bibnamefont {Nickel}}, \bibinfo {author} {\bibfnamefont
  {G.}~\bibnamefont {M\"unzenberg}}, \bibinfo {author} {\bibfnamefont
  {W.}~\bibnamefont {Schwab}}, \bibinfo {author} {\bibfnamefont
  {A.}~\bibnamefont {Weiss}}, \bibinfo {author} {\bibfnamefont
  {J.}~\bibnamefont {Friese}}, \bibinfo {author} {\bibfnamefont
  {A.}~\bibnamefont {Gillitzer}}, \bibinfo {author} {\bibfnamefont {H.~J.}\
  \bibnamefont {K\"orner}}, \bibinfo {author} {\bibfnamefont {M.}~\bibnamefont
  {Peter}}, \bibinfo {author} {\bibfnamefont {W.~F.}\ \bibnamefont {Henning}},
  \bibinfo {author} {\bibfnamefont {J.~P.}\ \bibnamefont {Schiffer}}, \bibinfo
  {author} {\bibfnamefont {J.~V.}\ \bibnamefont {Kratz}}, \bibinfo {author}
  {\bibfnamefont {L.}~\bibnamefont {Chulkov}}, \bibinfo {author} {\bibfnamefont
  {M.}~\bibnamefont {Golovkov}}, \bibinfo {author} {\bibfnamefont
  {A.}~\bibnamefont {Ogloblin}},\ and\ \bibinfo {author} {\bibfnamefont
  {B.~A.}\ \bibnamefont {Brown}},\ }\bibfield  {title} {\bibinfo {title}
  {{Proton Inelastic Scattering on $^{56}\mathrm{Ni}$ in Inverse Kinematics}},\
  }\href {https://doi.org/10.1103/PhysRevLett.73.1773} {\bibfield  {journal}
  {\bibinfo  {journal} {Phys. Rev. Lett.}\ }\textbf {\bibinfo {volume} {73}},\
  \bibinfo {pages} {1773} (\bibinfo {year} {1994})}\BibitemShut {NoStop}%
\bibitem [{\citenamefont {Arnswald}\ \emph {et~al.}(2021)\citenamefont
  {Arnswald}, \citenamefont {Blazhev}, \citenamefont {Nowacki}, \citenamefont
  {Petkov}, \citenamefont {Reiter}, \citenamefont {Braunroth}, \citenamefont
  {Dewald}, \citenamefont {Droste}, \citenamefont {Fransen}, \citenamefont
  {Hirsch}, \citenamefont {Karayonchev}, \citenamefont {Kaya}, \citenamefont
  {Lewandowski}, \citenamefont {Müller-Gatermann}, \citenamefont {Seidlitz},
  \citenamefont {Siebeck}, \citenamefont {Vogt}, \citenamefont {Werner},\ and\
  \citenamefont {Zell}}]{Arnswald2021}%
  \BibitemOpen
  \bibfield  {author} {\bibinfo {author} {\bibfnamefont {K.}~\bibnamefont
  {Arnswald}}, \bibinfo {author} {\bibfnamefont {A.}~\bibnamefont {Blazhev}},
  \bibinfo {author} {\bibfnamefont {F.}~\bibnamefont {Nowacki}}, \bibinfo
  {author} {\bibfnamefont {P.}~\bibnamefont {Petkov}}, \bibinfo {author}
  {\bibfnamefont {P.}~\bibnamefont {Reiter}}, \bibinfo {author} {\bibfnamefont
  {T.}~\bibnamefont {Braunroth}}, \bibinfo {author} {\bibfnamefont
  {A.}~\bibnamefont {Dewald}}, \bibinfo {author} {\bibfnamefont
  {M.}~\bibnamefont {Droste}}, \bibinfo {author} {\bibfnamefont
  {C.}~\bibnamefont {Fransen}}, \bibinfo {author} {\bibfnamefont
  {R.}~\bibnamefont {Hirsch}}, \bibinfo {author} {\bibfnamefont
  {V.}~\bibnamefont {Karayonchev}}, \bibinfo {author} {\bibfnamefont
  {L.}~\bibnamefont {Kaya}}, \bibinfo {author} {\bibfnamefont {L.}~\bibnamefont
  {Lewandowski}}, \bibinfo {author} {\bibfnamefont {C.}~\bibnamefont
  {Müller-Gatermann}}, \bibinfo {author} {\bibfnamefont {M.}~\bibnamefont
  {Seidlitz}}, \bibinfo {author} {\bibfnamefont {B.}~\bibnamefont {Siebeck}},
  \bibinfo {author} {\bibfnamefont {A.}~\bibnamefont {Vogt}}, \bibinfo {author}
  {\bibfnamefont {D.}~\bibnamefont {Werner}},\ and\ \bibinfo {author}
  {\bibfnamefont {K.}~\bibnamefont {Zell}},\ }\bibfield  {title} {\bibinfo
  {title} {Enhanced quadrupole collectivity in doubly-magic $^{56}${Ni}:
  Lifetime measurements of the 4$_1^+$ and 6$_1^+$ states},\ }\href
  {https://doi.org/10.1016/j.physletb.2021.136592} {\bibfield  {journal}
  {\bibinfo  {journal} {Phys. Lett. B}\ }\textbf {\bibinfo {volume} {820}},\
  \bibinfo {pages} {136592} (\bibinfo {year} {2021})}\BibitemShut {NoStop}%
\bibitem [{\citenamefont {Berryman}\ \emph {et~al.}(2009)\citenamefont
  {Berryman}, \citenamefont {Minamisono}, \citenamefont {Rogers}, \citenamefont
  {Brown}, \citenamefont {Crawford}, \citenamefont {Grinyer}, \citenamefont
  {Mantica}, \citenamefont {Stoker},\ and\ \citenamefont
  {Towner}}]{Berryman.2009}%
  \BibitemOpen
  \bibfield  {author} {\bibinfo {author} {\bibfnamefont {J.~S.}\ \bibnamefont
  {Berryman}}, \bibinfo {author} {\bibfnamefont {K.}~\bibnamefont
  {Minamisono}}, \bibinfo {author} {\bibfnamefont {W.~F.}\ \bibnamefont
  {Rogers}}, \bibinfo {author} {\bibfnamefont {B.~A.}\ \bibnamefont {Brown}},
  \bibinfo {author} {\bibfnamefont {H.~L.}\ \bibnamefont {Crawford}}, \bibinfo
  {author} {\bibfnamefont {G.~F.}\ \bibnamefont {Grinyer}}, \bibinfo {author}
  {\bibfnamefont {P.~F.}\ \bibnamefont {Mantica}}, \bibinfo {author}
  {\bibfnamefont {J.~B.}\ \bibnamefont {Stoker}},\ and\ \bibinfo {author}
  {\bibfnamefont {I.~S.}\ \bibnamefont {Towner}},\ }\bibfield  {title}
  {\bibinfo {title} {{Doubly-magic nature of \textsuperscript{56}Ni :
  Measurement of the ground state nuclear magnetic dipole moment of
  \textsuperscript{55}Ni}},\ }\href
  {https://doi.org/10.1103/PhysRevC.79.064305} {\bibfield  {journal} {\bibinfo
  {journal} {Phys.\ Rev.\ C}\ }\textbf {\bibinfo {volume} {79}},\ \bibinfo
  {pages} {064305} (\bibinfo {year} {2009})}\BibitemShut {NoStop}%
\bibitem [{\citenamefont {Callaghan}\ \emph {et~al.}(1973)\citenamefont
  {Callaghan}, \citenamefont {Kaplan},\ and\ \citenamefont
  {Stone}}]{Callaghan.1973}%
  \BibitemOpen
  \bibfield  {author} {\bibinfo {author} {\bibfnamefont {P.~T.}\ \bibnamefont
  {Callaghan}}, \bibinfo {author} {\bibfnamefont {M.}~\bibnamefont {Kaplan}},\
  and\ \bibinfo {author} {\bibfnamefont {N.~J.}\ \bibnamefont {Stone}},\
  }\bibfield  {title} {\bibinfo {title} {{The magnetic dipole moment of
  \textsuperscript{55}Co}},\ }\href
  {https://doi.org/10.1016/0375-9474(73)90320-5} {\bibfield  {journal}
  {\bibinfo  {journal} {Nucl. Phys. A}\ }\textbf {\bibinfo {volume} {201}},\
  \bibinfo {pages} {561--569} (\bibinfo {year} {1973})}\BibitemShut {NoStop}%
\bibitem [{\citenamefont {Cocolios}\ \emph {et~al.}(2009)\citenamefont
  {Cocolios}, \citenamefont {Andreyev}, \citenamefont {Bastin}, \citenamefont
  {Bree}, \citenamefont {B{\"u}scher}, \citenamefont {Elseviers}, \citenamefont
  {Gentens}, \citenamefont {Huyse}, \citenamefont {Kudryavtsev}, \citenamefont
  {Pauwels}, \citenamefont {Sonoda}, \citenamefont {{van den Bergh}},\ and\
  \citenamefont {{van Duppen}}}]{Cocolios.2009}%
  \BibitemOpen
  \bibfield  {author} {\bibinfo {author} {\bibfnamefont {T.~E.}\ \bibnamefont
  {Cocolios}}, \bibinfo {author} {\bibfnamefont {A.~N.}\ \bibnamefont
  {Andreyev}}, \bibinfo {author} {\bibfnamefont {B.}~\bibnamefont {Bastin}},
  \bibinfo {author} {\bibfnamefont {N.}~\bibnamefont {Bree}}, \bibinfo {author}
  {\bibfnamefont {J.}~\bibnamefont {B{\"u}scher}}, \bibinfo {author}
  {\bibfnamefont {J.}~\bibnamefont {Elseviers}}, \bibinfo {author}
  {\bibfnamefont {J.}~\bibnamefont {Gentens}}, \bibinfo {author} {\bibfnamefont
  {M.}~\bibnamefont {Huyse}}, \bibinfo {author} {\bibfnamefont
  {Y.}~\bibnamefont {Kudryavtsev}}, \bibinfo {author} {\bibfnamefont
  {D.}~\bibnamefont {Pauwels}}, \bibinfo {author} {\bibfnamefont
  {T.}~\bibnamefont {Sonoda}}, \bibinfo {author} {\bibfnamefont
  {P.}~\bibnamefont {{van den Bergh}}},\ and\ \bibinfo {author} {\bibfnamefont
  {P.}~\bibnamefont {{van Duppen}}},\ }\bibfield  {title} {\bibinfo {title}
  {{Magnetic dipole moment of \textsuperscript{57,59}Cu measured by in-gas-cell
  laser spectroscopy}},\ }\href
  {https://doi.org/10.1103/PhysRevLett.103.102501} {\bibfield  {journal}
  {\bibinfo  {journal} {Phys.\ Rev.\ Lett.}\ }\textbf {\bibinfo {volume}
  {103}},\ \bibinfo {pages} {102501} (\bibinfo {year} {2009})}\BibitemShut
  {NoStop}%
\bibitem [{\citenamefont {Ohtsubo}\ \emph {et~al.}(1996)\citenamefont
  {Ohtsubo}, \citenamefont {Cho}, \citenamefont {Yanagihashi}, \citenamefont
  {Ohya},\ and\ \citenamefont {Muto}}]{Ohtsubo.1996}%
  \BibitemOpen
  \bibfield  {author} {\bibinfo {author} {\bibnamefont {Ohtsubo}}, \bibinfo
  {author} {\bibnamefont {Cho}}, \bibinfo {author} {\bibnamefont
  {Yanagihashi}}, \bibinfo {author} {\bibnamefont {Ohya}},\ and\ \bibinfo
  {author} {\bibnamefont {Muto}},\ }\bibfield  {title} {\bibinfo {title}
  {{Measurement of the nuclear magnetic moments of \textsuperscript{57}Ni and
  \textsuperscript{59}Fe}},\ }\href {https://doi.org/10.1103/physrevc.54.554}
  {\bibfield  {journal} {\bibinfo  {journal} {Phys.\ Rev.\ C}\ }\textbf
  {\bibinfo {volume} {54}},\ \bibinfo {pages} {554--558} (\bibinfo {year}
  {1996})}\BibitemShut {NoStop}%
\bibitem [{\citenamefont {Honma}\ \emph {et~al.}(2004)\citenamefont {Honma},
  \citenamefont {Otsuka}, \citenamefont {Brown},\ and\ \citenamefont
  {Mizusaki}}]{Honma.2004}%
  \BibitemOpen
  \bibfield  {author} {\bibinfo {author} {\bibfnamefont {M.}~\bibnamefont
  {Honma}}, \bibinfo {author} {\bibfnamefont {T.}~\bibnamefont {Otsuka}},
  \bibinfo {author} {\bibfnamefont {B.~A.}\ \bibnamefont {Brown}},\ and\
  \bibinfo {author} {\bibfnamefont {T.}~\bibnamefont {Mizusaki}},\ }\bibfield
  {title} {\bibinfo {title} {New effective interaction for pf -shell nuclei and
  its implications for the stability of the {$N=Z=28$} closed core},\ }\href
  {https://doi.org/10.1103/PhysRevC.69.034335} {\bibfield  {journal} {\bibinfo
  {journal} {Phys.\ Rev.\ C}\ }\textbf {\bibinfo {volume} {69}},\ \bibinfo
  {pages} {034335} (\bibinfo {year} {2004})}\BibitemShut {NoStop}%
\bibitem [{\citenamefont {Vingerhoets}\ \emph {et~al.}(2010)\citenamefont
  {Vingerhoets}, \citenamefont {Flanagan}, \citenamefont {Avgoulea},
  \citenamefont {Billowes}, \citenamefont {Bissell}, \citenamefont {Blaum},
  \citenamefont {Brown}, \citenamefont {Cheal}, \citenamefont {de~Rydt},
  \citenamefont {Forest}, \citenamefont {Geppert}, \citenamefont {Honma},
  \citenamefont {Kowalska}, \citenamefont {Kr{\"a}mer}, \citenamefont
  {Krieger}, \citenamefont {Man{\'e}}, \citenamefont {Neugart}, \citenamefont
  {Neyens}, \citenamefont {N{\"o}rtersh{\"a}user}, \citenamefont {Otsuka},
  \citenamefont {Schug}, \citenamefont {Stroke}, \citenamefont {Tungate},\ and\
  \citenamefont {Yordanov}}]{Vingerhoets.2010}%
  \BibitemOpen
  \bibfield  {author} {\bibinfo {author} {\bibfnamefont {P.}~\bibnamefont
  {Vingerhoets}}, \bibinfo {author} {\bibfnamefont {K.~T.}\ \bibnamefont
  {Flanagan}}, \bibinfo {author} {\bibfnamefont {M.}~\bibnamefont {Avgoulea}},
  \bibinfo {author} {\bibfnamefont {J.}~\bibnamefont {Billowes}}, \bibinfo
  {author} {\bibfnamefont {M.~L.}\ \bibnamefont {Bissell}}, \bibinfo {author}
  {\bibfnamefont {K.}~\bibnamefont {Blaum}}, \bibinfo {author} {\bibfnamefont
  {B.~A.}\ \bibnamefont {Brown}}, \bibinfo {author} {\bibfnamefont
  {B.}~\bibnamefont {Cheal}}, \bibinfo {author} {\bibfnamefont
  {M.}~\bibnamefont {de~Rydt}}, \bibinfo {author} {\bibfnamefont {D.~H.}\
  \bibnamefont {Forest}}, \bibinfo {author} {\bibfnamefont {C.}~\bibnamefont
  {Geppert}}, \bibinfo {author} {\bibfnamefont {M.}~\bibnamefont {Honma}},
  \bibinfo {author} {\bibfnamefont {M.}~\bibnamefont {Kowalska}}, \bibinfo
  {author} {\bibfnamefont {J.}~\bibnamefont {Kr{\"a}mer}}, \bibinfo {author}
  {\bibfnamefont {A.}~\bibnamefont {Krieger}}, \bibinfo {author} {\bibfnamefont
  {E.}~\bibnamefont {Man{\'e}}}, \bibinfo {author} {\bibfnamefont
  {R.}~\bibnamefont {Neugart}}, \bibinfo {author} {\bibfnamefont
  {G.}~\bibnamefont {Neyens}}, \bibinfo {author} {\bibfnamefont
  {W.}~\bibnamefont {N{\"o}rtersh{\"a}user}}, \bibinfo {author} {\bibfnamefont
  {T.}~\bibnamefont {Otsuka}}, \bibinfo {author} {\bibfnamefont
  {M.}~\bibnamefont {Schug}}, \bibinfo {author} {\bibfnamefont {H.~H.}\
  \bibnamefont {Stroke}}, \bibinfo {author} {\bibfnamefont {G.}~\bibnamefont
  {Tungate}},\ and\ \bibinfo {author} {\bibfnamefont {D.~T.}\ \bibnamefont
  {Yordanov}},\ }\bibfield  {title} {\bibinfo {title} {Nuclear spins, magnetic
  moments, and quadrupole moments of {Cu} isotopes from {$N$=28} to {$N$=46}:
  Probes for core polarization effects},\ }\href
  {https://doi.org/10.1103/PhysRevC.82.064311} {\bibfield  {journal} {\bibinfo
  {journal} {Phys. Rev. C}\ }\textbf {\bibinfo {volume} {82}},\ \bibinfo
  {pages} {064311} (\bibinfo {year} {2010})}\BibitemShut {NoStop}%
\bibitem [{\citenamefont {Vingerhoets}\ \emph {et~al.}(2011)\citenamefont
  {Vingerhoets}, \citenamefont {Flanagan}, \citenamefont {Billowes},
  \citenamefont {Bissell}, \citenamefont {Blaum}, \citenamefont {Cheal},
  \citenamefont {de~Rydt}, \citenamefont {Forest}, \citenamefont {Geppert},
  \citenamefont {Honma}, \citenamefont {Kowalska}, \citenamefont {Kr{\"a}mer},
  \citenamefont {Kreim}, \citenamefont {Krieger}, \citenamefont {Neugart},
  \citenamefont {Neyens}, \citenamefont {N{\"o}rtersh{\"a}user}, \citenamefont
  {Papuga}, \citenamefont {Procter}, \citenamefont {Rajabali}, \citenamefont
  {S{\'a}nchez}, \citenamefont {Stroke},\ and\ \citenamefont
  {Yordanov}}]{Vingerhoets.2011}%
  \BibitemOpen
  \bibfield  {author} {\bibinfo {author} {\bibfnamefont {P.}~\bibnamefont
  {Vingerhoets}}, \bibinfo {author} {\bibfnamefont {K.~T.}\ \bibnamefont
  {Flanagan}}, \bibinfo {author} {\bibfnamefont {J.}~\bibnamefont {Billowes}},
  \bibinfo {author} {\bibfnamefont {M.~L.}\ \bibnamefont {Bissell}}, \bibinfo
  {author} {\bibfnamefont {K.}~\bibnamefont {Blaum}}, \bibinfo {author}
  {\bibfnamefont {B.}~\bibnamefont {Cheal}}, \bibinfo {author} {\bibfnamefont
  {M.}~\bibnamefont {de~Rydt}}, \bibinfo {author} {\bibfnamefont {D.~H.}\
  \bibnamefont {Forest}}, \bibinfo {author} {\bibfnamefont {C.}~\bibnamefont
  {Geppert}}, \bibinfo {author} {\bibfnamefont {M.}~\bibnamefont {Honma}},
  \bibinfo {author} {\bibfnamefont {M.}~\bibnamefont {Kowalska}}, \bibinfo
  {author} {\bibfnamefont {J.}~\bibnamefont {Kr{\"a}mer}}, \bibinfo {author}
  {\bibfnamefont {K.}~\bibnamefont {Kreim}}, \bibinfo {author} {\bibfnamefont
  {A.}~\bibnamefont {Krieger}}, \bibinfo {author} {\bibfnamefont
  {R.}~\bibnamefont {Neugart}}, \bibinfo {author} {\bibfnamefont
  {G.}~\bibnamefont {Neyens}}, \bibinfo {author} {\bibfnamefont
  {W.}~\bibnamefont {N{\"o}rtersh{\"a}user}}, \bibinfo {author} {\bibfnamefont
  {J.}~\bibnamefont {Papuga}}, \bibinfo {author} {\bibfnamefont {T.~J.}\
  \bibnamefont {Procter}}, \bibinfo {author} {\bibfnamefont {M.~M.}\
  \bibnamefont {Rajabali}}, \bibinfo {author} {\bibfnamefont {R.}~\bibnamefont
  {S{\'a}nchez}}, \bibinfo {author} {\bibfnamefont {H.~H.}\ \bibnamefont
  {Stroke}},\ and\ \bibinfo {author} {\bibfnamefont {D.~T.}\ \bibnamefont
  {Yordanov}},\ }\bibfield  {title} {\bibinfo {title} {Magnetic and quadrupole
  moments of neutron deficient $^{58-62}${Cu} isotopes},\ }\href
  {https://doi.org/10.1016/j.physletb.2011.07.050} {\bibfield  {journal}
  {\bibinfo  {journal} {Phys. Lett. B}\ }\textbf {\bibinfo {volume} {703}},\
  \bibinfo {pages} {34} (\bibinfo {year} {2011})}\BibitemShut {NoStop}%
\bibitem [{\citenamefont {Pritychenko}\ \emph {et~al.}(2016)\citenamefont
  {Pritychenko}, \citenamefont {Birch}, \citenamefont {Singh},\ and\
  \citenamefont {Horoi}}]{Pritychenko.2016}%
  \BibitemOpen
  \bibfield  {author} {\bibinfo {author} {\bibfnamefont {B.}~\bibnamefont
  {Pritychenko}}, \bibinfo {author} {\bibfnamefont {M.}~\bibnamefont {Birch}},
  \bibinfo {author} {\bibfnamefont {B.}~\bibnamefont {Singh}},\ and\ \bibinfo
  {author} {\bibfnamefont {M.}~\bibnamefont {Horoi}},\ }\bibfield  {title}
  {\bibinfo {title} {Tables of {E2} transition probabilities from the first
  2$^+$ states in even--even nuclei},\ }\href
  {https://doi.org/10.1016/j.adt.2015.10.001} {\bibfield  {journal} {\bibinfo
  {journal} {At. Data Nucl. Data Tables}\ }\textbf {\bibinfo {volume} {107}},\
  \bibinfo {pages} {1--139} (\bibinfo {year} {2016})}\BibitemShut {NoStop}%
\bibitem [{\citenamefont {Otsuka}\ \emph {et~al.}(2005)\citenamefont {Otsuka},
  \citenamefont {Suzuki}, \citenamefont {Fujimoto}, \citenamefont {Grawe},\
  and\ \citenamefont {Akaishi}}]{Otsuka.2005}%
  \BibitemOpen
  \bibfield  {author} {\bibinfo {author} {\bibfnamefont {T.}~\bibnamefont
  {Otsuka}}, \bibinfo {author} {\bibfnamefont {T.}~\bibnamefont {Suzuki}},
  \bibinfo {author} {\bibfnamefont {R.}~\bibnamefont {Fujimoto}}, \bibinfo
  {author} {\bibfnamefont {H.}~\bibnamefont {Grawe}},\ and\ \bibinfo {author}
  {\bibfnamefont {Y.}~\bibnamefont {Akaishi}},\ }\bibfield  {title} {\bibinfo
  {title} {Evolution of nuclear shells due to the tensor force},\ }\href@noop
  {} {\bibfield  {journal} {\bibinfo  {journal} {Phys.\ Rev.\ Lett.}\ }\textbf
  {\bibinfo {volume} {95}},\ \bibinfo {pages} {232502} (\bibinfo {year}
  {2005})}\BibitemShut {NoStop}%
\bibitem [{\citenamefont {Kaufmann}\ \emph {et~al.}(2020)\citenamefont
  {Kaufmann}, \citenamefont {Simonis}, \citenamefont {Bacca}, \citenamefont
  {Billowes}, \citenamefont {Bissell}, \citenamefont {Blaum}, \citenamefont
  {Cheal}, \citenamefont {Ruiz}, \citenamefont {Gins}, \citenamefont {Gorges},
  \citenamefont {Hagen}, \citenamefont {Heylen}, \citenamefont
  {Kanellakopoulos}, \citenamefont {Malbrunot-Ettenauer}, \citenamefont
  {Miorelli}, \citenamefont {Neugart}, \citenamefont {Neyens}, \citenamefont
  {N{\"o}rtersh{\"a}user}, \citenamefont {S{\'a}nchez}, \citenamefont {Sailer},
  \citenamefont {Schwenk}, \citenamefont {Ratajczyk}, \citenamefont
  {Rodr{\'i}guez}, \citenamefont {Wehner}, \citenamefont {Wraith},
  \citenamefont {Xie}, \citenamefont {Xu}, \citenamefont {Yang},\ and\
  \citenamefont {Yordanov}}]{Kaufmann.2020}%
  \BibitemOpen
  \bibfield  {author} {\bibinfo {author} {\bibfnamefont {S.}~\bibnamefont
  {Kaufmann}}, \bibinfo {author} {\bibfnamefont {J.}~\bibnamefont {Simonis}},
  \bibinfo {author} {\bibfnamefont {S.}~\bibnamefont {Bacca}}, \bibinfo
  {author} {\bibfnamefont {J.}~\bibnamefont {Billowes}}, \bibinfo {author}
  {\bibfnamefont {M.~L.}\ \bibnamefont {Bissell}}, \bibinfo {author}
  {\bibfnamefont {K.}~\bibnamefont {Blaum}}, \bibinfo {author} {\bibfnamefont
  {B.}~\bibnamefont {Cheal}}, \bibinfo {author} {\bibfnamefont {R.~F.~G.}\
  \bibnamefont {Ruiz}}, \bibinfo {author} {\bibfnamefont {W.}~\bibnamefont
  {Gins}}, \bibinfo {author} {\bibfnamefont {C.}~\bibnamefont {Gorges}},
  \bibinfo {author} {\bibfnamefont {G.}~\bibnamefont {Hagen}}, \bibinfo
  {author} {\bibfnamefont {H.}~\bibnamefont {Heylen}}, \bibinfo {author}
  {\bibfnamefont {A.}~\bibnamefont {Kanellakopoulos}}, \bibinfo {author}
  {\bibfnamefont {S.}~\bibnamefont {Malbrunot-Ettenauer}}, \bibinfo {author}
  {\bibfnamefont {M.}~\bibnamefont {Miorelli}}, \bibinfo {author}
  {\bibfnamefont {R.}~\bibnamefont {Neugart}}, \bibinfo {author} {\bibfnamefont
  {G.}~\bibnamefont {Neyens}}, \bibinfo {author} {\bibfnamefont
  {W.}~\bibnamefont {N{\"o}rtersh{\"a}user}}, \bibinfo {author} {\bibfnamefont
  {R.}~\bibnamefont {S{\'a}nchez}}, \bibinfo {author} {\bibfnamefont
  {S.}~\bibnamefont {Sailer}}, \bibinfo {author} {\bibfnamefont
  {A.}~\bibnamefont {Schwenk}}, \bibinfo {author} {\bibfnamefont
  {T.}~\bibnamefont {Ratajczyk}}, \bibinfo {author} {\bibfnamefont {L.~V.}\
  \bibnamefont {Rodr{\'i}guez}}, \bibinfo {author} {\bibfnamefont
  {L.}~\bibnamefont {Wehner}}, \bibinfo {author} {\bibfnamefont
  {C.}~\bibnamefont {Wraith}}, \bibinfo {author} {\bibfnamefont
  {L.}~\bibnamefont {Xie}}, \bibinfo {author} {\bibfnamefont {Z.~Y.}\
  \bibnamefont {Xu}}, \bibinfo {author} {\bibfnamefont {X.~F.}\ \bibnamefont
  {Yang}},\ and\ \bibinfo {author} {\bibfnamefont {D.~T.}\ \bibnamefont
  {Yordanov}},\ }\bibfield  {title} {\bibinfo {title} {{Charge Radius of the
  Short-Lived \textsuperscript{68}Ni and Correlation with the Dipole
  Polarizability}},\ }\href {https://doi.org/10.1103/PhysRevLett.124.132502}
  {\bibfield  {journal} {\bibinfo  {journal} {Phys.\ Rev.\ Lett.}\ }\textbf
  {\bibinfo {volume} {124}},\ \bibinfo {pages} {132502} (\bibinfo {year}
  {2020})}\BibitemShut {NoStop}%
\bibitem [{\citenamefont {Pineda}\ \emph {et~al.}(2021)\citenamefont {Pineda},
  \citenamefont {K{\"o}nig}, \citenamefont {Rossi}, \citenamefont {Brown},
  \citenamefont {Incorvati}, \citenamefont {Lantis}, \citenamefont
  {Minamisono}, \citenamefont {N{\"o}rtersh{\"a}user}, \citenamefont
  {Piekarewicz}, \citenamefont {Powel},\ and\ \citenamefont
  {Sommer}}]{Pineda.2021}%
  \BibitemOpen
  \bibfield  {author} {\bibinfo {author} {\bibfnamefont {S.~V.}\ \bibnamefont
  {Pineda}}, \bibinfo {author} {\bibfnamefont {K.}~\bibnamefont {K{\"o}nig}},
  \bibinfo {author} {\bibfnamefont {D.~M.}\ \bibnamefont {Rossi}}, \bibinfo
  {author} {\bibfnamefont {B.~A.}\ \bibnamefont {Brown}}, \bibinfo {author}
  {\bibfnamefont {A.}~\bibnamefont {Incorvati}}, \bibinfo {author}
  {\bibfnamefont {J.}~\bibnamefont {Lantis}}, \bibinfo {author} {\bibfnamefont
  {K.}~\bibnamefont {Minamisono}}, \bibinfo {author} {\bibfnamefont
  {W.}~\bibnamefont {N{\"o}rtersh{\"a}user}}, \bibinfo {author} {\bibfnamefont
  {J.}~\bibnamefont {Piekarewicz}}, \bibinfo {author} {\bibfnamefont
  {R.}~\bibnamefont {Powel}},\ and\ \bibinfo {author} {\bibfnamefont
  {F.}~\bibnamefont {Sommer}},\ }\bibfield  {title} {\bibinfo {title} {{Charge
  Radius of Neutron-Deficient $^{54}$Ni and Symmetry Energy Constraints Using
  the Difference in Mirror Pair Charge Radii}},\ }\href
  {https://doi.org/10.1103/PhysRevLett.127.182503} {\bibfield  {journal}
  {\bibinfo  {journal} {Phys.\ Rev.\ Lett.}\ }\textbf {\bibinfo {volume}
  {127}},\ \bibinfo {pages} {182503} (\bibinfo {year} {2021})}\BibitemShut
  {NoStop}%
\bibitem [{\citenamefont {Malbrunot-Ettenauer}\ \emph
  {et~al.}(2022)\citenamefont {Malbrunot-Ettenauer}, \citenamefont {Kaufmann},
  \citenamefont {Bacca}, \citenamefont {Barbieri}, \citenamefont {Billowes},
  \citenamefont {Bissell}, \citenamefont {Blaum}, \citenamefont {Cheal},
  \citenamefont {Duguet}, \citenamefont {Ruiz}, \citenamefont {Gins},
  \citenamefont {Gorges}, \citenamefont {Hagen}, \citenamefont {Heylen},
  \citenamefont {Holt}, \citenamefont {Jansen}, \citenamefont
  {Kanellakopoulos}, \citenamefont {Kortelainen}, \citenamefont {Miyagi},
  \citenamefont {Navr\'atil}, \citenamefont {Nazarewicz}, \citenamefont
  {Neugart}, \citenamefont {Neyens}, \citenamefont {N\"ortersh\"auser},
  \citenamefont {Novario}, \citenamefont {Papenbrock}, \citenamefont
  {Ratajczyk}, \citenamefont {Reinhard}, \citenamefont {Rodr\'{\i}guez},
  \citenamefont {S\'anchez}, \citenamefont {Sailer}, \citenamefont {Schwenk},
  \citenamefont {Simonis}, \citenamefont {Som\`a}, \citenamefont {Stroberg},
  \citenamefont {Wehner}, \citenamefont {Wraith}, \citenamefont {Xie},
  \citenamefont {Xu}, \citenamefont {Yang},\ and\ \citenamefont
  {Yordanov}}]{Malbrunot.2022}%
  \BibitemOpen
  \bibfield  {author} {\bibinfo {author} {\bibfnamefont {S.}~\bibnamefont
  {Malbrunot-Ettenauer}}, \bibinfo {author} {\bibfnamefont {S.}~\bibnamefont
  {Kaufmann}}, \bibinfo {author} {\bibfnamefont {S.}~\bibnamefont {Bacca}},
  \bibinfo {author} {\bibfnamefont {C.}~\bibnamefont {Barbieri}}, \bibinfo
  {author} {\bibfnamefont {J.}~\bibnamefont {Billowes}}, \bibinfo {author}
  {\bibfnamefont {M.~L.}\ \bibnamefont {Bissell}}, \bibinfo {author}
  {\bibfnamefont {K.}~\bibnamefont {Blaum}}, \bibinfo {author} {\bibfnamefont
  {B.}~\bibnamefont {Cheal}}, \bibinfo {author} {\bibfnamefont
  {T.}~\bibnamefont {Duguet}}, \bibinfo {author} {\bibfnamefont {R.~F.~G.}\
  \bibnamefont {Ruiz}}, \bibinfo {author} {\bibfnamefont {W.}~\bibnamefont
  {Gins}}, \bibinfo {author} {\bibfnamefont {C.}~\bibnamefont {Gorges}},
  \bibinfo {author} {\bibfnamefont {G.}~\bibnamefont {Hagen}}, \bibinfo
  {author} {\bibfnamefont {H.}~\bibnamefont {Heylen}}, \bibinfo {author}
  {\bibfnamefont {J.~D.}\ \bibnamefont {Holt}}, \bibinfo {author}
  {\bibfnamefont {G.~R.}\ \bibnamefont {Jansen}}, \bibinfo {author}
  {\bibfnamefont {A.}~\bibnamefont {Kanellakopoulos}}, \bibinfo {author}
  {\bibfnamefont {M.}~\bibnamefont {Kortelainen}}, \bibinfo {author}
  {\bibfnamefont {T.}~\bibnamefont {Miyagi}}, \bibinfo {author} {\bibfnamefont
  {P.}~\bibnamefont {Navr\'atil}}, \bibinfo {author} {\bibfnamefont
  {W.}~\bibnamefont {Nazarewicz}}, \bibinfo {author} {\bibfnamefont
  {R.}~\bibnamefont {Neugart}}, \bibinfo {author} {\bibfnamefont
  {G.}~\bibnamefont {Neyens}}, \bibinfo {author} {\bibfnamefont
  {W.}~\bibnamefont {N\"ortersh\"auser}}, \bibinfo {author} {\bibfnamefont
  {S.~J.}\ \bibnamefont {Novario}}, \bibinfo {author} {\bibfnamefont
  {T.}~\bibnamefont {Papenbrock}}, \bibinfo {author} {\bibfnamefont
  {T.}~\bibnamefont {Ratajczyk}}, \bibinfo {author} {\bibfnamefont {P.-G.}\
  \bibnamefont {Reinhard}}, \bibinfo {author} {\bibfnamefont {L.~V.}\
  \bibnamefont {Rodr\'{\i}guez}}, \bibinfo {author} {\bibfnamefont
  {R.}~\bibnamefont {S\'anchez}}, \bibinfo {author} {\bibfnamefont
  {S.}~\bibnamefont {Sailer}}, \bibinfo {author} {\bibfnamefont
  {A.}~\bibnamefont {Schwenk}}, \bibinfo {author} {\bibfnamefont
  {J.}~\bibnamefont {Simonis}}, \bibinfo {author} {\bibfnamefont
  {V.}~\bibnamefont {Som\`a}}, \bibinfo {author} {\bibfnamefont {S.~R.}\
  \bibnamefont {Stroberg}}, \bibinfo {author} {\bibfnamefont {L.}~\bibnamefont
  {Wehner}}, \bibinfo {author} {\bibfnamefont {C.}~\bibnamefont {Wraith}},
  \bibinfo {author} {\bibfnamefont {L.}~\bibnamefont {Xie}}, \bibinfo {author}
  {\bibfnamefont {Z.~Y.}\ \bibnamefont {Xu}}, \bibinfo {author} {\bibfnamefont
  {X.~F.}\ \bibnamefont {Yang}},\ and\ \bibinfo {author} {\bibfnamefont
  {D.~T.}\ \bibnamefont {Yordanov}},\ }\bibfield  {title} {\bibinfo {title}
  {{Nuclear Charge Radii of the Nickel Isotopes
  $^{58\ensuremath{-}68,70}\mathrm{Ni}$}},\ }\href
  {https://doi.org/10.1103/PhysRevLett.128.022502} {\bibfield  {journal}
  {\bibinfo  {journal} {Phys. Rev. Lett.}\ }\textbf {\bibinfo {volume} {128}},\
  \bibinfo {pages} {022502} (\bibinfo {year} {2022})}\BibitemShut {NoStop}%
\bibitem [{\citenamefont {Hagen}\ \emph
  {et~al.}(2016{\natexlab{a}})\citenamefont {Hagen}, \citenamefont
  {Ekstr{\"o}m}, \citenamefont {Forss{\'e}n}, \citenamefont {Jansen},
  \citenamefont {Nazarewicz}, \citenamefont {Papenbrock}, \citenamefont
  {Wendt}, \citenamefont {Bacca}, \citenamefont {Barnea}, \citenamefont
  {Carlsson}, \citenamefont {Drischler}, \citenamefont {Hebeler}, \citenamefont
  {Hjorth-Jensen}, \citenamefont {Miorelli}, \citenamefont {Orlandini},
  \citenamefont {Schwenk},\ and\ \citenamefont {Simonis}}]{Hagen.2016b}%
  \BibitemOpen
  \bibfield  {author} {\bibinfo {author} {\bibfnamefont {G.}~\bibnamefont
  {Hagen}}, \bibinfo {author} {\bibfnamefont {A.}~\bibnamefont {Ekstr{\"o}m}},
  \bibinfo {author} {\bibfnamefont {C.}~\bibnamefont {Forss{\'e}n}}, \bibinfo
  {author} {\bibfnamefont {G.~R.}\ \bibnamefont {Jansen}}, \bibinfo {author}
  {\bibfnamefont {W.}~\bibnamefont {Nazarewicz}}, \bibinfo {author}
  {\bibfnamefont {T.}~\bibnamefont {Papenbrock}}, \bibinfo {author}
  {\bibfnamefont {K.~A.}\ \bibnamefont {Wendt}}, \bibinfo {author}
  {\bibfnamefont {S.}~\bibnamefont {Bacca}}, \bibinfo {author} {\bibfnamefont
  {N.}~\bibnamefont {Barnea}}, \bibinfo {author} {\bibfnamefont
  {B.}~\bibnamefont {Carlsson}}, \bibinfo {author} {\bibfnamefont
  {C.}~\bibnamefont {Drischler}}, \bibinfo {author} {\bibfnamefont
  {K.}~\bibnamefont {Hebeler}}, \bibinfo {author} {\bibfnamefont
  {M.}~\bibnamefont {Hjorth-Jensen}}, \bibinfo {author} {\bibfnamefont
  {M.}~\bibnamefont {Miorelli}}, \bibinfo {author} {\bibfnamefont
  {G.}~\bibnamefont {Orlandini}}, \bibinfo {author} {\bibfnamefont
  {A.}~\bibnamefont {Schwenk}},\ and\ \bibinfo {author} {\bibfnamefont
  {J.}~\bibnamefont {Simonis}},\ }\bibfield  {title} {\bibinfo {title}
  {{Neutron and weak-charge distributions of the \textsuperscript{48}Ca
  nucleus}},\ }\href {https://doi.org/10.1038/nphys3529} {\bibfield  {journal}
  {\bibinfo  {journal} {Nat. Phys.}\ }\textbf {\bibinfo {volume} {12}},\
  \bibinfo {pages} {186--190} (\bibinfo {year}
  {2016}{\natexlab{a}})}\BibitemShut {NoStop}%
\bibitem [{\citenamefont {Hagen}\ \emph
  {et~al.}(2016{\natexlab{b}})\citenamefont {Hagen}, \citenamefont
  {Hjorth-Jensen}, \citenamefont {Jansen},\ and\ \citenamefont
  {Papenbrock}}]{Hagen.2016}%
  \BibitemOpen
  \bibfield  {author} {\bibinfo {author} {\bibfnamefont {G.}~\bibnamefont
  {Hagen}}, \bibinfo {author} {\bibfnamefont {M.}~\bibnamefont
  {Hjorth-Jensen}}, \bibinfo {author} {\bibfnamefont {G.~R.}\ \bibnamefont
  {Jansen}},\ and\ \bibinfo {author} {\bibfnamefont {T.}~\bibnamefont
  {Papenbrock}},\ }\bibfield  {title} {\bibinfo {title} {Emergent properties of
  nuclei from ab initio coupled-cluster calculations},\ }\href
  {https://doi.org/10.1088/0031-8949/91/6/063006} {\bibfield  {journal}
  {\bibinfo  {journal} {Phys.\ Scr.}\ }\textbf {\bibinfo {volume} {91}},\
  \bibinfo {pages} {063006} (\bibinfo {year} {2016}{\natexlab{b}})}\BibitemShut
  {NoStop}%
\bibitem [{\citenamefont {Hoppe}\ \emph {et~al.}(2019)\citenamefont {Hoppe},
  \citenamefont {Drischler}, \citenamefont {Hebeler}, \citenamefont {Schwenk},\
  and\ \citenamefont {Simonis}}]{Hoppe.2019}%
  \BibitemOpen
  \bibfield  {author} {\bibinfo {author} {\bibfnamefont {J.}~\bibnamefont
  {Hoppe}}, \bibinfo {author} {\bibfnamefont {C.}~\bibnamefont {Drischler}},
  \bibinfo {author} {\bibfnamefont {K.}~\bibnamefont {Hebeler}}, \bibinfo
  {author} {\bibfnamefont {A.}~\bibnamefont {Schwenk}},\ and\ \bibinfo {author}
  {\bibfnamefont {J.}~\bibnamefont {Simonis}},\ }\bibfield  {title} {\bibinfo
  {title} {{Probing chiral interactions up to next-to-next-to-next-to-leading
  order in medium-mass nuclei}},\ }\href
  {https://doi.org/10.1103/PhysRevC.100.024318} {\bibfield  {journal} {\bibinfo
   {journal} {Phys. Rev. C}\ }\textbf {\bibinfo {volume} {100}},\ \bibinfo
  {pages} {024318} (\bibinfo {year} {2019})}\BibitemShut {NoStop}%
\bibitem [{\citenamefont {H{\"u}ther}\ \emph {et~al.}(2020)\citenamefont
  {H{\"u}ther}, \citenamefont {Vobig}, \citenamefont {Hebeler}, \citenamefont
  {Machleidt},\ and\ \citenamefont {Roth}}]{Huther.2020}%
  \BibitemOpen
  \bibfield  {author} {\bibinfo {author} {\bibfnamefont {T.}~\bibnamefont
  {H{\"u}ther}}, \bibinfo {author} {\bibfnamefont {K.}~\bibnamefont {Vobig}},
  \bibinfo {author} {\bibfnamefont {K.}~\bibnamefont {Hebeler}}, \bibinfo
  {author} {\bibfnamefont {R.}~\bibnamefont {Machleidt}},\ and\ \bibinfo
  {author} {\bibfnamefont {R.}~\bibnamefont {Roth}},\ }\bibfield  {title}
  {\bibinfo {title} {Family of chiral two- plus three-nucleon interactions for
  accurate nuclear structure studies},\ }\href
  {https://doi.org/10.1016/j.physletb.2020.135651} {\bibfield  {journal}
  {\bibinfo  {journal} {Phys.\ Lett.\ B}\ }\textbf {\bibinfo {volume} {808}},\
  \bibinfo {pages} {135651} (\bibinfo {year} {2020})}\BibitemShut {NoStop}%
\bibitem [{\citenamefont {Minamisono}\ \emph {et~al.}(2013)\citenamefont
  {Minamisono}, \citenamefont {Mantica}, \citenamefont {Klose}, \citenamefont
  {Vinnikova}, \citenamefont {Schneider}, \citenamefont {Johnson},\ and\
  \citenamefont {Barquest}}]{Minamisono.2013}%
  \BibitemOpen
  \bibfield  {author} {\bibinfo {author} {\bibfnamefont {K.}~\bibnamefont
  {Minamisono}}, \bibinfo {author} {\bibfnamefont {P.~F.}\ \bibnamefont
  {Mantica}}, \bibinfo {author} {\bibfnamefont {A.}~\bibnamefont {Klose}},
  \bibinfo {author} {\bibfnamefont {S.}~\bibnamefont {Vinnikova}}, \bibinfo
  {author} {\bibfnamefont {A.}~\bibnamefont {Schneider}}, \bibinfo {author}
  {\bibfnamefont {B.}~\bibnamefont {Johnson}},\ and\ \bibinfo {author}
  {\bibfnamefont {B.~R.}\ \bibnamefont {Barquest}},\ }\bibfield  {title}
  {\bibinfo {title} {{Commissioning of the collinear laser spectroscopy system
  in the BECOLA facility at NSCL}},\ }\href
  {https://doi.org/10.1016/j.nima.2013.01.038} {\bibfield  {journal} {\bibinfo
  {journal} {Nucl. Instr. Meth. Phys. Res. A}\ }\textbf {\bibinfo {volume}
  {709}},\ \bibinfo {pages} {85} (\bibinfo {year} {2013})}\BibitemShut
  {NoStop}%
\bibitem [{\citenamefont {Morrissey}\ \emph {et~al.}(2003)\citenamefont
  {Morrissey}, \citenamefont {Sherrill}, \citenamefont {Steiner}, \citenamefont
  {Stolz},\ and\ \citenamefont {Wiedenhoever}}]{Morrisey.2003}%
  \BibitemOpen
  \bibfield  {author} {\bibinfo {author} {\bibfnamefont {D.~J.}\ \bibnamefont
  {Morrissey}}, \bibinfo {author} {\bibfnamefont {B.~M.}\ \bibnamefont
  {Sherrill}}, \bibinfo {author} {\bibfnamefont {M.}~\bibnamefont {Steiner}},
  \bibinfo {author} {\bibfnamefont {A.}~\bibnamefont {Stolz}},\ and\ \bibinfo
  {author} {\bibfnamefont {I.}~\bibnamefont {Wiedenhoever}},\ }\bibfield
  {title} {\bibinfo {title} {{Commissioning the A1900 projectile fragment
  separator}},\ }\href {https://doi.org/10.1016/S0168-583X(02)01895-5}
  {\bibfield  {journal} {\bibinfo  {journal} {Nucl. Instr. Meth. Phys. Res. B}\
  }\textbf {\bibinfo {volume} {204}},\ \bibinfo {pages} {90} (\bibinfo {year}
  {2003})}\BibitemShut {NoStop}%
\bibitem [{\citenamefont {Sumithrarachchi}\ \emph {et~al.}(2020)\citenamefont
  {Sumithrarachchi}, \citenamefont {Morrissey}, \citenamefont {Schwarz},
  \citenamefont {Lund}, \citenamefont {Bollen}, \citenamefont {Ringle},
  \citenamefont {Savard},\ and\ \citenamefont {Villari}}]{sum20}%
  \BibitemOpen
  \bibfield  {author} {\bibinfo {author} {\bibfnamefont {C.}~\bibnamefont
  {Sumithrarachchi}}, \bibinfo {author} {\bibfnamefont {D.}~\bibnamefont
  {Morrissey}}, \bibinfo {author} {\bibfnamefont {S.}~\bibnamefont {Schwarz}},
  \bibinfo {author} {\bibfnamefont {K.}~\bibnamefont {Lund}}, \bibinfo {author}
  {\bibfnamefont {G.}~\bibnamefont {Bollen}}, \bibinfo {author} {\bibfnamefont
  {R.}~\bibnamefont {Ringle}}, \bibinfo {author} {\bibfnamefont
  {G.}~\bibnamefont {Savard}},\ and\ \bibinfo {author} {\bibfnamefont
  {A.}~\bibnamefont {Villari}},\ }\bibfield  {title} {\bibinfo {title} {Beam
  thermalization in a large gas catcher},\ }\href
  {https://doi.org/https://doi.org/10.1016/j.nimb.2019.04.077} {\bibfield
  {journal} {\bibinfo  {journal} {Nucl. Instr. Meth. Phys. Res. B}\ }\textbf
  {\bibinfo {volume} {463}},\ \bibinfo {pages} {305} (\bibinfo {year}
  {2020})}\BibitemShut {NoStop}%
\bibitem [{SM()}]{SM}%
  \BibitemOpen
  \href@noop {} {}\bibinfo {note} {See Supplemental Material at
  \textcolor{red}{(link)} for supplementary information about the nuclear
  magnetic moment of $^{55}$Ni.}\BibitemShut {Stop}%
\bibitem [{\citenamefont {Barquest}\ \emph {et~al.}(2017)\citenamefont
  {Barquest}, \citenamefont {Bollen}, \citenamefont {Mantica}, \citenamefont
  {Minamisono}, \citenamefont {Ringle},\ and\ \citenamefont {Schwarz}}]{bar17}%
  \BibitemOpen
  \bibfield  {author} {\bibinfo {author} {\bibfnamefont {B.~R.}\ \bibnamefont
  {Barquest}}, \bibinfo {author} {\bibfnamefont {G.}~\bibnamefont {Bollen}},
  \bibinfo {author} {\bibfnamefont {P.~F.}\ \bibnamefont {Mantica}}, \bibinfo
  {author} {\bibfnamefont {K.}~\bibnamefont {Minamisono}}, \bibinfo {author}
  {\bibfnamefont {R.}~\bibnamefont {Ringle}},\ and\ \bibinfo {author}
  {\bibfnamefont {S.}~\bibnamefont {Schwarz}},\ }\bibfield  {title} {\bibinfo
  {title} {{RFQ} beam cooler and buncher for collinear laser spectroscopy of
  rare isotopes},\ }\href {https://doi.org/10.1016/j.nima.2017.05.036}
  {\bibfield  {journal} {\bibinfo  {journal} {Nucl. Instrum. Meth. Phys. Res.
  A}\ }\textbf {\bibinfo {volume} {866}},\ \bibinfo {pages} {18} (\bibinfo
  {year} {2017})}\BibitemShut {NoStop}%
\bibitem [{\citenamefont {Nouri}\ \emph {et~al.}(2010)\citenamefont {Nouri},
  \citenamefont {Li}, \citenamefont {Holt},\ and\ \citenamefont
  {Rosner}}]{Nouri.2010}%
  \BibitemOpen
  \bibfield  {author} {\bibinfo {author} {\bibfnamefont {Z.}~\bibnamefont
  {Nouri}}, \bibinfo {author} {\bibfnamefont {R.}~\bibnamefont {Li}}, \bibinfo
  {author} {\bibfnamefont {R.~A.}\ \bibnamefont {Holt}},\ and\ \bibinfo
  {author} {\bibfnamefont {S.~D.}\ \bibnamefont {Rosner}},\ }\bibfield  {title}
  {\bibinfo {title} {A penning sputter ion source with very low energy
  spread},\ }\href {https://doi.org/10.1016/j.nima.2009.12.060} {\bibfield
  {journal} {\bibinfo  {journal} {Nucl. Instr. Meth. Phys. Res. A}\ }\textbf
  {\bibinfo {volume} {614}},\ \bibinfo {pages} {174--178} (\bibinfo {year}
  {2010})}\BibitemShut {NoStop}%
\bibitem [{\citenamefont {Klose}\ \emph {et~al.}(2012)\citenamefont {Klose},
  \citenamefont {Minamisono}, \citenamefont {Geppert}, \citenamefont
  {Fr{\"o}mmgen}, \citenamefont {Hammen}, \citenamefont {Kr{\"a}mer},
  \citenamefont {Krieger}, \citenamefont {Levy}, \citenamefont {Mantica},
  \citenamefont {N{\"o}rtersh{\"a}user},\ and\ \citenamefont
  {Vinnikova}}]{Klose.2012}%
  \BibitemOpen
  \bibfield  {author} {\bibinfo {author} {\bibfnamefont {A.}~\bibnamefont
  {Klose}}, \bibinfo {author} {\bibfnamefont {K.}~\bibnamefont {Minamisono}},
  \bibinfo {author} {\bibfnamefont {C.}~\bibnamefont {Geppert}}, \bibinfo
  {author} {\bibfnamefont {N.}~\bibnamefont {Fr{\"o}mmgen}}, \bibinfo {author}
  {\bibfnamefont {M.}~\bibnamefont {Hammen}}, \bibinfo {author} {\bibfnamefont
  {J.}~\bibnamefont {Kr{\"a}mer}}, \bibinfo {author} {\bibfnamefont
  {A.}~\bibnamefont {Krieger}}, \bibinfo {author} {\bibfnamefont
  {C.}~\bibnamefont {Levy}}, \bibinfo {author} {\bibfnamefont {P.~F.}\
  \bibnamefont {Mantica}}, \bibinfo {author} {\bibfnamefont {W.}~\bibnamefont
  {N{\"o}rtersh{\"a}user}},\ and\ \bibinfo {author} {\bibfnamefont
  {S.}~\bibnamefont {Vinnikova}},\ }\bibfield  {title} {\bibinfo {title} {Tests
  of atomic charge-exchange cells for collinear laser spectroscopy},\ }\href
  {https://doi.org/10.1016/j.nima.2012.03.006} {\bibfield  {journal} {\bibinfo
  {journal} {Nucl. Instr. Meth. Phys. Res. A}\ }\textbf {\bibinfo {volume}
  {678}},\ \bibinfo {pages} {114--121} (\bibinfo {year} {2012})}\BibitemShut
  {NoStop}%
\bibitem [{\citenamefont {Ryder}\ \emph {et~al.}(2015)\citenamefont {Ryder},
  \citenamefont {Minamisono}, \citenamefont {Asberry}, \citenamefont
  {Isherwood}, \citenamefont {Mantica}, \citenamefont {Miller}, \citenamefont
  {Rossi},\ and\ \citenamefont {Strum}}]{Ryder.2015}%
  \BibitemOpen
  \bibfield  {author} {\bibinfo {author} {\bibfnamefont {C.~A.}\ \bibnamefont
  {Ryder}}, \bibinfo {author} {\bibfnamefont {K.}~\bibnamefont {Minamisono}},
  \bibinfo {author} {\bibfnamefont {H.~B.}\ \bibnamefont {Asberry}}, \bibinfo
  {author} {\bibfnamefont {B.}~\bibnamefont {Isherwood}}, \bibinfo {author}
  {\bibfnamefont {P.~F.}\ \bibnamefont {Mantica}}, \bibinfo {author}
  {\bibfnamefont {A.}~\bibnamefont {Miller}}, \bibinfo {author} {\bibfnamefont
  {D.~M.}\ \bibnamefont {Rossi}},\ and\ \bibinfo {author} {\bibfnamefont
  {R.}~\bibnamefont {Strum}},\ }\bibfield  {title} {\bibinfo {title}
  {{Population distribution subsequent to charge exchange of 29.85\,keV
  Ni\textsuperscript{+} on sodium vapor}},\ }\href
  {https://doi.org/10.1016/j.sab.2015.08.004} {\bibfield  {journal} {\bibinfo
  {journal} {Spectrochim. Acta B}\ }\textbf {\bibinfo {volume} {113}},\
  \bibinfo {pages} {16--21} (\bibinfo {year} {2015})}\BibitemShut {NoStop}%
\bibitem [{\citenamefont {Minamisono}\ \emph {et~al.}(2015)\citenamefont
  {Minamisono}, \citenamefont {Barquest}, \citenamefont {Bollen}, \citenamefont
  {Cooper}, \citenamefont {Hammerton}, \citenamefont {Hughes}, \citenamefont
  {Mantica}, \citenamefont {Morrissey}, \citenamefont {Ringle}, \citenamefont
  {Rodriguez}, \citenamefont {Ryder}, \citenamefont {Rossi}, \citenamefont
  {Schwarz}, \citenamefont {Strum}, \citenamefont {Sumithrarachchi},\ and\
  \citenamefont {Tarazona}}]{Minamisono.2015}%
  \BibitemOpen
  \bibfield  {author} {\bibinfo {author} {\bibfnamefont {K.}~\bibnamefont
  {Minamisono}}, \bibinfo {author} {\bibfnamefont {B.~R.}\ \bibnamefont
  {Barquest}}, \bibinfo {author} {\bibfnamefont {G.}~\bibnamefont {Bollen}},
  \bibinfo {author} {\bibfnamefont {K.}~\bibnamefont {Cooper}}, \bibinfo
  {author} {\bibfnamefont {K.}~\bibnamefont {Hammerton}}, \bibinfo {author}
  {\bibfnamefont {M.}~\bibnamefont {Hughes}}, \bibinfo {author} {\bibfnamefont
  {P.~F.}\ \bibnamefont {Mantica}}, \bibinfo {author} {\bibfnamefont {D.~J.}\
  \bibnamefont {Morrissey}}, \bibinfo {author} {\bibfnamefont {R.}~\bibnamefont
  {Ringle}}, \bibinfo {author} {\bibfnamefont {J.~A.}\ \bibnamefont
  {Rodriguez}}, \bibinfo {author} {\bibfnamefont {C.~A.}\ \bibnamefont
  {Ryder}}, \bibinfo {author} {\bibfnamefont {D.~M.}\ \bibnamefont {Rossi}},
  \bibinfo {author} {\bibfnamefont {S.}~\bibnamefont {Schwarz}}, \bibinfo
  {author} {\bibfnamefont {R.}~\bibnamefont {Strum}}, \bibinfo {author}
  {\bibfnamefont {C.}~\bibnamefont {Sumithrarachchi}},\ and\ \bibinfo {author}
  {\bibfnamefont {D.}~\bibnamefont {Tarazona}},\ }\bibfield  {title} {\bibinfo
  {title} {{Commissioning of the collinear laser spectroscopy facility BECOLA
  at NSCL/MSU}},\ }\href {https://doi.org/10.1007/s10751-014-1089-5} {\bibfield
   {journal} {\bibinfo  {journal} {Hyperf. Int.}\ }\textbf {\bibinfo {volume}
  {230}},\ \bibinfo {pages} {57--63} (\bibinfo {year} {2015})}\BibitemShut
  {NoStop}%
\bibitem [{\citenamefont {Maa{\ss}}\ \emph {et~al.}(2020)\citenamefont
  {Maa{\ss}}, \citenamefont {K{\"o}nig}, \citenamefont {Kr{\"a}mer},
  \citenamefont {Miller}, \citenamefont {Minamisono}, \citenamefont
  {N{\"o}rtersh{\"a}user},\ and\ \citenamefont {Sommer}}]{Maass.2020}%
  \BibitemOpen
  \bibfield  {author} {\bibinfo {author} {\bibfnamefont {B.}~\bibnamefont
  {Maa{\ss}}}, \bibinfo {author} {\bibfnamefont {K.}~\bibnamefont {K{\"o}nig}},
  \bibinfo {author} {\bibfnamefont {J.}~\bibnamefont {Kr{\"a}mer}}, \bibinfo
  {author} {\bibfnamefont {A.~J.}\ \bibnamefont {Miller}}, \bibinfo {author}
  {\bibfnamefont {K.}~\bibnamefont {Minamisono}}, \bibinfo {author}
  {\bibfnamefont {W.}~\bibnamefont {N{\"o}rtersh{\"a}user}},\ and\ \bibinfo
  {author} {\bibfnamefont {F.}~\bibnamefont {Sommer}},\ }\href@noop {}
  {\bibinfo {title} {A $4\pi$ fluorescence detection region for collinear laser
  spectroscopy}} (\bibinfo {year} {2020}),\ \bibinfo {note} {to be published},\
  \Eprint {https://arxiv.org/abs/2007.02658} {arXiv:2007.02658
  [physics.ins-det]} \BibitemShut {NoStop}%
\bibitem [{\citenamefont {K{\"o}nig}\ \emph
  {et~al.}(2021{\natexlab{a}})\citenamefont {K{\"o}nig}, \citenamefont
  {Minamisono}, \citenamefont {Lantis}, \citenamefont {Pineda},\ and\
  \citenamefont {Powel}}]{Konig.2021}%
  \BibitemOpen
  \bibfield  {author} {\bibinfo {author} {\bibfnamefont {K.}~\bibnamefont
  {K{\"o}nig}}, \bibinfo {author} {\bibfnamefont {K.}~\bibnamefont
  {Minamisono}}, \bibinfo {author} {\bibfnamefont {J.}~\bibnamefont {Lantis}},
  \bibinfo {author} {\bibfnamefont {S.}~\bibnamefont {Pineda}},\ and\ \bibinfo
  {author} {\bibfnamefont {R.}~\bibnamefont {Powel}},\ }\bibfield  {title}
  {\bibinfo {title} {Beam energy determination via collinear laser
  spectroscopy},\ }\href {https://doi.org/10.1103/PhysRevA.103.032806}
  {\bibfield  {journal} {\bibinfo  {journal} {Phys.\ Rev\ A}\ }\textbf
  {\bibinfo {volume} {103}},\ \bibinfo {pages} {032806} (\bibinfo {year}
  {2021}{\natexlab{a}})}\BibitemShut {NoStop}%
\bibitem [{\citenamefont {K{\"o}nig}\ \emph
  {et~al.}(2021{\natexlab{b}})\citenamefont {K{\"o}nig}, \citenamefont
  {Sommer}, \citenamefont {Lantis}, \citenamefont {Minamisono}, \citenamefont
  {N{\"o}rtersh{\"a}user}, \citenamefont {Pineda},\ and\ \citenamefont
  {Powel}}]{Konig.2021b}%
  \BibitemOpen
  \bibfield  {author} {\bibinfo {author} {\bibfnamefont {K.}~\bibnamefont
  {K{\"o}nig}}, \bibinfo {author} {\bibfnamefont {F.}~\bibnamefont {Sommer}},
  \bibinfo {author} {\bibfnamefont {J.}~\bibnamefont {Lantis}}, \bibinfo
  {author} {\bibfnamefont {K.}~\bibnamefont {Minamisono}}, \bibinfo {author}
  {\bibfnamefont {W.}~\bibnamefont {N{\"o}rtersh{\"a}user}}, \bibinfo {author}
  {\bibfnamefont {S.}~\bibnamefont {Pineda}},\ and\ \bibinfo {author}
  {\bibfnamefont {R.}~\bibnamefont {Powel}},\ }\bibfield  {title} {\bibinfo
  {title} {{Isotope-shift measurements and King-fit analysis in nickel
  isotopes}},\ }\href {https://doi.org/10.1103/PhysRevC.103.054305} {\bibfield
  {journal} {\bibinfo  {journal} {Phys.\ Rev.\ C}\ }\textbf {\bibinfo {volume}
  {103}},\ \bibinfo {pages} {054305} (\bibinfo {year}
  {2021}{\natexlab{b}})}\BibitemShut {NoStop}%
\bibitem [{\citenamefont {Hammen}\ \emph {et~al.}(2018)\citenamefont {Hammen},
  \citenamefont {N{\"o}rtersh{\"a}user}, \citenamefont {Balabanski},
  \citenamefont {Bissell}, \citenamefont {Blaum}, \citenamefont
  {Budin{\v{c}}evi{\'c}}, \citenamefont {Cheal}, \citenamefont {Flanagan},
  \citenamefont {Fr{\"o}mmgen}, \citenamefont {Georgiev}, \citenamefont
  {Geppert}, \citenamefont {Kowalska}, \citenamefont {Kreim}, \citenamefont
  {Krieger}, \citenamefont {Nazarewicz}, \citenamefont {Neugart}, \citenamefont
  {Neyens}, \citenamefont {Papuga}, \citenamefont {Reinhard}, \citenamefont
  {Rajabali}, \citenamefont {Schmidt},\ and\ \citenamefont
  {Yordanov}}]{Hammen.2018}%
  \BibitemOpen
  \bibfield  {author} {\bibinfo {author} {\bibfnamefont {M.}~\bibnamefont
  {Hammen}}, \bibinfo {author} {\bibfnamefont {W.}~\bibnamefont
  {N{\"o}rtersh{\"a}user}}, \bibinfo {author} {\bibfnamefont {D.~L.}\
  \bibnamefont {Balabanski}}, \bibinfo {author} {\bibfnamefont {M.~L.}\
  \bibnamefont {Bissell}}, \bibinfo {author} {\bibfnamefont {K.}~\bibnamefont
  {Blaum}}, \bibinfo {author} {\bibfnamefont {I.}~\bibnamefont
  {Budin{\v{c}}evi{\'c}}}, \bibinfo {author} {\bibfnamefont {B.}~\bibnamefont
  {Cheal}}, \bibinfo {author} {\bibfnamefont {K.~T.}\ \bibnamefont {Flanagan}},
  \bibinfo {author} {\bibfnamefont {N.}~\bibnamefont {Fr{\"o}mmgen}}, \bibinfo
  {author} {\bibfnamefont {G.}~\bibnamefont {Georgiev}}, \bibinfo {author}
  {\bibfnamefont {C.}~\bibnamefont {Geppert}}, \bibinfo {author} {\bibfnamefont
  {M.}~\bibnamefont {Kowalska}}, \bibinfo {author} {\bibfnamefont
  {K.}~\bibnamefont {Kreim}}, \bibinfo {author} {\bibfnamefont
  {A.}~\bibnamefont {Krieger}}, \bibinfo {author} {\bibfnamefont
  {W.}~\bibnamefont {Nazarewicz}}, \bibinfo {author} {\bibfnamefont
  {R.}~\bibnamefont {Neugart}}, \bibinfo {author} {\bibfnamefont
  {G.}~\bibnamefont {Neyens}}, \bibinfo {author} {\bibfnamefont
  {J.}~\bibnamefont {Papuga}}, \bibinfo {author} {\bibfnamefont {P.-G.}\
  \bibnamefont {Reinhard}}, \bibinfo {author} {\bibfnamefont {M.~M.}\
  \bibnamefont {Rajabali}}, \bibinfo {author} {\bibfnamefont {S.}~\bibnamefont
  {Schmidt}},\ and\ \bibinfo {author} {\bibfnamefont {D.~T.}\ \bibnamefont
  {Yordanov}},\ }\bibfield  {title} {\bibinfo {title} {{From Calcium to
  Cadmium: Testing the Pairing Functional through Charge Radii Measurements of
  \textsuperscript{100-130}Cd}},\ }\href
  {https://doi.org/10.1103/PhysRevLett.121.102501} {\bibfield  {journal}
  {\bibinfo  {journal} {Phys.\ Rev.\ Lett.}\ }\textbf {\bibinfo {volume}
  {121}},\ \bibinfo {pages} {102501} (\bibinfo {year} {2018})}\BibitemShut
  {NoStop}%
\bibitem [{\citenamefont {Fricke}\ and\ \citenamefont
  {Heilig}(2004)}]{FrickeHeilig.2004}%
  \BibitemOpen
  \bibfield  {author} {\bibinfo {author} {\bibfnamefont {G.}~\bibnamefont
  {Fricke}}\ and\ \bibinfo {author} {\bibfnamefont {K.}~\bibnamefont
  {Heilig}},\ }\bibfield  {title} {\bibinfo {title} {{Nuclear Charge Radii ·
  28-Ni Nickel}},\ }in\ \href {https://doi.org/10.1007/10856314_30} {\emph
  {\bibinfo {booktitle} {Nuclear Charge Radii: New Series}}},\ \bibinfo
  {series} {Landolt-B{\"o}rnstein - Numerical data and functional relationships
  in science and technology}, Vol.~\bibinfo {volume} {20},\ \bibinfo {editor}
  {edited by\ \bibinfo {editor} {\bibfnamefont {G.}~\bibnamefont {Fricke}},
  \bibinfo {editor} {\bibfnamefont {K.}~\bibnamefont {Heilig}}, \bibinfo
  {editor} {\bibfnamefont {H.}~\bibnamefont {Schopper}}, \bibinfo {editor}
  {\bibfnamefont {H.}~\bibnamefont {Landolt}}, \bibinfo {editor} {\bibfnamefont
  {R.}~\bibnamefont {B{\"o}rnstein}},\ and\ \bibinfo {editor} {\bibfnamefont
  {W.}~\bibnamefont {Martienssen}}}\ (\bibinfo  {publisher} {Springer},\
  \bibinfo {address} {Berlin},\ \bibinfo {year} {2004})\BibitemShut {NoStop}%
\bibitem [{\citenamefont {Steudel}\ \emph {et~al.}(1980)\citenamefont
  {Steudel}, \citenamefont {Triebe},\ and\ \citenamefont
  {Wendlandt}}]{Steudel.1980}%
  \BibitemOpen
  \bibfield  {author} {\bibinfo {author} {\bibfnamefont {A.}~\bibnamefont
  {Steudel}}, \bibinfo {author} {\bibfnamefont {U.}~\bibnamefont {Triebe}},\
  and\ \bibinfo {author} {\bibfnamefont {D.}~\bibnamefont {Wendlandt}},\
  }\bibfield  {title} {\bibinfo {title} {{Isotope shift in Ni I and changes in
  mean-square nuclear charge radii of the stable Ni isotopes}},\ }\href
  {https://doi.org/10.1007/BF01415832} {\bibfield  {journal} {\bibinfo
  {journal} {Zeitschr. f. Phys. A}\ }\textbf {\bibinfo {volume} {296}},\
  \bibinfo {pages} {189} (\bibinfo {year} {1980})}\BibitemShut {NoStop}%
\bibitem [{\citenamefont {{\"A}yst{\"o}}\ \emph {et~al.}(1984)\citenamefont
  {{\"A}yst{\"o}}, \citenamefont {{\"A}rje}, \citenamefont {Koponen},
  \citenamefont {Taskinen}, \citenamefont {Hyv{\"o}nen}, \citenamefont
  {Hautoj{\"a}rvi},\ and\ \citenamefont {Vierinen}}]{Aysto.1984}%
  \BibitemOpen
  \bibfield  {author} {\bibinfo {author} {\bibfnamefont {J.}~\bibnamefont
  {{\"A}yst{\"o}}}, \bibinfo {author} {\bibfnamefont {J.}~\bibnamefont
  {{\"A}rje}}, \bibinfo {author} {\bibfnamefont {V.}~\bibnamefont {Koponen}},
  \bibinfo {author} {\bibfnamefont {P.}~\bibnamefont {Taskinen}}, \bibinfo
  {author} {\bibfnamefont {H.}~\bibnamefont {Hyv{\"o}nen}}, \bibinfo {author}
  {\bibfnamefont {A.}~\bibnamefont {Hautoj{\"a}rvi}},\ and\ \bibinfo {author}
  {\bibfnamefont {K.}~\bibnamefont {Vierinen}},\ }\bibfield  {title} {\bibinfo
  {title} {{Beta decay of nuclides \textsuperscript{51}Fe and
  \textsuperscript{55}Ni: A new approach to on-line isotope separation}},\
  }\href {https://doi.org/10.1016/0370-2693(84)91919-1} {\bibfield  {journal}
  {\bibinfo  {journal} {Phys.\ Lett.\ B}\ }\textbf {\bibinfo {volume} {138}},\
  \bibinfo {pages} {369--372} (\bibinfo {year} {1984})}\BibitemShut {NoStop}%
\bibitem [{\citenamefont {Kaufmann}(2019)}]{Kaufmann.PhD}%
  \BibitemOpen
  \bibfield  {author} {\bibinfo {author} {\bibfnamefont {S.}~\bibnamefont
  {Kaufmann}},\ }\emph {\bibinfo {title} {Laser Spectroscopy of Nickel Isotopes
  with a new Data Acquisition System at ISOLDE}},\ \href
  {http://tuprints.ulb.tu-darmstadt.de/9286} {\bibinfo {type} {Dissertation}},\
  \bibinfo  {school} {{Technische Universit{\"a}t Darmstadt}}, \bibinfo
  {address} {Darmstadt} (\bibinfo {year} {2019})\BibitemShut {NoStop}%
\bibitem [{\citenamefont {Stone}(2019)}]{Stone.2019}%
  \BibitemOpen
  \bibfield  {author} {\bibinfo {author} {\bibfnamefont {N.~J.}\ \bibnamefont
  {Stone}},\ }\bibfield  {title} {\bibinfo {title} {{Table of Recommended
  Nuclear Magnetic Dipole Moments: Part I, Long-Lived States}},\ }\href
  {https://www-nds.iaea.org/publications/indc/indc-nds-0794/} {\bibfield
  {journal} {\bibinfo  {journal} {IAEA Nuclear Data Section, INDC(NDS)-0794}\ }
  (\bibinfo {year} {2019})}\BibitemShut {NoStop}%
\bibitem [{\citenamefont {Stone}(2020)}]{Stone.2020}%
  \BibitemOpen
  \bibfield  {author} {\bibinfo {author} {\bibfnamefont {N.~J.}\ \bibnamefont
  {Stone}},\ }\bibfield  {title} {\bibinfo {title} {{Table of Recommended
  Nuclear Magnetic Dipole Moments - Part II, Short-lived States}},\ }\href
  {https://www-nds.iaea.org/publications/indc/indc-nds-0816/} {\bibfield
  {journal} {\bibinfo  {journal} {IAEA Nuclear Data Section, INDC(NDS)-0816}\ }
  (\bibinfo {year} {2020})}\BibitemShut {NoStop}%
\bibitem [{\citenamefont {Georgiev}\ \emph {et~al.}(2002)\citenamefont
  {Georgiev}, \citenamefont {Neyens}, \citenamefont {Hass}, \citenamefont
  {Balabanski}, \citenamefont {Bingham}, \citenamefont {Borcea}, \citenamefont
  {Coulier}, \citenamefont {Coussement}, \citenamefont {Daugas}, \citenamefont
  {France}, \citenamefont {de~Oliveira~Santos}, \citenamefont {rska},
  \citenamefont {Grawe}, \citenamefont {Grzywacz}, \citenamefont {Lewitowicz},
  \citenamefont {Mach}, \citenamefont {Matea}, \citenamefont {Page},
  \citenamefont {tzner}, \citenamefont {Penionzhkevich}, \citenamefont {k},
  \citenamefont {Regan}, \citenamefont {Rykaczewski}, \citenamefont {Sawicka},
  \citenamefont {Smirnova}, \citenamefont {Sobolev}, \citenamefont {Stanoiu},
  \citenamefont {Teughels},\ and\ \citenamefont {Vyvey}}]{Georgiev.2002}%
  \BibitemOpen
  \bibfield  {author} {\bibinfo {author} {\bibfnamefont {G.}~\bibnamefont
  {Georgiev}}, \bibinfo {author} {\bibfnamefont {G.}~\bibnamefont {Neyens}},
  \bibinfo {author} {\bibfnamefont {M.}~\bibnamefont {Hass}}, \bibinfo {author}
  {\bibfnamefont {D.~L.}\ \bibnamefont {Balabanski}}, \bibinfo {author}
  {\bibfnamefont {C.}~\bibnamefont {Bingham}}, \bibinfo {author} {\bibfnamefont
  {C.}~\bibnamefont {Borcea}}, \bibinfo {author} {\bibfnamefont
  {N.}~\bibnamefont {Coulier}}, \bibinfo {author} {\bibfnamefont
  {R.}~\bibnamefont {Coussement}}, \bibinfo {author} {\bibfnamefont {J.~M.}\
  \bibnamefont {Daugas}}, \bibinfo {author} {\bibfnamefont {G.~D.}\
  \bibnamefont {France}}, \bibinfo {author} {\bibfnamefont {F.}~\bibnamefont
  {de~Oliveira~Santos}}, \bibinfo {author} {\bibfnamefont {M.~G.}\ \bibnamefont
  {rska}}, \bibinfo {author} {\bibfnamefont {H.}~\bibnamefont {Grawe}},
  \bibinfo {author} {\bibfnamefont {R.}~\bibnamefont {Grzywacz}}, \bibinfo
  {author} {\bibfnamefont {M.}~\bibnamefont {Lewitowicz}}, \bibinfo {author}
  {\bibfnamefont {H.}~\bibnamefont {Mach}}, \bibinfo {author} {\bibfnamefont
  {I.}~\bibnamefont {Matea}}, \bibinfo {author} {\bibfnamefont {R.~D.}\
  \bibnamefont {Page}}, \bibinfo {author} {\bibfnamefont {M.~P.}\ \bibnamefont
  {tzner}}, \bibinfo {author} {\bibfnamefont {Y.~E.}\ \bibnamefont
  {Penionzhkevich}}, \bibinfo {author} {\bibfnamefont {Z.~P.}\ \bibnamefont
  {k}}, \bibinfo {author} {\bibfnamefont {P.~H.}\ \bibnamefont {Regan}},
  \bibinfo {author} {\bibfnamefont {K.}~\bibnamefont {Rykaczewski}}, \bibinfo
  {author} {\bibfnamefont {M.}~\bibnamefont {Sawicka}}, \bibinfo {author}
  {\bibfnamefont {N.~A.}\ \bibnamefont {Smirnova}}, \bibinfo {author}
  {\bibfnamefont {Y.~G.}\ \bibnamefont {Sobolev}}, \bibinfo {author}
  {\bibfnamefont {M.}~\bibnamefont {Stanoiu}}, \bibinfo {author} {\bibfnamefont
  {S.}~\bibnamefont {Teughels}},\ and\ \bibinfo {author} {\bibfnamefont
  {K.}~\bibnamefont {Vyvey}},\ }\bibfield  {title} {\bibinfo {title}
  {$g$-factor measurements of $\mu$s isomeric states in neutron-rich nuclei
  around $^{68}${Ni} produced in projectile-fragmentation reactions},\ }\href
  {https://doi.org/10.1088/0954-3899/28/12/308} {\bibfield  {journal} {\bibinfo
   {journal} {J. Phys. G}\ }\textbf {\bibinfo {volume} {28}},\ \bibinfo {pages}
  {2993} (\bibinfo {year} {2002})}\BibitemShut {NoStop}%
\bibitem [{\citenamefont {{Garcia Ruiz}}\ \emph {et~al.}(2015)\citenamefont
  {{Garcia Ruiz}}, \citenamefont {Bissell}, \citenamefont {Blaum},
  \citenamefont {Fr{\"o}mmgen}, \citenamefont {Hammen}, \citenamefont {Holt},
  \citenamefont {Kowalska}, \citenamefont {Kreim}, \citenamefont
  {Men{\'e}ndez}, \citenamefont {Neugart}, \citenamefont {Neyens},
  \citenamefont {N{\"o}rtersh{\"a}user}, \citenamefont {Nowacki}, \citenamefont
  {Papuga}, \citenamefont {Poves}, \citenamefont {Schwenk}, \citenamefont
  {Simonis},\ and\ \citenamefont {Yordanov}}]{GarciaRuiz.2015}%
  \BibitemOpen
  \bibfield  {author} {\bibinfo {author} {\bibfnamefont {R.~F.}\ \bibnamefont
  {{Garcia Ruiz}}}, \bibinfo {author} {\bibfnamefont {M.~L.}\ \bibnamefont
  {Bissell}}, \bibinfo {author} {\bibfnamefont {K.}~\bibnamefont {Blaum}},
  \bibinfo {author} {\bibfnamefont {N.}~\bibnamefont {Fr{\"o}mmgen}}, \bibinfo
  {author} {\bibfnamefont {M.}~\bibnamefont {Hammen}}, \bibinfo {author}
  {\bibfnamefont {J.~D.}\ \bibnamefont {Holt}}, \bibinfo {author}
  {\bibfnamefont {M.}~\bibnamefont {Kowalska}}, \bibinfo {author}
  {\bibfnamefont {K.}~\bibnamefont {Kreim}}, \bibinfo {author} {\bibfnamefont
  {J.}~\bibnamefont {Men{\'e}ndez}}, \bibinfo {author} {\bibfnamefont
  {R.}~\bibnamefont {Neugart}}, \bibinfo {author} {\bibfnamefont
  {G.}~\bibnamefont {Neyens}}, \bibinfo {author} {\bibfnamefont
  {W.}~\bibnamefont {N{\"o}rtersh{\"a}user}}, \bibinfo {author} {\bibfnamefont
  {F.}~\bibnamefont {Nowacki}}, \bibinfo {author} {\bibfnamefont
  {J.}~\bibnamefont {Papuga}}, \bibinfo {author} {\bibfnamefont
  {A.}~\bibnamefont {Poves}}, \bibinfo {author} {\bibfnamefont
  {A.}~\bibnamefont {Schwenk}}, \bibinfo {author} {\bibfnamefont
  {J.}~\bibnamefont {Simonis}},\ and\ \bibinfo {author} {\bibfnamefont {D.~T.}\
  \bibnamefont {Yordanov}},\ }\bibfield  {title} {\bibinfo {title}
  {Ground-state electromagnetic moments of calcium isotopes},\ }\href
  {https://doi.org/10.1103/PhysRevC.91.041304} {\bibfield  {journal} {\bibinfo
  {journal} {Phys. Rev. C}\ }\textbf {\bibinfo {volume} {91}},\ \bibinfo
  {pages} {041304} (\bibinfo {year} {2015})}\BibitemShut {NoStop}%
\bibitem [{\citenamefont {Stone}(2014)}]{Stone.2014}%
  \BibitemOpen
  \bibfield  {author} {\bibinfo {author} {\bibfnamefont {N.~J.}\ \bibnamefont
  {Stone}},\ }\bibfield  {title} {\bibinfo {title} {Table of nuclear magnetic
  dipole and electric quadrupole moments},\ }\href
  {https://www-nds.iaea.org/publications/indc/indc-nds-0658.pdf} {\bibfield
  {journal} {\bibinfo  {journal} {IAEA Nuclear Data Section, INDC(NDS)-0658}\ }
  (\bibinfo {year} {2014})}\BibitemShut {NoStop}%
\bibitem [{\citenamefont {Stroberg}\ \emph {et~al.}(2017)\citenamefont
  {Stroberg}, \citenamefont {Calci}, \citenamefont {Hergert}, \citenamefont
  {Holt}, \citenamefont {Bogner}, \citenamefont {Roth},\ and\ \citenamefont
  {Schwenk}}]{Stroberg.2017}%
  \BibitemOpen
  \bibfield  {author} {\bibinfo {author} {\bibfnamefont {S.~R.}\ \bibnamefont
  {Stroberg}}, \bibinfo {author} {\bibfnamefont {A.}~\bibnamefont {Calci}},
  \bibinfo {author} {\bibfnamefont {H.}~\bibnamefont {Hergert}}, \bibinfo
  {author} {\bibfnamefont {J.~D.}\ \bibnamefont {Holt}}, \bibinfo {author}
  {\bibfnamefont {S.~K.}\ \bibnamefont {Bogner}}, \bibinfo {author}
  {\bibfnamefont {R.}~\bibnamefont {Roth}},\ and\ \bibinfo {author}
  {\bibfnamefont {A.}~\bibnamefont {Schwenk}},\ }\bibfield  {title} {\bibinfo
  {title} {{A nucleus-dependent valence-space approach to nuclear structure}},\
  }\href {https://doi.org/10.1103/PhysRevLett.118.032502} {\bibfield  {journal}
  {\bibinfo  {journal} {Phys. Rev. Lett.}\ }\textbf {\bibinfo {volume} {118}},\
  \bibinfo {pages} {032502} (\bibinfo {year} {2017})}\BibitemShut {NoStop}%
\bibitem [{\citenamefont {Stroberg}\ \emph {et~al.}(2021)\citenamefont
  {Stroberg}, \citenamefont {Holt}, \citenamefont {Schwenk},\ and\
  \citenamefont {Simonis}}]{Stroberg.limits}%
  \BibitemOpen
  \bibfield  {author} {\bibinfo {author} {\bibfnamefont {S.~R.}\ \bibnamefont
  {Stroberg}}, \bibinfo {author} {\bibfnamefont {J.~D.}\ \bibnamefont {Holt}},
  \bibinfo {author} {\bibfnamefont {A.}~\bibnamefont {Schwenk}},\ and\ \bibinfo
  {author} {\bibfnamefont {J.}~\bibnamefont {Simonis}},\ }\bibfield  {title}
  {\bibinfo {title} {{Ab Initio Limits of Atomic Nuclei}},\ }\href
  {https://doi.org/10.1103/PhysRevLett.126.022501} {\bibfield  {journal}
  {\bibinfo  {journal} {Phys. Rev. Lett.}\ }\textbf {\bibinfo {volume} {126}},\
  \bibinfo {pages} {022501} (\bibinfo {year} {2021})}\BibitemShut {NoStop}%
\bibitem [{\citenamefont {Stroberg}\ \emph {et~al.}(2019)\citenamefont
  {Stroberg}, \citenamefont {Bogner}, \citenamefont {Hergert},\ and\
  \citenamefont {Holt}}]{Stroberg.2019}%
  \BibitemOpen
  \bibfield  {author} {\bibinfo {author} {\bibfnamefont {S.~R.}\ \bibnamefont
  {Stroberg}}, \bibinfo {author} {\bibfnamefont {S.~K.}\ \bibnamefont
  {Bogner}}, \bibinfo {author} {\bibfnamefont {H.}~\bibnamefont {Hergert}},\
  and\ \bibinfo {author} {\bibfnamefont {J.~D.}\ \bibnamefont {Holt}},\
  }\bibfield  {title} {\bibinfo {title} {{Nonempirical Interactions for the
  Nuclear Shell Model: An Update}},\ }\href
  {https://doi.org/10.1146/annurev-nucl-101917-021120} {\bibfield  {journal}
  {\bibinfo  {journal} {Ann. Rev. Nucl. Part. Sci.}\ }\textbf {\bibinfo
  {volume} {69}},\ \bibinfo {pages} {307--362} (\bibinfo {year}
  {2019})}\BibitemShut {NoStop}%
\bibitem [{\citenamefont {Stroberg}()}]{Stro17imsrg++}%
  \BibitemOpen
  \bibfield  {author} {\bibinfo {author} {\bibfnamefont {S.~R.}\ \bibnamefont
  {Stroberg}},\ }\href {https://github.com/ragnarstroberg/imsrg} {\bibinfo
  {title} {https://github.com/ragnarstroberg/imsrg}}\BibitemShut {NoStop}%
\bibitem [{\citenamefont {Miyagi}\ \emph {et~al.}(2022)\citenamefont {Miyagi},
  \citenamefont {Stroberg}, \citenamefont {Navr\'atil}, \citenamefont
  {Hebeler},\ and\ \citenamefont {Holt}}]{Miyagi.2022}%
  \BibitemOpen
  \bibfield  {author} {\bibinfo {author} {\bibfnamefont {T.}~\bibnamefont
  {Miyagi}}, \bibinfo {author} {\bibfnamefont {S.~R.}\ \bibnamefont
  {Stroberg}}, \bibinfo {author} {\bibfnamefont {P.}~\bibnamefont
  {Navr\'atil}}, \bibinfo {author} {\bibfnamefont {K.}~\bibnamefont
  {Hebeler}},\ and\ \bibinfo {author} {\bibfnamefont {J.~D.}\ \bibnamefont
  {Holt}},\ }\bibfield  {title} {\bibinfo {title} {{Converged ab initio
  calculations of heavy nuclei}},\ }\href
  {https://doi.org/10.1103/PhysRevC.105.014302} {\bibfield  {journal} {\bibinfo
   {journal} {Phys.\ Rev.\ C}\ }\textbf {\bibinfo {volume} {105}},\ \bibinfo
  {pages} {014302} (\bibinfo {year} {2022})}\BibitemShut {NoStop}%
\bibitem [{\citenamefont {Jiang}\ \emph {et~al.}(2020)\citenamefont {Jiang},
  \citenamefont {Ekstr{\"{o}}m}, \citenamefont {Forss{\'{e}}n}, \citenamefont
  {Hagen}, \citenamefont {Jansen},\ and\ \citenamefont
  {Papenbrock}}]{Jiang.2020}%
  \BibitemOpen
  \bibfield  {author} {\bibinfo {author} {\bibfnamefont {W.~G.}\ \bibnamefont
  {Jiang}}, \bibinfo {author} {\bibfnamefont {A.}~\bibnamefont
  {Ekstr{\"{o}}m}}, \bibinfo {author} {\bibfnamefont {C.}~\bibnamefont
  {Forss{\'{e}}n}}, \bibinfo {author} {\bibfnamefont {G.}~\bibnamefont
  {Hagen}}, \bibinfo {author} {\bibfnamefont {G.~R.}\ \bibnamefont {Jansen}},\
  and\ \bibinfo {author} {\bibfnamefont {T.}~\bibnamefont {Papenbrock}},\
  }\bibfield  {title} {\bibinfo {title} {{Accurate bulk properties of nuclei
  from $A=2$ to $\infty$ from potentials with $\Delta$ isobars}},\ }\href
  {https://doi.org/10.1103/PhysRevC.102.054301} {\bibfield  {journal} {\bibinfo
   {journal} {Phys.\ Rev.\ C}\ }\textbf {\bibinfo {volume} {102}},\ \bibinfo
  {pages} {054301} (\bibinfo {year} {2020})}\BibitemShut {NoStop}%
\bibitem [{\citenamefont {de~Groote}\ \emph {et~al.}(2020)\citenamefont
  {de~Groote}, \citenamefont {Billowes}, \citenamefont {Binnersley},
  \citenamefont {Bissell}, \citenamefont {Cocolios}, \citenamefont {{Day
  Goodacre}}, \citenamefont {Farooq-Smith}, \citenamefont {Fedorov},
  \citenamefont {Flanagan}, \citenamefont {Franchoo}, \citenamefont {{Garcia
  Ruiz}}, \citenamefont {Gins}, \citenamefont {Holt}, \citenamefont
  {Koszor{\'u}s}, \citenamefont {Lynch}, \citenamefont {Miyagi}, \citenamefont
  {Nazarewicz}, \citenamefont {Neyens}, \citenamefont {Reinhard}, \citenamefont
  {Rothe}, \citenamefont {Stroke}, \citenamefont {Vernon}, \citenamefont
  {Wendt}, \citenamefont {Wilkins}, \citenamefont {Xu},\ and\ \citenamefont
  {Yang}}]{Groote.2020}%
  \BibitemOpen
  \bibfield  {author} {\bibinfo {author} {\bibfnamefont {R.~P.}\ \bibnamefont
  {de~Groote}}, \bibinfo {author} {\bibfnamefont {J.}~\bibnamefont {Billowes}},
  \bibinfo {author} {\bibfnamefont {C.~L.}\ \bibnamefont {Binnersley}},
  \bibinfo {author} {\bibfnamefont {M.~L.}\ \bibnamefont {Bissell}}, \bibinfo
  {author} {\bibfnamefont {T.~E.}\ \bibnamefont {Cocolios}}, \bibinfo {author}
  {\bibfnamefont {T.}~\bibnamefont {{Day Goodacre}}}, \bibinfo {author}
  {\bibfnamefont {G.~J.}\ \bibnamefont {Farooq-Smith}}, \bibinfo {author}
  {\bibfnamefont {D.~V.}\ \bibnamefont {Fedorov}}, \bibinfo {author}
  {\bibfnamefont {K.~T.}\ \bibnamefont {Flanagan}}, \bibinfo {author}
  {\bibfnamefont {S.}~\bibnamefont {Franchoo}}, \bibinfo {author}
  {\bibfnamefont {R.~F.}\ \bibnamefont {{Garcia Ruiz}}}, \bibinfo {author}
  {\bibfnamefont {W.}~\bibnamefont {Gins}}, \bibinfo {author} {\bibfnamefont
  {J.~D.}\ \bibnamefont {Holt}}, \bibinfo {author} {\bibfnamefont
  {{\'A}.}~\bibnamefont {Koszor{\'u}s}}, \bibinfo {author} {\bibfnamefont
  {K.~M.}\ \bibnamefont {Lynch}}, \bibinfo {author} {\bibfnamefont
  {T.}~\bibnamefont {Miyagi}}, \bibinfo {author} {\bibfnamefont
  {W.}~\bibnamefont {Nazarewicz}}, \bibinfo {author} {\bibfnamefont
  {G.}~\bibnamefont {Neyens}}, \bibinfo {author} {\bibfnamefont {P.-G.}\
  \bibnamefont {Reinhard}}, \bibinfo {author} {\bibfnamefont {S.}~\bibnamefont
  {Rothe}}, \bibinfo {author} {\bibfnamefont {H.~H.}\ \bibnamefont {Stroke}},
  \bibinfo {author} {\bibfnamefont {A.~R.}\ \bibnamefont {Vernon}}, \bibinfo
  {author} {\bibfnamefont {K.~D.~A.}\ \bibnamefont {Wendt}}, \bibinfo {author}
  {\bibfnamefont {S.~G.}\ \bibnamefont {Wilkins}}, \bibinfo {author}
  {\bibfnamefont {Z.~Y.}\ \bibnamefont {Xu}},\ and\ \bibinfo {author}
  {\bibfnamefont {X.~F.}\ \bibnamefont {Yang}},\ }\bibfield  {title} {\bibinfo
  {title} {Measurement and microscopic description of odd--even staggering of
  charge radii of exotic copper isotopes},\ }\href
  {https://doi.org/10.1038/s41567-020-0868-y} {\bibfield  {journal} {\bibinfo
  {journal} {Nat. Phys.}\ }\textbf {\bibinfo {volume} {16}},\ \bibinfo {pages}
  {620--624} (\bibinfo {year} {2020})}\BibitemShut {NoStop}%
\bibitem [{\citenamefont {Melendez}\ \emph {et~al.}(2019)\citenamefont
  {Melendez}, \citenamefont {Furnstahl}, \citenamefont {Phillips},
  \citenamefont {Pratola},\ and\ \citenamefont {Wesolowski}}]{Melendez.2019}%
  \BibitemOpen
  \bibfield  {author} {\bibinfo {author} {\bibfnamefont {J.~A.}\ \bibnamefont
  {Melendez}}, \bibinfo {author} {\bibfnamefont {R.~J.}\ \bibnamefont
  {Furnstahl}}, \bibinfo {author} {\bibfnamefont {D.~R.}\ \bibnamefont
  {Phillips}}, \bibinfo {author} {\bibfnamefont {M.~T.}\ \bibnamefont
  {Pratola}},\ and\ \bibinfo {author} {\bibfnamefont {S.}~\bibnamefont
  {Wesolowski}},\ }\bibfield  {title} {\bibinfo {title} {{Quantifying
  Correlated Truncation Errors in Effective Field Theory}},\ }\href
  {https://doi.org/10.1103/PhysRevC.100.044001} {\bibfield  {journal} {\bibinfo
   {journal} {Phys.\ Rev.\ C}\ }\textbf {\bibinfo {volume} {100}},\ \bibinfo
  {pages} {044001} (\bibinfo {year} {2019})}\BibitemShut {NoStop}%
\bibitem [{\citenamefont {Hu}\ \emph {et~al.}()\citenamefont {Hu},
  \citenamefont {Jiang}, \citenamefont {Miyagi}, \citenamefont {Sun},
  \citenamefont {Ekström}, \citenamefont {Forssén}, \citenamefont {Hagen},
  \citenamefont {Holt}, \citenamefont {Papenbrock}, \citenamefont {Stroberg},\
  and\ \citenamefont {Vernon}}]{Hu.2021}%
  \BibitemOpen
  \bibfield  {author} {\bibinfo {author} {\bibfnamefont {B.}~\bibnamefont
  {Hu}}, \bibinfo {author} {\bibfnamefont {W.}~\bibnamefont {Jiang}}, \bibinfo
  {author} {\bibfnamefont {T.}~\bibnamefont {Miyagi}}, \bibinfo {author}
  {\bibfnamefont {Z.}~\bibnamefont {Sun}}, \bibinfo {author} {\bibfnamefont
  {A.}~\bibnamefont {Ekström}}, \bibinfo {author} {\bibfnamefont
  {C.}~\bibnamefont {Forssén}}, \bibinfo {author} {\bibfnamefont
  {G.}~\bibnamefont {Hagen}}, \bibinfo {author} {\bibfnamefont {J.~D.}\
  \bibnamefont {Holt}}, \bibinfo {author} {\bibfnamefont {T.}~\bibnamefont
  {Papenbrock}}, \bibinfo {author} {\bibfnamefont {S.~R.}\ \bibnamefont
  {Stroberg}},\ and\ \bibinfo {author} {\bibfnamefont {I.}~\bibnamefont
  {Vernon}},\ }\bibfield  {title} {\bibinfo {title} {{Ab initio predictions
  link the neutron skin of ${}^{208}$Pb to nuclear forces}},\ }\href@noop {} {\
  }\Eprint {https://arxiv.org/abs/2112.01125} {arXiv:2112.01125 [nucl-th]}
  \BibitemShut {NoStop}%
\bibitem [{\citenamefont {Heinz}\ \emph {et~al.}(2021)\citenamefont {Heinz},
  \citenamefont {Tichai}, \citenamefont {Hoppe}, \citenamefont {Hebeler},\ and\
  \citenamefont {Schwenk}}]{Heinz.2021}%
  \BibitemOpen
  \bibfield  {author} {\bibinfo {author} {\bibfnamefont {M.}~\bibnamefont
  {Heinz}}, \bibinfo {author} {\bibfnamefont {A.}~\bibnamefont {Tichai}},
  \bibinfo {author} {\bibfnamefont {J.}~\bibnamefont {Hoppe}}, \bibinfo
  {author} {\bibfnamefont {K.}~\bibnamefont {Hebeler}},\ and\ \bibinfo {author}
  {\bibfnamefont {A.}~\bibnamefont {Schwenk}},\ }\bibfield  {title} {\bibinfo
  {title} {{In-medium similarity renormalization group with three-body
  operators}},\ }\href {https://doi.org/10.1103/PhysRevC.103.044318} {\bibfield
   {journal} {\bibinfo  {journal} {Phys.\ Rev.\ C}\ }\textbf {\bibinfo {volume}
  {103}},\ \bibinfo {pages} {044318} (\bibinfo {year} {2021})}\BibitemShut
  {NoStop}%
\bibitem [{\citenamefont {Gebrerufael}\ \emph {et~al.}(2017)\citenamefont
  {Gebrerufael}, \citenamefont {Vobig}, \citenamefont {Hergert},\ and\
  \citenamefont {Roth}}]{Gebrerufael.2016}%
  \BibitemOpen
  \bibfield  {author} {\bibinfo {author} {\bibfnamefont {E.}~\bibnamefont
  {Gebrerufael}}, \bibinfo {author} {\bibfnamefont {K.}~\bibnamefont {Vobig}},
  \bibinfo {author} {\bibfnamefont {H.}~\bibnamefont {Hergert}},\ and\ \bibinfo
  {author} {\bibfnamefont {R.}~\bibnamefont {Roth}},\ }\bibfield  {title}
  {\bibinfo {title} {{Ab Initio Description of Open-Shell Nuclei: Merging
  No-Core Shell Model and In-Medium Similarity Renormalization Group}},\ }\href
  {https://doi.org/10.1103/PhysRevLett.118.152503} {\bibfield  {journal}
  {\bibinfo  {journal} {Phys. Rev. Lett.}\ }\textbf {\bibinfo {volume} {118}},\
  \bibinfo {pages} {152503} (\bibinfo {year} {2017})}\BibitemShut {NoStop}%
\bibitem [{\citenamefont {Stumpf}\ \emph {et~al.}(2016)\citenamefont {Stumpf},
  \citenamefont {Braun},\ and\ \citenamefont {Roth}}]{Stumpf.2015}%
  \BibitemOpen
  \bibfield  {author} {\bibinfo {author} {\bibfnamefont {C.}~\bibnamefont
  {Stumpf}}, \bibinfo {author} {\bibfnamefont {J.}~\bibnamefont {Braun}},\ and\
  \bibinfo {author} {\bibfnamefont {R.}~\bibnamefont {Roth}},\ }\bibfield
  {title} {\bibinfo {title} {{Importance-Truncated Large-Scale Shell Model}},\
  }\href {https://doi.org/10.1103/PhysRevC.93.021301} {\bibfield  {journal}
  {\bibinfo  {journal} {Phys.\ Rev.\ C}\ }\textbf {\bibinfo {volume} {93}},\
  \bibinfo {pages} {021301} (\bibinfo {year} {2016})}\BibitemShut {NoStop}%
\bibitem [{\citenamefont {Tichai}\ \emph {et~al.}(2019)\citenamefont {Tichai},
  \citenamefont {M\"uller}, \citenamefont {Vobig},\ and\ \citenamefont
  {Roth}}]{Tichai.2018}%
  \BibitemOpen
  \bibfield  {author} {\bibinfo {author} {\bibfnamefont {A.}~\bibnamefont
  {Tichai}}, \bibinfo {author} {\bibfnamefont {J.}~\bibnamefont {M\"uller}},
  \bibinfo {author} {\bibfnamefont {K.}~\bibnamefont {Vobig}},\ and\ \bibinfo
  {author} {\bibfnamefont {R.}~\bibnamefont {Roth}},\ }\bibfield  {title}
  {\bibinfo {title} {{Natural orbitals for ab initio no-core shell model
  calculations}},\ }\href {https://doi.org/10.1103/PhysRevC.99.034321}
  {\bibfield  {journal} {\bibinfo  {journal} {Phys.\ Rev.\ C}\ }\textbf
  {\bibinfo {volume} {99}},\ \bibinfo {pages} {034321} (\bibinfo {year}
  {2019})}\BibitemShut {NoStop}%
\bibitem [{\citenamefont {Kl{\"{u}}pfel}\ \emph {et~al.}(2009)\citenamefont
  {Kl{\"{u}}pfel}, \citenamefont {Reinhard}, \citenamefont {B{\"{u}}rvenich},\
  and\ \citenamefont {Maruhn}}]{Kluepfel2009}%
  \BibitemOpen
  \bibfield  {author} {\bibinfo {author} {\bibfnamefont {P.}~\bibnamefont
  {Kl{\"{u}}pfel}}, \bibinfo {author} {\bibfnamefont {P.-G.}\ \bibnamefont
  {Reinhard}}, \bibinfo {author} {\bibfnamefont {T.~J.}\ \bibnamefont
  {B{\"{u}}rvenich}},\ and\ \bibinfo {author} {\bibfnamefont {J.~A.}\
  \bibnamefont {Maruhn}},\ }\bibfield  {title} {\bibinfo {title} {Variations on
  a theme by {Skyrme}: A systematic study of adjustments of model parameters},\
  }\href {https://doi.org/10.1103/PhysRevC.79.034310} {\bibfield  {journal}
  {\bibinfo  {journal} {Phys.\ Rev.\ C}\ }\textbf {\bibinfo {volume} {79}},\
  \bibinfo {pages} {034310} (\bibinfo {year} {2009})}\BibitemShut {NoStop}%
\bibitem [{\citenamefont {Miller}\ \emph
  {et~al.}(2019{\natexlab{b}})\citenamefont {Miller}, \citenamefont
  {Minamisono}, \citenamefont {Klose} \emph {et~al.}}]{Miller2019}%
  \BibitemOpen
  \bibfield  {author} {\bibinfo {author} {\bibfnamefont {A.~J.}\ \bibnamefont
  {Miller}}, \bibinfo {author} {\bibfnamefont {K.}~\bibnamefont {Minamisono}},
  \bibinfo {author} {\bibfnamefont {A.}~\bibnamefont {Klose}}, \emph {et~al.},\
  }\bibfield  {title} {\bibinfo {title} {Proton superfluidity and charge radii
  in proton-rich calcium isotopes},\ }\href
  {https://doi.org/10.1038/s41567-019-0416-9} {\bibfield  {journal} {\bibinfo
  {journal} {Nat. Phys.}\ }\textbf {\bibinfo {volume} {15}},\ \bibinfo {pages}
  {432} (\bibinfo {year} {2019}{\natexlab{b}})}\BibitemShut {NoStop}%
\bibitem [{\citenamefont {Reinhard}\ and\ \citenamefont
  {Nazarewicz}(2022)}]{ReinhardNazarewicz2022}%
  \BibitemOpen
  \bibfield  {author} {\bibinfo {author} {\bibfnamefont {P.-G.}\ \bibnamefont
  {Reinhard}}\ and\ \bibinfo {author} {\bibfnamefont {W.}~\bibnamefont
  {Nazarewicz}},\ }\bibfield  {title} {\bibinfo {title} {Information content of
  the differences in the charge radii of mirror nuclei},\ }\href
  {https://doi.org/10.1103/PhysRevC.105.L021301} {\bibfield  {journal}
  {\bibinfo  {journal} {Phys. Rev. C}\ }\textbf {\bibinfo {volume} {105}},\
  \bibinfo {pages} {L021301} (\bibinfo {year} {2022})}\BibitemShut {NoStop}%
\bibitem [{\citenamefont {Dobaczewski}\ \emph {et~al.}(2014)\citenamefont
  {Dobaczewski}, \citenamefont {Nazarewicz},\ and\ \citenamefont
  {Reinhard}}]{Dobaczewski2014}%
  \BibitemOpen
  \bibfield  {author} {\bibinfo {author} {\bibfnamefont {J.}~\bibnamefont
  {Dobaczewski}}, \bibinfo {author} {\bibfnamefont {W.}~\bibnamefont
  {Nazarewicz}},\ and\ \bibinfo {author} {\bibfnamefont {P.-G.}\ \bibnamefont
  {Reinhard}},\ }\bibfield  {title} {\bibinfo {title} {Error estimates of
  theoretical models: a guide},\ }\href
  {http://dx.doi.org/10.1088/0954-3899/41/7/074001} {\bibfield  {journal}
  {\bibinfo  {journal} {J. Phys. G}\ }\textbf {\bibinfo {volume} {41}},\
  \bibinfo {pages} {074001} (\bibinfo {year} {2014})}\BibitemShut {NoStop}%
\bibitem [{\citenamefont {Kl\"upfel}\ \emph {et~al.}(2008)\citenamefont
  {Kl\"upfel}, \citenamefont {Erler}, \citenamefont {Reinhard},\ and\
  \citenamefont {Maruhn}}]{Kluepfel2008}%
  \BibitemOpen
  \bibfield  {author} {\bibinfo {author} {\bibfnamefont {P.}~\bibnamefont
  {Kl\"upfel}}, \bibinfo {author} {\bibfnamefont {J.}~\bibnamefont {Erler}},
  \bibinfo {author} {\bibfnamefont {P.-G.}\ \bibnamefont {Reinhard}},\ and\
  \bibinfo {author} {\bibfnamefont {J.~A.}\ \bibnamefont {Maruhn}},\ }\bibfield
   {title} {\bibinfo {title} {Systematics of collective correlation energies
  from self-consistent mean-field calculations},\ }\href
  {http://dx.doi.org/10.1140/epja/i2008-10633-3} {\bibfield  {journal}
  {\bibinfo  {journal} {Eur. Phys. J A}\ }\textbf {\bibinfo {volume} {37}},\
  \bibinfo {pages} {343} (\bibinfo {year} {2008})}\BibitemShut {NoStop}%
\end{thebibliography}%


\begin{thebibliography}{6}%
\makeatletter
\providecommand \@ifxundefined [1]{%
 \@ifx{#1\undefined}
}%
\providecommand \@ifnum [1]{%
 \ifnum #1\expandafter \@firstoftwo
 \else \expandafter \@secondoftwo
 \fi
}%
\providecommand \@ifx [1]{%
 \ifx #1\expandafter \@firstoftwo
 \else \expandafter \@secondoftwo
 \fi
}%
\providecommand \natexlab [1]{#1}%
\providecommand \enquote  [1]{``#1''}%
\providecommand \bibnamefont  [1]{#1}%
\providecommand \bibfnamefont [1]{#1}%
\providecommand \citenamefont [1]{#1}%
\providecommand \href@noop [0]{\@secondoftwo}%
\providecommand \href [0]{\begingroup \@sanitize@url \@href}%
\providecommand \@href[1]{\@@startlink{#1}\@@href}%
\providecommand \@@href[1]{\endgroup#1\@@endlink}%
\providecommand \@sanitize@url [0]{\catcode `\\12\catcode `\$12\catcode
  `\&12\catcode `\#12\catcode `\^12\catcode `\_12\catcode `\%12\relax}%
\providecommand \@@startlink[1]{}%
\providecommand \@@endlink[0]{}%
\providecommand \url  [0]{\begingroup\@sanitize@url \@url }%
\providecommand \@url [1]{\endgroup\@href {#1}{\urlprefix }}%
\providecommand \urlprefix  [0]{URL }%
\providecommand \Eprint [0]{\href }%
\providecommand \doibase [0]{https://doi.org/}%
\providecommand \selectlanguage [0]{\@gobble}%
\providecommand \bibinfo  [0]{\@secondoftwo}%
\providecommand \bibfield  [0]{\@secondoftwo}%
\providecommand \translation [1]{[#1]}%
\providecommand \BibitemOpen [0]{}%
\providecommand \bibitemStop [0]{}%
\providecommand \bibitemNoStop [0]{.\EOS\space}%
\providecommand \EOS [0]{\spacefactor3000\relax}%
\providecommand \BibitemShut  [1]{\csname bibitem#1\endcsname}%
\let\auto@bib@innerbib\@empty
\bibitem [{\citenamefont {Kaufmann}(2019)}]{Kaufmann.PhD}%
  \BibitemOpen
  \bibfield  {author} {\bibinfo {author} {\bibfnamefont {S.}~\bibnamefont
  {Kaufmann}},\ }\emph {\bibinfo {title} {Laser Spectroscopy of Nickel Isotopes
  with a new Data Acquisition System at ISOLDE}},\ \href
  {http://tuprints.ulb.tu-darmstadt.de/9286} {\bibinfo {type} {Dissertation}},\
  \bibinfo  {school} {{Technische Universit{\"a}t Darmstadt}}, \bibinfo
  {address} {Darmstadt} (\bibinfo {year} {2019})\BibitemShut {NoStop}%
\bibitem [{\citenamefont {Stone}(2019)}]{Stone.2019}%
  \BibitemOpen
  \bibfield  {author} {\bibinfo {author} {\bibfnamefont {N.~J.}\ \bibnamefont
  {Stone}},\ }\bibfield  {title} {\bibinfo {title} {{Table of Recommended
  Nuclear Magnetic Dipole Moments: Part I, Long-Lived States}},\ }\href
  {https://www-nds.iaea.org/publications/indc/indc-nds-0794/} {\bibfield
  {journal} {\bibinfo  {journal} {IAEA Nuclear Data Section, INDC(NDS)-0794}\ }
  (\bibinfo {year} {2019})}\BibitemShut {NoStop}%
\bibitem [{\citenamefont {Berryman}\ \emph {et~al.}(2009)\citenamefont
  {Berryman}, \citenamefont {Minamisono}, \citenamefont {Rogers}, \citenamefont
  {Brown}, \citenamefont {Crawford}, \citenamefont {Grinyer}, \citenamefont
  {Mantica}, \citenamefont {Stoker},\ and\ \citenamefont
  {Towner}}]{Berryman.2009}%
  \BibitemOpen
  \bibfield  {author} {\bibinfo {author} {\bibfnamefont {J.~S.}\ \bibnamefont
  {Berryman}}, \bibinfo {author} {\bibfnamefont {K.}~\bibnamefont
  {Minamisono}}, \bibinfo {author} {\bibfnamefont {W.~F.}\ \bibnamefont
  {Rogers}}, \bibinfo {author} {\bibfnamefont {B.~A.}\ \bibnamefont {Brown}},
  \bibinfo {author} {\bibfnamefont {H.~L.}\ \bibnamefont {Crawford}}, \bibinfo
  {author} {\bibfnamefont {G.~F.}\ \bibnamefont {Grinyer}}, \bibinfo {author}
  {\bibfnamefont {P.~F.}\ \bibnamefont {Mantica}}, \bibinfo {author}
  {\bibfnamefont {J.~B.}\ \bibnamefont {Stoker}},\ and\ \bibinfo {author}
  {\bibfnamefont {I.~S.}\ \bibnamefont {Towner}},\ }\bibfield  {title}
  {\bibinfo {title} {{Doubly-magic nature of \textsuperscript{56}Ni :
  Measurement of the ground state nuclear magnetic dipole moment of
  \textsuperscript{55}Ni}},\ }\href
  {https://doi.org/10.1103/PhysRevC.79.064305} {\bibfield  {journal} {\bibinfo
  {journal} {Phys.\ Rev.\ C}\ }\textbf {\bibinfo {volume} {79}},\ \bibinfo
  {pages} {064305} (\bibinfo {year} {2009})}\BibitemShut {NoStop}%
\bibitem [{\citenamefont {Stone}(2020)}]{Stone.2020}%
  \BibitemOpen
  \bibfield  {author} {\bibinfo {author} {\bibfnamefont {N.~J.}\ \bibnamefont
  {Stone}},\ }\bibfield  {title} {\bibinfo {title} {{Table of Recommended
  Nuclear Magnetic Dipole Moments - Part II, Short-lived States}},\ }\href
  {https://www-nds.iaea.org/publications/indc/indc-nds-0816/} {\bibfield
  {journal} {\bibinfo  {journal} {IAEA Nuclear Data Section, INDC(NDS)-0816}\ }
  (\bibinfo {year} {2020})}\BibitemShut {NoStop}%
\bibitem [{\citenamefont {Georgiev}\ \emph {et~al.}(2002)\citenamefont
  {Georgiev}, \citenamefont {Neyens}, \citenamefont {Hass}, \citenamefont
  {Balabanski}, \citenamefont {Bingham}, \citenamefont {Borcea}, \citenamefont
  {Coulier}, \citenamefont {Coussement}, \citenamefont {Daugas}, \citenamefont
  {France}, \citenamefont {de~Oliveira~Santos}, \citenamefont {rska},
  \citenamefont {Grawe}, \citenamefont {Grzywacz}, \citenamefont {Lewitowicz},
  \citenamefont {Mach}, \citenamefont {Matea}, \citenamefont {Page},
  \citenamefont {tzner}, \citenamefont {Penionzhkevich}, \citenamefont {k},
  \citenamefont {Regan}, \citenamefont {Rykaczewski}, \citenamefont {Sawicka},
  \citenamefont {Smirnova}, \citenamefont {Sobolev}, \citenamefont {Stanoiu},
  \citenamefont {Teughels},\ and\ \citenamefont {Vyvey}}]{Georgiev.2002}%
  \BibitemOpen
  \bibfield  {author} {\bibinfo {author} {\bibfnamefont {G.}~\bibnamefont
  {Georgiev}}, \bibinfo {author} {\bibfnamefont {G.}~\bibnamefont {Neyens}},
  \bibinfo {author} {\bibfnamefont {M.}~\bibnamefont {Hass}}, \bibinfo {author}
  {\bibfnamefont {D.~L.}\ \bibnamefont {Balabanski}}, \bibinfo {author}
  {\bibfnamefont {C.}~\bibnamefont {Bingham}}, \bibinfo {author} {\bibfnamefont
  {C.}~\bibnamefont {Borcea}}, \bibinfo {author} {\bibfnamefont
  {N.}~\bibnamefont {Coulier}}, \bibinfo {author} {\bibfnamefont
  {R.}~\bibnamefont {Coussement}}, \bibinfo {author} {\bibfnamefont {J.~M.}\
  \bibnamefont {Daugas}}, \bibinfo {author} {\bibfnamefont {G.~D.}\
  \bibnamefont {France}}, \bibinfo {author} {\bibfnamefont {F.}~\bibnamefont
  {de~Oliveira~Santos}}, \bibinfo {author} {\bibfnamefont {M.~G.}\ \bibnamefont
  {rska}}, \bibinfo {author} {\bibfnamefont {H.}~\bibnamefont {Grawe}},
  \bibinfo {author} {\bibfnamefont {R.}~\bibnamefont {Grzywacz}}, \bibinfo
  {author} {\bibfnamefont {M.}~\bibnamefont {Lewitowicz}}, \bibinfo {author}
  {\bibfnamefont {H.}~\bibnamefont {Mach}}, \bibinfo {author} {\bibfnamefont
  {I.}~\bibnamefont {Matea}}, \bibinfo {author} {\bibfnamefont {R.~D.}\
  \bibnamefont {Page}}, \bibinfo {author} {\bibfnamefont {M.~P.}\ \bibnamefont
  {tzner}}, \bibinfo {author} {\bibfnamefont {Y.~E.}\ \bibnamefont
  {Penionzhkevich}}, \bibinfo {author} {\bibfnamefont {Z.~P.}\ \bibnamefont
  {k}}, \bibinfo {author} {\bibfnamefont {P.~H.}\ \bibnamefont {Regan}},
  \bibinfo {author} {\bibfnamefont {K.}~\bibnamefont {Rykaczewski}}, \bibinfo
  {author} {\bibfnamefont {M.}~\bibnamefont {Sawicka}}, \bibinfo {author}
  {\bibfnamefont {N.~A.}\ \bibnamefont {Smirnova}}, \bibinfo {author}
  {\bibfnamefont {Y.~G.}\ \bibnamefont {Sobolev}}, \bibinfo {author}
  {\bibfnamefont {M.}~\bibnamefont {Stanoiu}}, \bibinfo {author} {\bibfnamefont
  {S.}~\bibnamefont {Teughels}},\ and\ \bibinfo {author} {\bibfnamefont
  {K.}~\bibnamefont {Vyvey}},\ }\bibfield  {title} {\bibinfo {title}
  {$g$-factor measurements of $\mu$s isomeric states in neutron-rich nuclei
  around $^{68}${Ni} produced in projectile-fragmentation reactions},\ }\href
  {https://doi.org/10.1088/0954-3899/28/12/308} {\bibfield  {journal} {\bibinfo
   {journal} {J. Phys. G}\ }\textbf {\bibinfo {volume} {28}},\ \bibinfo {pages}
  {2993} (\bibinfo {year} {2002})}\BibitemShut {NoStop}%
\bibitem [{\citenamefont {Honma}\ \emph {et~al.}(2004)\citenamefont {Honma},
  \citenamefont {Otsuka}, \citenamefont {Brown},\ and\ \citenamefont
  {Mizusaki}}]{Honma.2004}%
  \BibitemOpen
  \bibfield  {author} {\bibinfo {author} {\bibfnamefont {M.}~\bibnamefont
  {Honma}}, \bibinfo {author} {\bibfnamefont {T.}~\bibnamefont {Otsuka}},
  \bibinfo {author} {\bibfnamefont {B.~A.}\ \bibnamefont {Brown}},\ and\
  \bibinfo {author} {\bibfnamefont {T.}~\bibnamefont {Mizusaki}},\ }\bibfield
  {title} {\bibinfo {title} {New effective interaction for pf -shell nuclei and
  its implications for the stability of the {$N=Z=28$} closed core},\ }\href
  {https://doi.org/10.1103/PhysRevC.69.034335} {\bibfield  {journal} {\bibinfo
  {journal} {Phys.\ Rev.\ C}\ }\textbf {\bibinfo {volume} {69}},\ \bibinfo
  {pages} {034335} (\bibinfo {year} {2004})}\BibitemShut {NoStop}%
\end{thebibliography}%

\end{document}